\documentclass[12pt]{article}
\pdfoutput=1

\usepackage{draft,hyperref,tikz,float,subfig,cite}

\usetikzlibrary{snakes}
\usetikzlibrary{shapes.misc}

\begin{document}

\begin{titlepage}

\preprint{CALT-TH-2019-009}

\begin{center}

\hfill \\
\hfill \\
\vskip 1cm

\title{
Anomalies and Bounds on Charged Operators
}

\author{Ying-Hsuan Lin$^{a}$ and Shu-Heng Shao$^b$
}

\address{${}^a$Walter Burke Institute for Theoretical Physics,\\ California Institute of Technology,
Pasadena, CA 91125, USA}
\address{${}^b$School of Natural Sciences, Institute for Advanced Study,\\
Princeton, NJ 08540, USA}

\email{yhlin@caltech.edu, shao@ias.edu}

\end{center}

\vfill

\abstract{
We study the implications of 't Hooft anomaly ({\it i.e.} obstruction to gauging) on conformal field theory, focusing on the case when the global symmetry is $\bZ_2$.   
Using the modular bootstrap, universal bounds on (1+1)-dimensional bosonic conformal field theories with an internal $\bZ_2$ global symmetry are derived. 
The bootstrap bounds depend dramatically on the 't Hooft anomaly. 
In particular, there is a universal upper bound on the lightest $\bZ_2$ odd operator if the symmetry is anomalous, but there is no bound if the symmetry is non-anomalous. 
In the non-anomalous case, we find that the  lightest $\bZ_2$ odd state  and the defect  ground state cannot both be arbitrarily heavy. 
We also consider theories with a  $U(1)$ global symmetry, and comment that there is no bound on the lightest $U(1)$ charged operator if the symmetry is non-anomalous.
}

\vfill

\end{titlepage}

\eject

\tableofcontents

\section{Introduction and Summary of Results}

It is well known that 't Hooft anomalies ({\it i.e.} obstruction to gauging) of global symmetries have dramatic consequences on the gapped phases of quantum systems.\footnote{Throughout this paper, the term ``anomaly" will always refer to the 't Hooft anomaly of a global symmetry.  We emphasize that  a   symmetry with 't Hooft anomaly is still  a true global symmetry, but there is an obstruction to gauging it.  This is to be contrasted with a different but related concept, the Adler-Bell-Jackiw anomaly \cite{Bell:1969ts,Adler:1969gk}, where the axial ``symmetry" is not a true global symmetry because the associated current is not conserved.}  
For example, a non-trivial 't Hooft anomaly implies that, in a gapped phase, the symmetry must either be spontaneously broken, or there is a topological quantum field theory (TQFT) matching the anomaly. 
In this paper, we study the constraints from 't Hooft anomalies on the gapless phases of quantum systems -- described by conformal field theory (CFT) -- employing the techniques of the conformal bootstrap   (see \cite{Rychkov:2016iqz,Simmons-Duffin:2016gjk,Poland:2016chs,Poland:2018epd} for reviews).

\begin{figure}
\centering
\includegraphics[width=.5\textwidth]{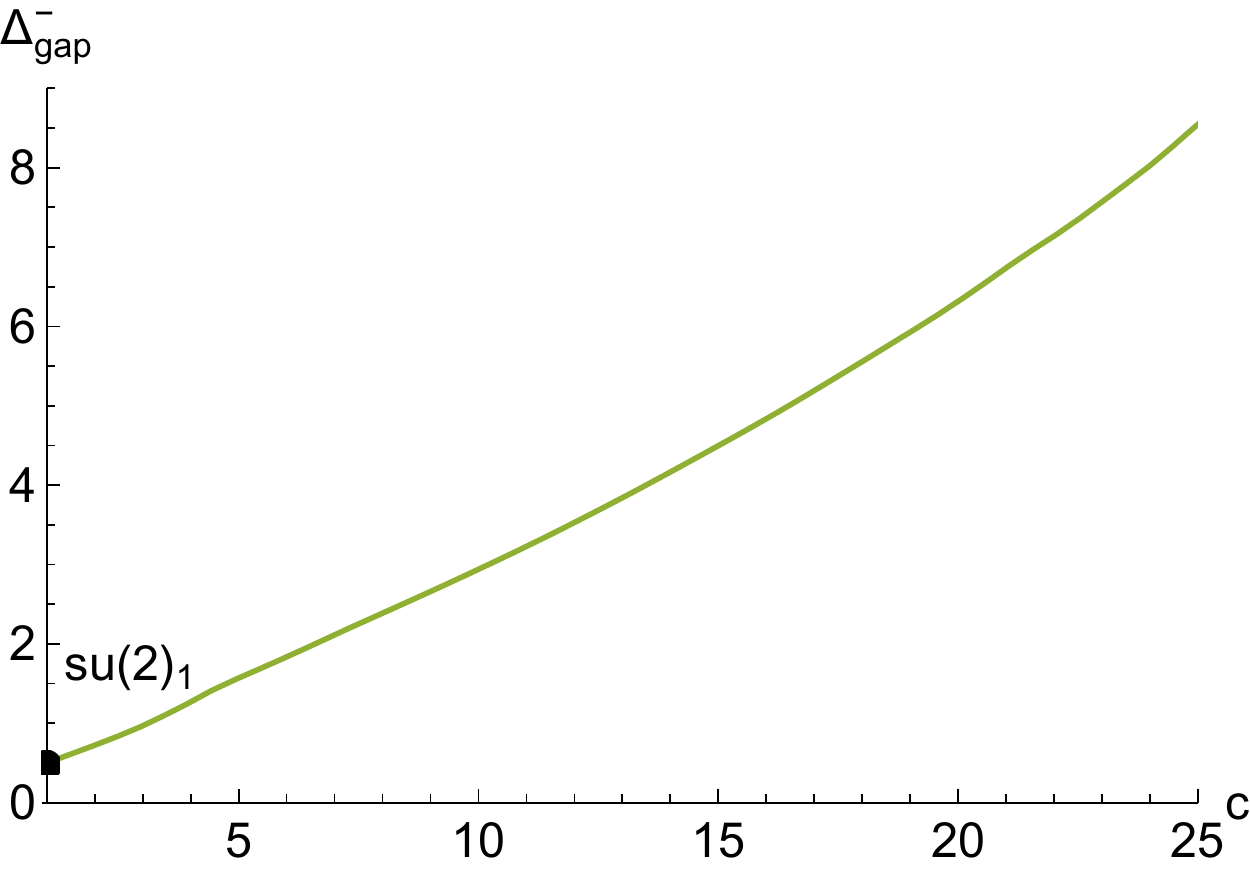}
\caption{Upper bound on the lightest $\bZ_2$ odd operator in a 2d CFT with an anomalous $\bZ_2$ symmetry, as a function of the central charge $c$ for $c \ge 1$. The region below the curve is allowed. The $\widehat{\mathfrak{su}(2)}_1$ WZW model with $\Delta_\text{gap}^- = {1\over2}$ saturates the bound at $c=1$.}
\label{Fig:IntroOdd}
\end{figure}

An intrinsic set of observables in a CFT are the local operators and their correlation functions.  
Using the operator-state map, the local operators are in one-to-one correspondence with the states in the Hilbert space ${\cal H}$ quantized on the sphere. 
The scaling dimension $\Delta$ of the local operator is mapped to the energy of the state on the sphere, whose finite size renders the spectrum discrete. 
Does the 't Hooft anomaly constrain the spectrum of local operators in any way? 
Specifically, given a global symmetry $G$ with 't Hooft anomaly $\A$, we ask: 
\begin{enumerate}
\item Is there a universal upper bound on the scaling dimension $\Delta$ of the lightest $G$-charged local operator?
\item How does the bound, if exists, depend on the 't Hooft anomaly $\A$?
\end{enumerate}
We approach these general questions from the simplest possible setup.  
We consider a bosonic, unitary CFT in (1+1) spacetime dimensions with an internal, unitary $\bZ_2$ global symmetry, either with or without 't Hooft anomaly.\footnote{We use (1+1)d and 2d interchangeably.}
We find that the bound depends dramatically on the 't Hooft anomaly of the global symmetry.  Our key finding is that:
\begin{itemize}
\item There is a universal upper bound on the scaling dimension $\Delta$ of the lightest $\bZ_2$ odd operator if the $\bZ_2$ is anomalous, but not otherwise.
\end{itemize}
This result is another manifestation of the moral that an anomalous global symmetry is harder to ``hide" in the infrared: it either implies the vacuum cannot be trivially gapped in a gapped phase, or it constrains the light charged operator spectrum in a gapless phase.\footnote{The scaling dimension of the lightest non-vacuum operator in a given sector of the CFT is referred to as the ``gap" in that sector. Equivalently, this is the gap in the Hilbert space quantized on a spatial circle, whose finite size renders the spectrum discrete. The gapless phase of the system is described by the CFT on a real line $\mathbb{R}$, where the gap vanishes.}    
This universal upper bound for an anomalous $\bZ_2$ is shown in Figure~\ref{Fig:IntroOdd}.

We argue that the same relation between the existence of a universal bound and the 't Hooft anomaly also holds true if the symmetry group is $U(1)$. 
Indeed, previous universal bounds for the lightest $U(1)$ charged operator in the literature \cite{Benjamin:2016fhe,Montero:2016tif,Bae:2018qym} are restricted to $U(1)$ global symmetries generated by  {\it holomorphic currents}, which are always anomalous ({\it i.e.} cannot be gauged). 
However, there are more general $U(1)$ global symmetries that are not generated by holomorphic currents ({\it e.g.} the momentum and the winding symmetry in the free compact boson theory), and they can be non-anomalous. 
For any such non-anomalous $U(1)$, we point out that there is {\it no} bound on the lightest $U(1)$ charged operator. 
We further discuss the interpretation of our bounds from the weak gravity conjecture \cite{ArkaniHamed:2006dz} in AdS$_3$/CFT$_2$.

Below, we provide an overview of our methods for deriving the above universal bounds.

\subsubsection*{Topological Defect Lines and Anomalies}

An invariant way to characterize a global symmetry and its 't Hooft anomaly is by the associated (invertible) {\it topological defect lines} $\cal L$ \cite{Verlinde:1988sn,Petkova:2000ip,Fuchs:2002cm,Frohlich:2004ef,Fuchs:2003id,Fuchs:2004dz,Fuchs:2004xi,Fjelstad:2005ua,Frohlich:2006ch,Runkel:2005qw,Fuchs:2007vk,Frohlich:2009gb,Davydov:2010rm}.  See \cite{Bhardwaj:2017xup,Chang:2018iay} for modern applications of topological defect lines to renormalization group flows and gauging.\footnote{In this paper, we focus on invertible topological defect lines, which are associated to global symmetries. There are also non-invertible (``non-symmetry") topological defect lines that  have interesting consequences on the dynamics of quantum field theory (QFT) under renormalization group (RG) flows \cite{Chang:2018iay}.
}
These are extended objects in quantum field theory whose contraction of a loop around a local operator $\phi(x)$ implements the symmetry transformation. 
In the case  of a $U(1)$ continuous global symmetry  that is associated to a conserved Noether current $J_\mu(x)$, the contour integral of the latter along a curve $\cal L$ 
defines a continuous  family of topological defect lines $e^{i \theta \int_{\cal L} ds^\mu J_\mu}$ labeled by an angle $\theta$.  
 The topological property of the line follows from the conservation of the current $J_\mu$. 

The topological defect lines obey a fusion relation that is simply the group multiplication law of the associated global symmetry. 
Furthermore, the locality property of the topological defect lines implies that they obey crossing relations, such as the one depicted in Figure~\ref{fig:crossing}. The more general structure of the crossing relations is described by the mathematics of fusion categories \cite{Etingof:aa,etingof2016tensor}. The 't Hooft anomalies of the global symmetry are encoded in and classified by the crossing relations.

In the presence of a global symmetry, a theory can be quantized with twisted boundary conditions on a spatial circle. 
The twisting can be understood as the insertion of a topological defect line associated to the global symmetry. 
This defines a Hilbert space which will be called the {\it defect Hilbert space} ${\cal H}_{\cal L}$.   
The 't Hooft anomaly constrains the spin content of ${\cal H}_{\cal L}$ as follows:
\begin{align}
s \in \begin{cases}
& {\bZ\over2}\,,~~~~~~~~~~~(\text{non-anomalous}~~\bZ_2)\,,\\
&\frac 14 + {\bZ\over2}\,,~~~~~~(\text{anomalous}~~\bZ_2)\,,
\end{cases}
\end{align}
Note that even though we start with a bosonic CFT, there are anyonic or fermionic operators living at the end of the $\bZ_2$ line, depending on whether the $\bZ_2$ is anomalous or not. 
In the non-anomalous case, this is analogous to the emergent fermionic excitations of lattice spin models \cite{PhysRevB.67.245316}, and  we will  review  a general CFT derivation in \cite{Chang:2018iay}. 
This spin selection rule in turn constrains the light operator spectrum in the Hilbert space of local operators $\cal H$ via modular transformations, as we will discuss below.

\subsubsection*{Modular Bootstrap}

Our method for deriving universal bounds is by exploiting the general consistency condition of 2d CFTs on a torus.  
In particular, the invariance the torus partition function $Z(\tau,\bar \tau)$ under modular transformations puts strong constraints on the operator content of the theory. 
 For example, by considering a  high/low temperature limit of the partition function, Cardy famously derived a universal formula for the asymptotic density of heavy local operators \cite{Cardy:1986ie}. 
Extending this success, the modern modular bootstrap program has been developed to study, among others, a medium temperature expansion of the torus partition function and its consistency with modular invariance. 
It generalizes the Cardy constraints on the heavy operators to, in particular, the gap and degeneracies in the spectrum in any 2d CFT \cite{Hellerman:2009bu,Friedan:2013cba,Qualls:2013eha,Collier:2016cls,Collier:2017shs,Anous:2018hjh,Afkhami-Jeddi:2019zci}.
It has been proven that the lightest primary operator above the vacuum is universally bounded from above by ${c \over 6} + 0.474$ for all $c>1$ CFTs \cite{Hellerman:2009bu}.
In the large $c$ limit, this bound has recently been improved to ${c \over 9.1} + {\cal O}(1)$ \cite{Afkhami-Jeddi:2019zci}.

In the presence of a $\bZ_2$ topological defect line $\cal L$, we can consider torus partition functions with $\cal L$ extending along the time or spatial direction, which we denote as $Z_{\cal L}(\tau,\bar\tau)$ and $Z^{\cal L}(\tau,\bar\tau)$, respectively. They admit the following interpretation as sums over different Hilbert spaces:
\ie
Z_{\cal L}  (\tau,\bar\tau) = \text{Tr}_{{\cal H}_{\cal L}}[q^{L_0 -c/24}\bar q^{\bar L_0-c/24}]\,,~~~~
Z^{\cal L}  (\tau,\bar\tau) = \text{Tr}_{{\cal H}}[\, \widehat{\cal L}\,q^{L_0 -c/24}\bar q^{\bar L_0-c/24}]\,,
\fe
where $\widehat{\cal L}$ is the generator of the $\bZ_2$.  The modular crossing equation we are to explore is
\ie\label{introcrossing}
\text{Crossing:}~~~~&Z(-1/\tau ,-1/\bar\tau) = Z(\tau,\bar\tau)\,,\\
~~~&Z_{\cal L} (-1/\tau ,-1/\bar\tau)  =  Z^{\cal L}(\tau,\bar\tau) \,.
\fe
On the other hand, the positivity statement is that the expansions of $Z^\pm (\tau,\bar\tau)$ and $Z_{\cal L}(\tau,\bar\tau)$ in Virasoro characters $\chi_h(\tau)\chi_{\bar h}(\bar\tau)$ have non-negative coefficients:
\ie
&Z^\pm (\tau,\bar\tau) \equiv \frac 12 [Z(\tau,\bar\tau)\pm Z^{\cal L}(\tau,\bar\tau)] = \sum_{(h,\bar h )\in {\cal H}} n_{h,\bar h}^\pm \,\chi_h(\tau)\chi_{\bar h}(\bar\tau)\,,\\
&Z_{\cal L}(\tau,\bar\tau) =  \sum_{(h,\bar h) \in {\cal H}_{\cal L}}  (n_{\cal L})_{h,\bar h}\,  \chi_h(\tau)\chi_{\bar h}(\bar\tau)\,,\\
\text{Positivity:}~~~~ & n^\pm_{h,\bar h}, ~~ (n_{\cal L})_{h,\bar h}\in  \bZ_{\ge0}.
\label{intropositivity}
\fe
The information of the 't Hooft anomaly enters through the spin content of the defect Hilbert space ${\cal H}_{\cal L}$. 
It follows that the partition function $Z^{\cal L}$ is invariant under the $\Gamma_0(2)$ congruence subgroup if the $\bZ_2$ is non-anomalous, and invariant under $\Gamma_0(4)$ if  anomalous. 
The relation between the modular crossing equations and anomalies has been discussed extensively in \cite{Freed:1987qk,Felder:1988sd,Gaiotto:2008jt,Sule:2013qla,Numasawa:2017crf}.

\subsubsection*{Summary of Results}

The conformal bootstrap program has produced drastic improvements to computational techniques that make possible the precision study of constraints from modular invariance \cite{Simmons-Duffin:2015qma}.
Employing these techniques, we find precise bounds in various sectors of the Hilbert space as functions of the central charge $c$.  
At small values of $c$, our bounds are saturated by a number of theories, including the free compact boson, the $(E_7)_1$ WZW model, and several $B$- and $D$-series WZW models.  
We highlight our results below:
\begin{itemize}
\item For either the non-anomalous or anomalous case, there is a bound on the lightest $\bZ_2$ even primary. The bounds for $c < 25$ are presented in Figures~\ref{Fig:NonanomalousGeneral} and~\ref{Fig:AnomalousGeneral}.
\item When the $\bZ_2$ is anomalous, we find an analytic bound for the lightest $\bZ_2$ odd operator for $1<c<3$: 
\ie
{\bf \Delta}^-_{\rm gap}\le (\hat y+1){c\over 12}\,,
\fe
where $\hat y$ is the largest root of a cubic polynomial $\alpha [{\bf M}^-(\Delta,t)]$  \eqref{ChargedAct}, whose coefficients depend on $c$. More refined numerical bounds that apply to a larger range of values of $c$ are presented in Figure~\ref{Fig:AnomalousGeneral} (also Figure~\ref{Fig:IntroOdd}). 
\item When the $\bZ_2$ is non-anomalous, we find that the lightest $\bZ_2$ odd state (``order") and the defect Hilbert space ground state (``disorder") cannot both be arbitrarily heavy relative to $c$.\footnote{In the anomalous case, the same statement is also true, but follows trivially from the existence of an upper bound on the lightest $\bZ_2$ odd state alone.
}
This ``order-disorder" bound is presented in Figure~\ref{Fig:NonanomalousGeneral-ord-dis}.
\item RG flows preserving \textit{only} a $\bZ_2$ symmetry generically do not end at gapless fixed points with $1 < c < 7$ without fine-tuning. 
In other words, there is no  $\bZ_2$-protected gapless phase for $1<c<7$ without fine-tuning. 
If the $\bZ_2$ is non-anomalous, then the range is further extended to $1 < c < 7.81$.
\end{itemize}

This paper is organized as follows. 
In Section~2, we review the formulation of a $\bZ_2$ symmetry and its 't Hooft anomaly using topological defect lines, and derive the modular properties of the torus partition functions with defect insertions. In Section~3, we set up the modular bootstrap equations, and introduce the linear functional method that we use to derive bounds. In Section~4, we present an analytic functional that implies a simple bound on the $\bZ_2$-odd operators in the presence of 't Hooft anomaly, and argue for the non-existence of a bound when the $\bZ_2$ is non-anomalous. Section~5 presents further refined bounds and discusses the physical implications. Finally, Section~\ref{Sec:U1} discusses how our results for $\bZ_2$ extend to theories with $U(1)$ symmetry. In the appendices, we review a list of theories with $\bZ_2$ symmetry, determine the 't Hooft anomalies, and compute the lightest scaling dimension in each sector of each theory.

\section{$\bZ_2$ Symmetry and Its Anomaly in Two Dimensions}

We consider unitary, bosonic 2d conformal field theories with an internal, unitary global symmetry group. We assume that the symmetry is unbroken, and there is a unique weight $(h=0, \bar h=0)$ operator, namely, the identity operator. 
In 2d, the unitary operator implementing a global symmetry transformation is a topological defect line.
The definition and general properties of these  topological defect lines and their 't Hooft anomalies are discussed extensively in \cite{Bhardwaj:2017xup,Chang:2018iay} from a modern viewpoint. 
In this section, we present a self-contained review of \cite{Chang:2018iay} specialized to the case of $\bZ_2$ symmetry.

\subsection{Topological Defect Lines and the Defect Hilbert Space}
\label{Sec:TDL}

For quantum field theory in any spacetime dimension, a (0-form)  global symmetry transformation is implemented by a codimension-one topological defect \cite{Kapustin:2014gua,Gaiotto:2014kfa}. 
Physical observables, including correlation functions, can be dressed with these topological defects. 
The basic property of topological defects is that correlation functions are invariant under any continuous deformation of the defects that preserves their junctions. 
This implies that the stress tensor commutes with the topological defect up to contact terms. Throughout this paper, we assume that the conformal field theory is on a 2d manifold with vanishing Ricci scalar.\footnote{On a curved manifold, there is an interesting orientation-reversal anomaly for an anomalous $\bZ_2$ line, coming from the contact term between the stress tensor and the line. See Section~2.4 of \cite{Chang:2018iay}.  This results in a phase change as we deform an anomalous $\bZ_2$ line across a manifold with nonzero Ricci scalar curvature.}

When the global symmetry transformation is  $U(1)$, the defect associated to a rotation by angle $\theta$ is $U_\theta(\Sigma) = \exp[i \theta \int_\Sigma \star J]$, where $J_\mu(x)$ is the Noether current and   $\Sigma$ is a codimension-one manifold (sometimes taken to be a constant time slice).  The topological property follows from the conservation of the Noether current, $d\star J=0$. 

\begin{figure}[h]
\centering
\includegraphics[width =.25\textwidth]{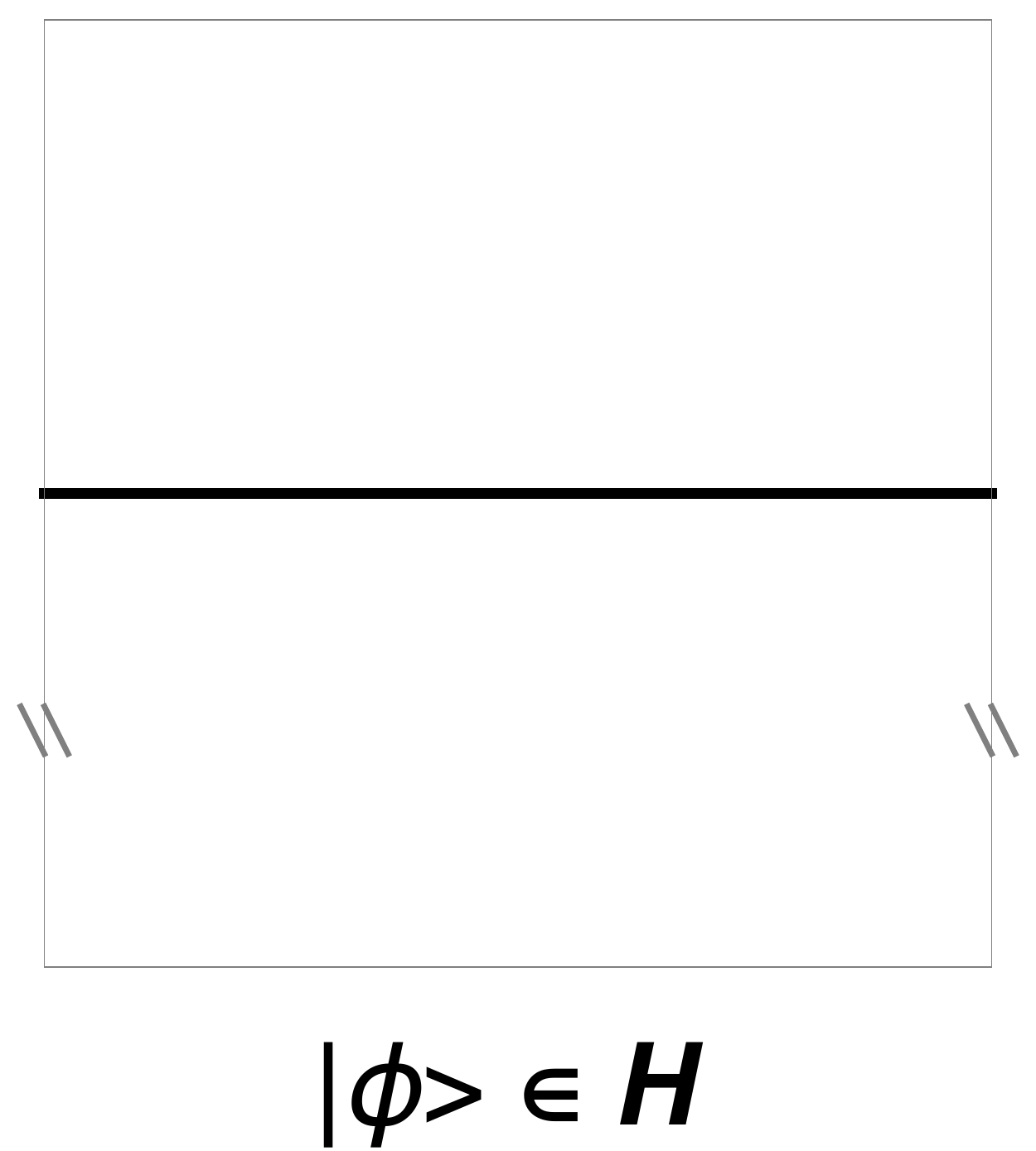}
\caption{The black line depicts the topological defect line $\cal L$ for the global symmetry $\bZ_2$. The $\bZ_2$ action on the Hilbert space can be realized by wrapping the line around the compact circle on the cylinder.}\label{fig:L}
\end{figure}

\begin{figure}[h]
\centering
\includegraphics[width =.3\textwidth]{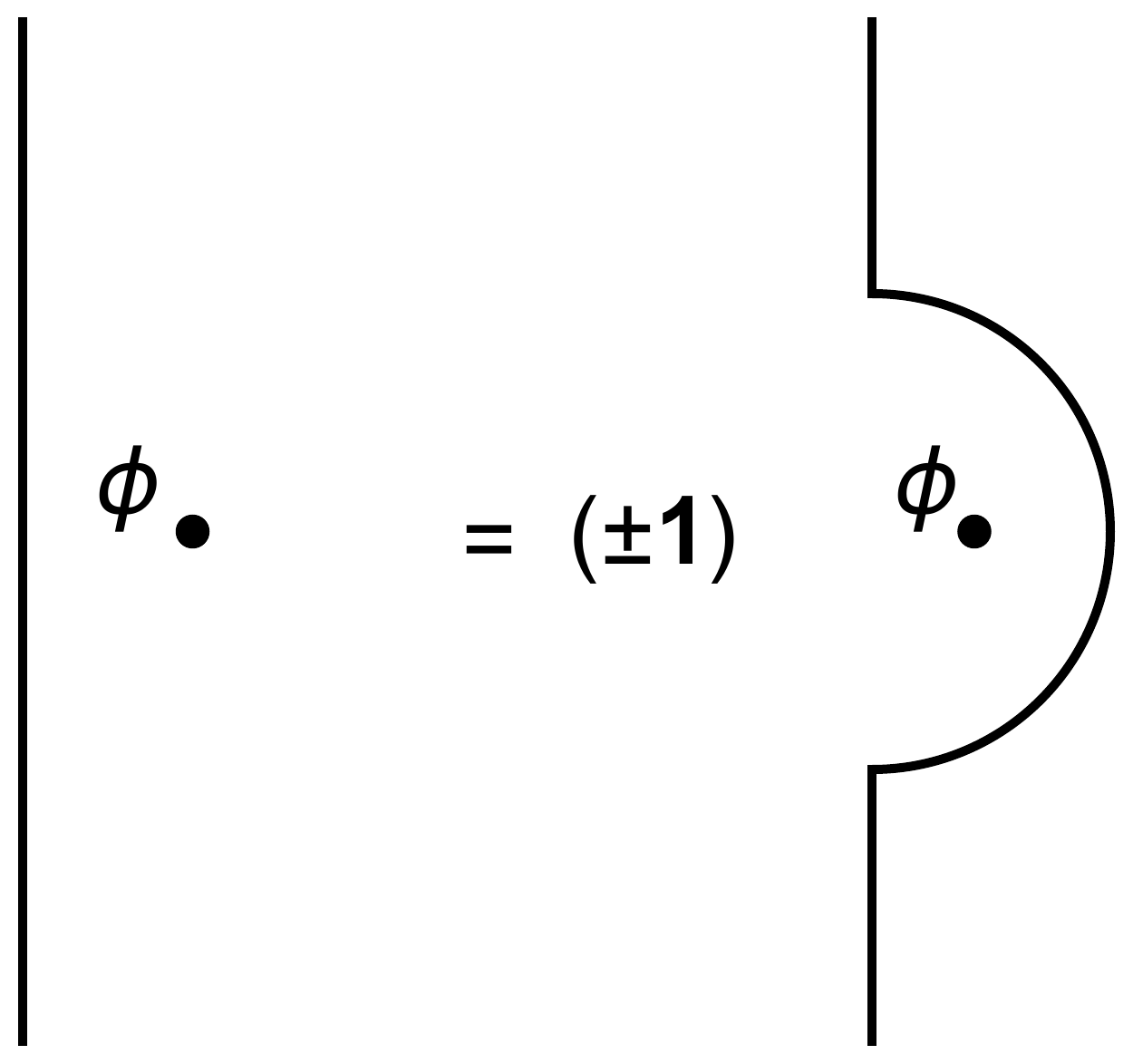}
\caption{As we sweep the $\bZ_2$ line past a local operator $\phi$, the correlation function might change by a sign.}
\label{fig:0Z2}
\end{figure}

In 2d, such codimension-one topological defects are lines. 
Consider the topological line $\cal L$ associated to an internal unitary $\bZ_2$ symmetry in a bosonic 2d CFT.  
The $\bZ_2$ line implements a  $\bZ_2$ action on the Hilbert space $\cal H$ when quantized on a circle $S^1$. 
This action can be realized by wrapping the $\bZ_2$ line along the compact $S^1$ direction at a fixed time on the cylinder $S^1\times \bR$, acting on a state $|\phi\rangle\in {\cal H}$ prepared at an earlier time (see Figure \ref{fig:L}).  We will denote this $\bZ_2$ unitary operator as
\begin{align}
\widehat{\cal L}:~ {\cal H} \to {\cal H}\,. 
\end{align}
Via the operator-state correspondence, the topological line also implements the $\bZ_2$ action on local operators. As we sweep the $\bZ_2$ line past a $\bZ_2$ even/odd local operator $\phi(x)$, the correlation function changes by a $\pm$ sign (see Figure \ref{fig:0Z2}). 

\begin{figure}
\centering
\includegraphics[width =.6\textwidth]{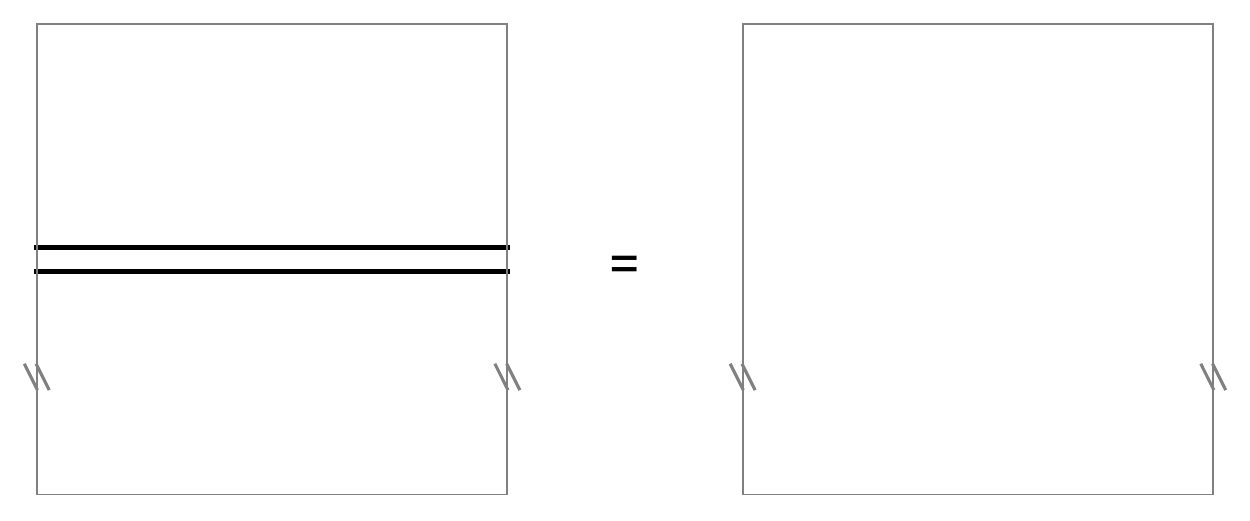}
\caption{The topological lines obey the group multiplication law under fusion.}\label{fig:LL}
\end{figure}

The fusion of topological lines obeys the group multiplication law. 
Namely, as we bring two parallel $\bZ_2$ circles  together, they fuse to a trivial line (see Figure \ref{fig:LL}).  
Thus $\widehat{\cal L}^2=+1$. 
Since the $\bZ_2$ line is its own inverse, we do not need an orientation for the line. 
We can decompose $\cal H$ into the $\bZ_2$ even and odd subsectors under $\widehat{\cal L}$:
\ie
{\cal H} = {\cal H}^+ \oplus {\cal H}^-\,.
\fe

\begin{figure}[h!]
\centering
\includegraphics[width=.5\textwidth]{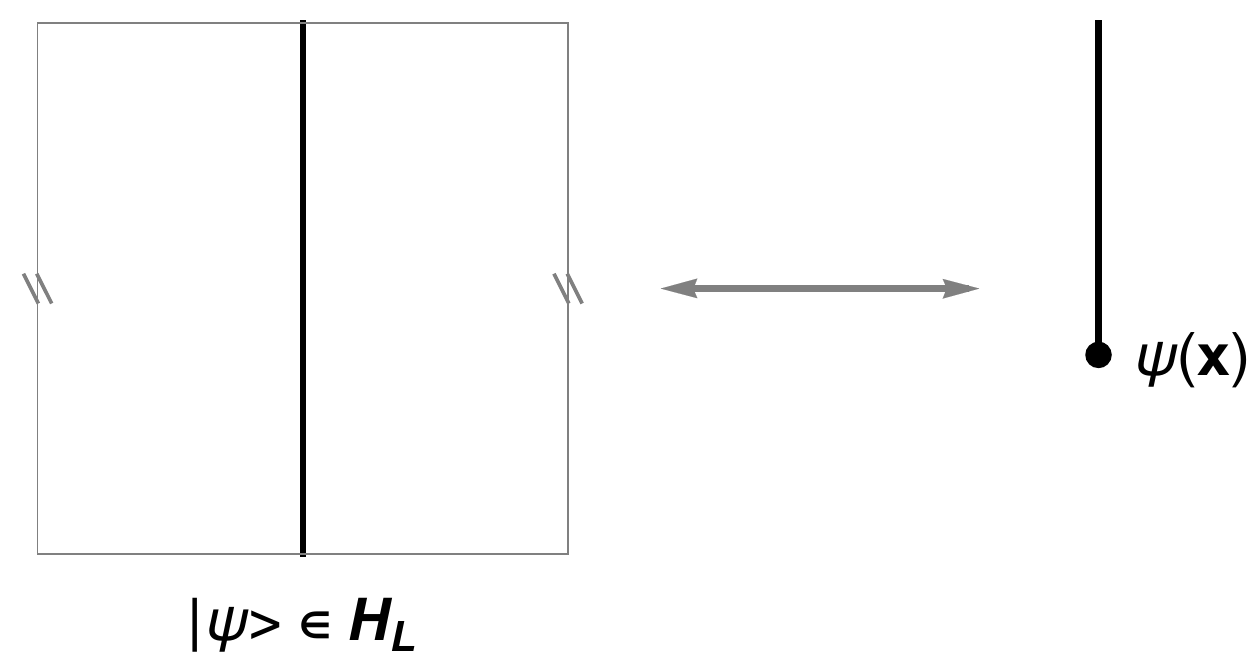}
\caption{The defect Hilbert space ${\cal H}_{\cal L}$ of a $\bZ_2$ line quantized on a circle $S^1$.  A state in the defect Hilbert space is mapped to an operator living at the end of the $\bZ_2$ line via the operator-state correspondence.}
\label{fig:HL}
\end{figure}

Consider placing the theory on a cylinder $S^1 \times \mathbb{R}$ with $\cal L$ running along the time $\mathbb{R}$ direction  (see Figure \ref{fig:HL}).  
The topological line $\cal L$ intersects with the spatial $S^1$, and therefore modifies the quantization by a twisted periodic boundary condition. 
This defines a {\it defect Hilbert space} denoted by ${\cal H}_{\cal L}$.\footnote{When the $\bZ_2$ is non-anomalous, the $\bZ_2$ even sector ${\cal H}_{\cal L}^+$ of the defect Hilbert space ${\cal H}_{\cal L}$ before gauging is the twisted sector of the orbifold theory.} 
 Via the operator-state correspondence, a defect Hilbert space state $|\psi\rangle\in {\cal H}_{\cal L}$ is mapped to an operator living at the end of the $\bZ_2$ line.\footnote{It was shown in Section 2.2.4 of \cite{Chang:2018iay} that the defect Hilbert space ${\cal H}_{\cal L}$ of a topological defect line $\cal L$ is never empty in a 2d unitary, compact, bosonic CFT with a unique vacuum.  
By the operator-state correspondence, it means that every topological defect line must be able to end on some (non-topological) point-like operator(s), {\it i.e.} all lines are breakable. 

On the other hand,  in a 2d theory with {\it degenerate} vacua, the defect Hilbert space ${\cal H}_{\cal L}$ might be empty.  For example, this is the case in the $\bZ_2$ gauge theory describing the spontaneously broken phase. By the operator-state correspondence, such a line is not breakable.
}

Since the topological line commutes with the stress tensor,  the states in the defect Hilbert space ${\cal H}_{\cal L}$  are organized into representations of the left and right Virasoro algebras. In particular,  the defect Hilbert space states can be diagonalized to have definite conformal weights $(h,\bar h)$. 

We can generalize the above construction by inserting multiple vertical topological lines along the time direction on the cylinder. This defines a more general defect Hilbert space ${\cal H}_{\cal L L{\cdots}L}$.  In the case of $\bZ_2$ lines, we can fuse these vertical lines pairwise to the trivial line.  This shows that the defect Hilbert space of an even number of lines is isomorphic to the Hilbert space $\cal H$ of local operators, while that of an odd number of lines is isomorphic to the defect Hilbert space ${\cal H}_{\cal L}$. 
In particular, the defect Hilbert space ${\cal H}_{\cal LL}$, which via the operator-state correspondence are the operators living \textit{on} the line, is isomorphic to the Hilbert space $\cal H$ of local operators.

Finally, we should require that the global symmetry acts faithfully on local operators, {\it i.e.} the only topological defect line that commutes with all local operators is the trivial line.  
It follows that the defect Hilbert space ${\cal H}_{\cal L}$ contains no weight-$(0,0)$ state, otherwise the topological line can be ``opened up" to commute with all local operators. See Section 2.2.5 of \cite{Chang:2018iay} for a detailed discussion.

\subsection{'t Hooft Anomaly}

In a bosonic 2d theory, the 't Hooft anomaly of a unitary $\bZ_2$ symmetry is classified by the group cohomology $H^3(\bZ_2,U(1)) = \bZ_2$, which  manifests in the crossing relation of  $\cal L$. The 3d Symmetry-Protected Topological phase is $\exp[{2\pi i\over 2}\int_{3d} A\cup A\cup A]$ where $A$ is the discrete background one-form gauge field.

\begin{figure}[H]
\centering
\includegraphics[width=.45\textwidth]{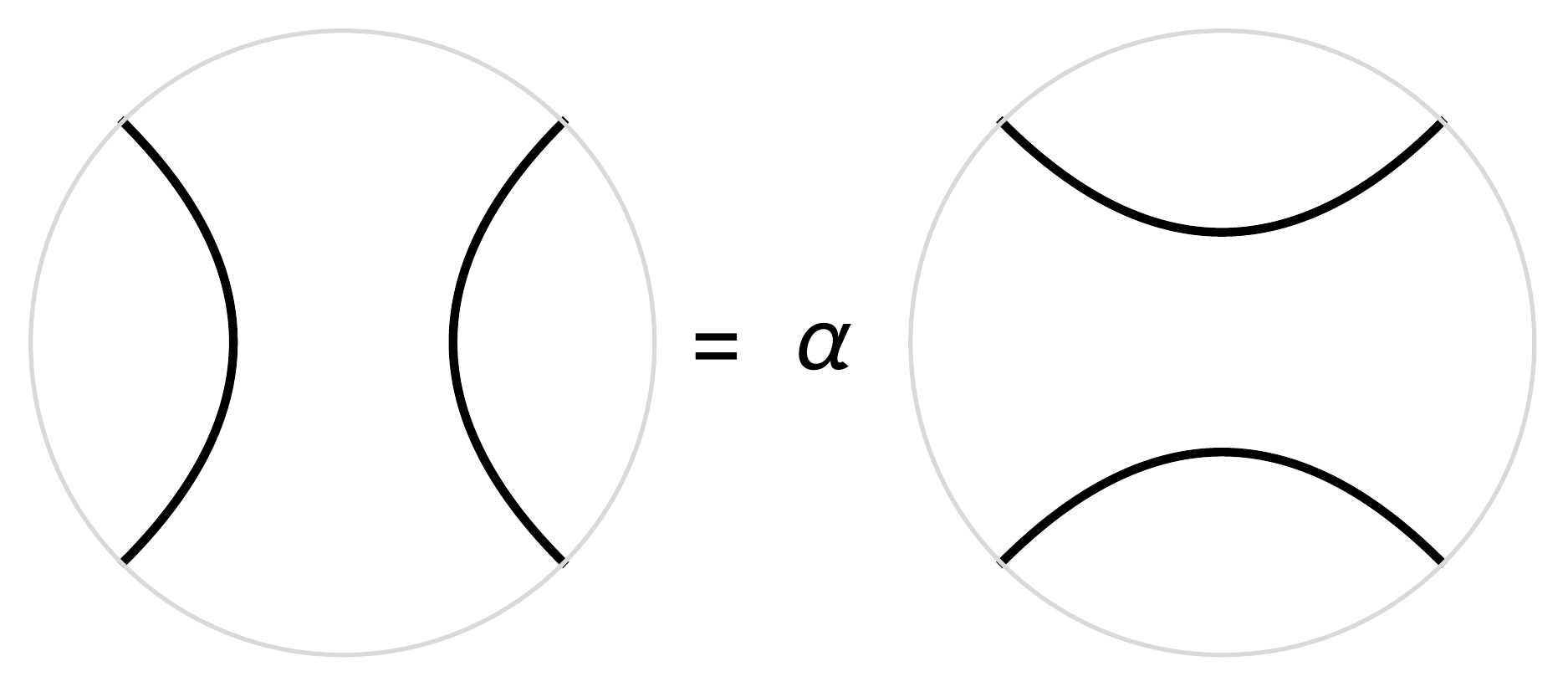}
\caption{The crossing relation of a $\bZ_2$ line $\cal L$ (shown in black) on a local patch of a 2-manifold.  By applying this crossing relation twice, we conclude that $\alpha$ has to be either +1 or $-1$. }\label{fig:crossing}
\end{figure}

Consider a general correlation function of local operators and topological lines.   
Let us focus on a local patch depicted by the gray circle on the left in Figure~\ref{fig:crossing}, where there are two segments of $\bZ_2$ lines.
This defines a state of weight $(h=0,\bar h=0)$ in the defect Hilbert space ${\cal H}_{\cal LLL L}$ on  the boundary of this patch.  
Since ${\cal H}_{\cal LLLL}\simeq {\cal H}$ by fusing the four vertical $\bZ_2$ lines, the subspace of such weight-$(0,0)$ states, denoted by $V_{\cal LLLL}$, is one-dimensional and generated by the identity operator. 

Next, we perform a crossing in that local patch without modifying the configuration outside the patch, so that we end up with the configuration on the right in Figure \ref{fig:crossing}.
 How is the new correlation function related to previous one? 
 Since $V_{\cal LLLL}$ is one-dimensional, the state corresponding to the right figure must be proportional to the state on the left.  
 We will denote this proportionality constant by $\alpha$. 
 
What is the constraint on $\alpha$?  By applying the crossing in Figure \ref{fig:crossing} twice, we return to the original configuration in the local patch, multiplied by $\alpha^2$.  Thus we conclude that $\alpha^2=1$. 
In more general terms, this consistency condition is the cocycle condition of $H^3(G,U(1))$, which classifies the bosonic anomaly of a global symmetry $G$ \cite{Freed:1987qk,Chen:2011bcp,Chen:2011pg}, in the case of $G=\bZ_2$.  

As we will argue now, a non-anomalous $\bZ_2$ line has $\alpha=+1$, while an anomalous $\bZ_2$ line has $\alpha=-1$.  Indeed, a configuration of $\bZ_2$ lines can be thought of as a background $\bZ_2$ gauge field on the manifold.  
The crossing in Figure \ref{fig:crossing} can be achieved by performing a $\bZ_2$ gauge transformation in the area between the two lines.  If the correlation function is not invariant under the gauge transformation in the presence of background gauge fields ({\it i.e.} if $\alpha=-1$), then it is by definition anomalous.

Another pragmatic way to detect the anomaly ($\alpha=-1$) is by the ambiguity/inconsistency in constructing the torus partition function of the $\bZ_2$ orbifold theory.\footnote{Since $H^2(\bZ_2,U(1))=1$, we do not have to consider the discrete torsion \cite{Vafa:1986wx}.
} 
Let us attempt to compute the torus partition function of the would-be orbifold theory, which can be written as a sum of four terms (times a factor of $1\over2$).  
The first two terms account for the contributions from the $\bZ_2$ even states in the $\cal H$, while the last two terms are from  the $\bZ_2$ even states in the defect Hilbert space ${\cal H}_{\cal L}$. 
A potential ambiguity arises for the last term, which is shown in Figure \ref{fig:noanom}. 
When $\alpha=+1$,  there is no ambiguity in resolving the cross ``$+$" of two $\bZ_2$ lines, hence there is no 't Hooft anomaly ({\it i.e.} no obstruction to orbifolding). On the other hand, if $\alpha=-1$, then the cross ``$+$" in Figure \ref{fig:noanom} is ambiguous and depends on the choice of resolution. 
In particular, the two resolutions in Figure~\ref{fig:Lpm} differ by a sign, and neither yields a modular invariant torus partition function. 
Thus we cannot consistently compute the torus partition function of the would-be orbifold theory when $\alpha=-1$, which means that the $\bZ_2$ is anomalous.

\subsection{Spin Selection Rule}
\label{Sec:SpinSelection}

All local operators in a bosonic 2d CFT have integer spins $s \equiv h-\bar h$, by the requirement of mutual locality.
However, an operator $\psi$ living at the end of  a topological defect line, which by the operator-state correspondence maps to a state in the defect Hilbert space ${\cal H}_{\cal L}$, need not obey this rule. 
This is because as we circle a local operator $\phi$ around $\psi$, the former will be acted on by the topological defect line attached to $\psi$.
 
In this subsection, we derive the constraints on the spins $h-\bar h$ of states in the defect Hilbert space ${\cal H}_{\cal L}$.  Indeed, we find that the states in the defect Hilbert space ${\cal H}_{\cal L}$ generally do not have integer spins. 
Along the way, we also discuss an interesting spin-charge relation for states in the defect Hilbert space.  
The constraints on the spins for a $\bZ_n$ line can be found in Section 4.4 of \cite{Chang:2018iay} (see also \cite{Hung:2013cda}).

\begin{figure}[t]
\centering
\includegraphics[width=.24\textwidth]{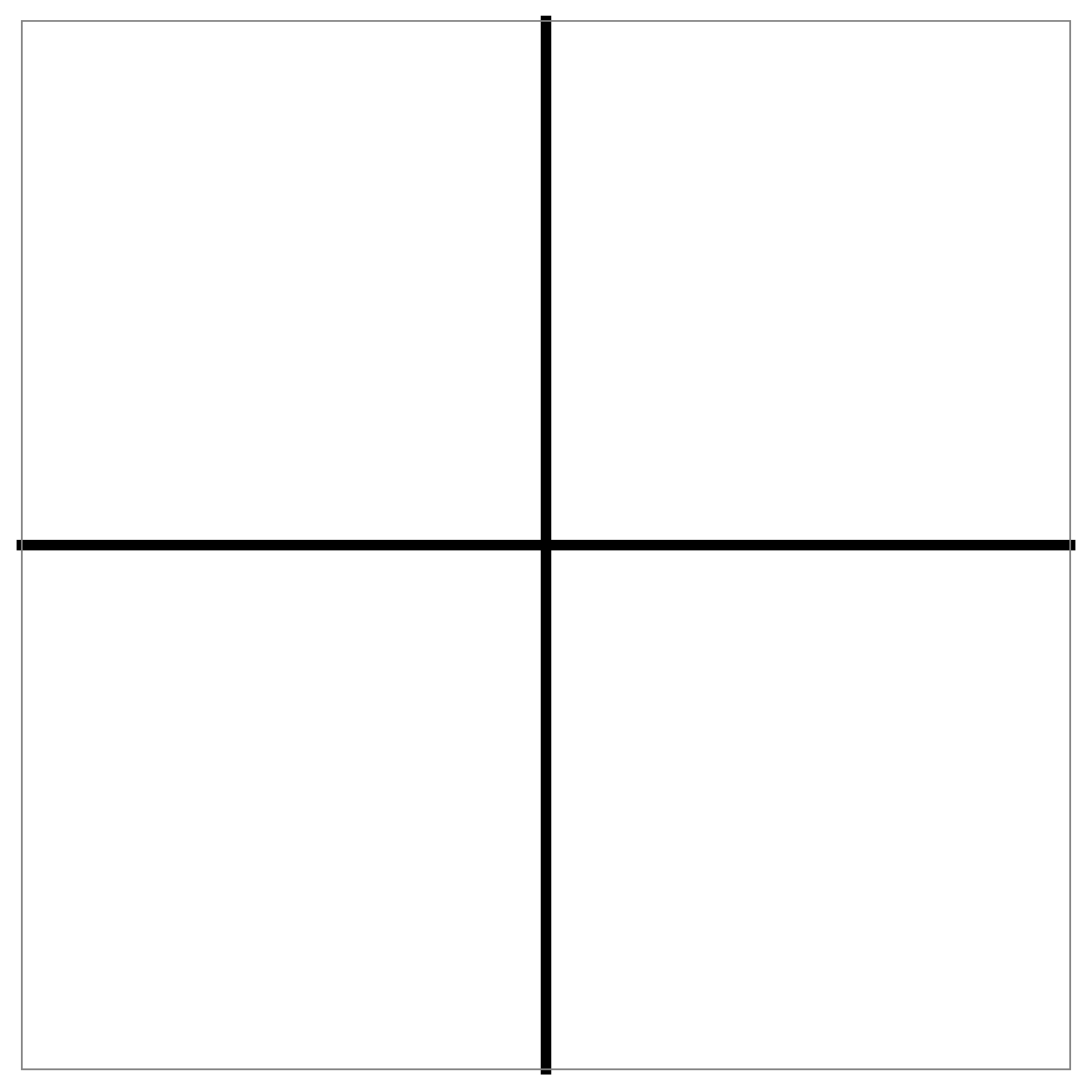}
\caption{A $\bZ_2$ action can be defined unambiguously on the defect Hilbert space ${\cal H}_{\cal L}$ when the bulk $\bZ_2$ is non-anomalous.  Because $\alpha=+1$ in Figure \ref{fig:crossing}, there is no ambiguity in resolving the cross ``$+$" of the two lines. }\label{fig:noanom}
\end{figure}

\begin{figure}[t]
\centering
\includegraphics[width=.25\textwidth]{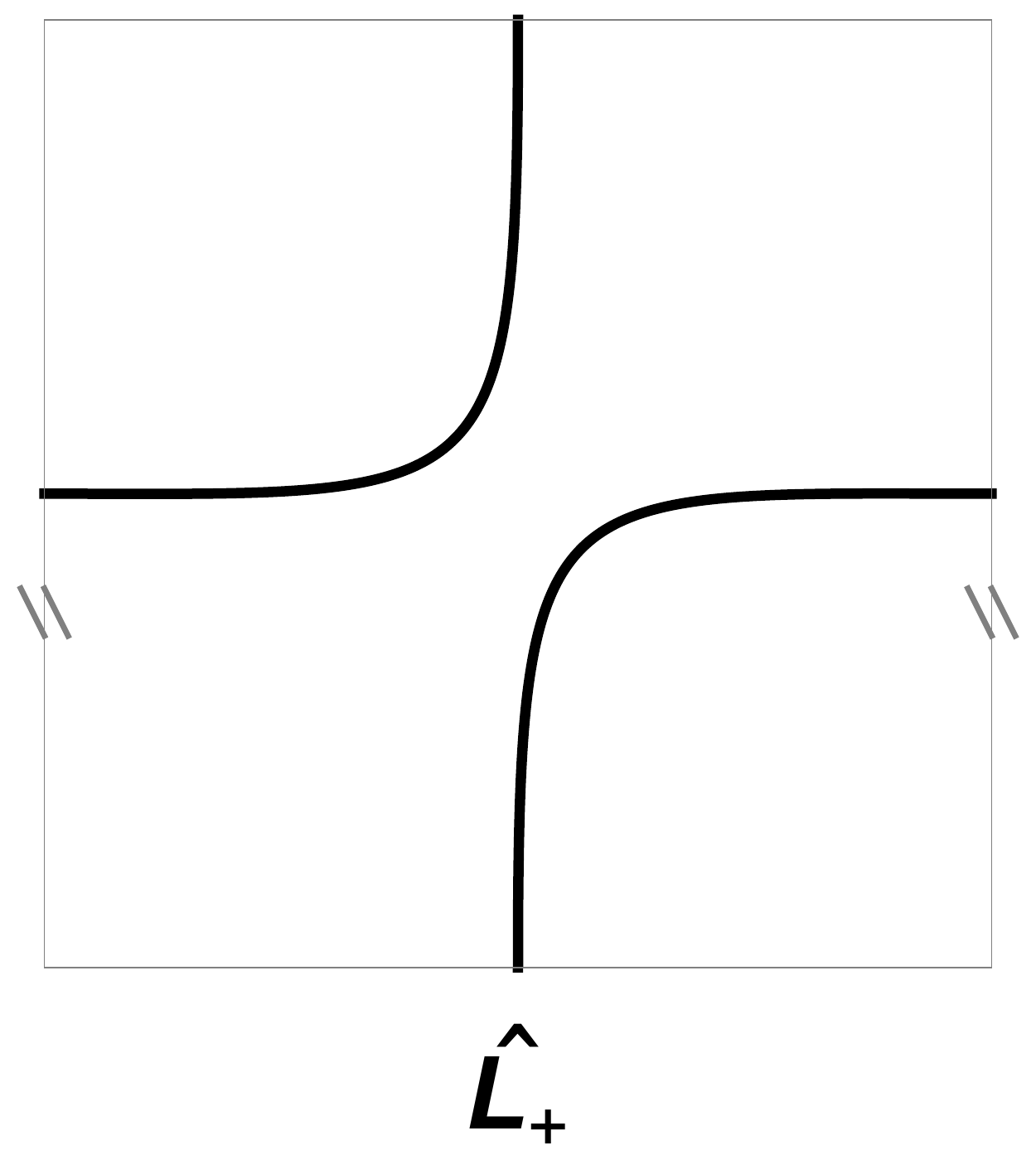}~~~~~~~~~~~~~~~~~
\includegraphics[width=.25\textwidth]{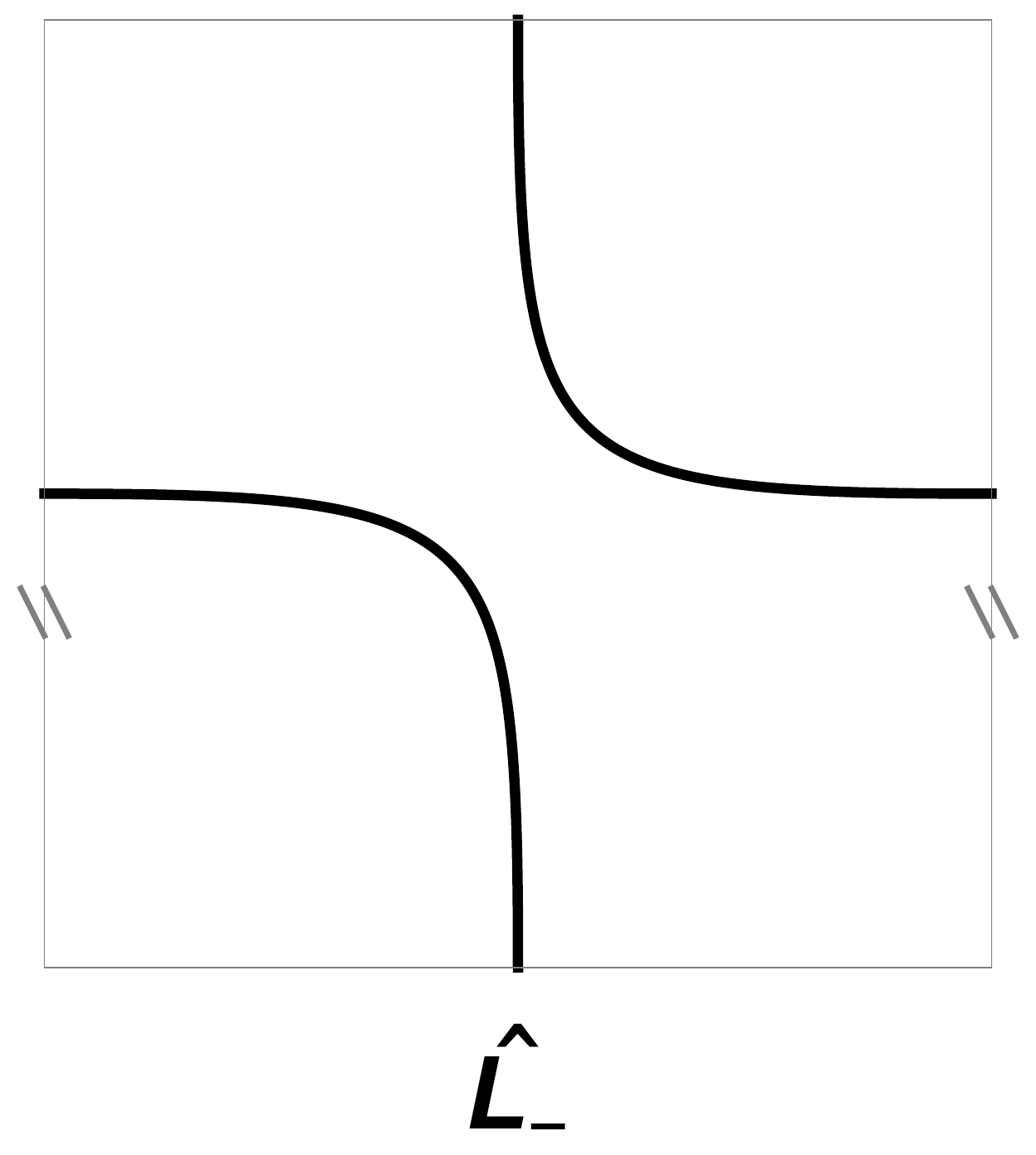}
\caption{When the $\bZ_2$ is anomalous ({\it i.e.} $\alpha=-1$), the two resolutions $\widehat{\cal L}_\pm$ of the cross ``$+$" lead to different actions on the defect Hilbert space ${\cal H}_{\cal L}$.  From Figure \ref{fig:crossing}, we see that $\widehat{\cal L}_+ = \alpha\widehat{\cal L}_-$.  
}\label{fig:Lpm}
\end{figure}

Let us start by defining a  $\bZ_2$ action on the defect Hilbert space ${\cal H}_{\cal L}$, analogous to the $\widehat{\cal L}$ action on the Hilbert space $\cal H$ of local operators. 
The first  attempt is to represent such an action on ${\cal H}_{\cal L}$ as in  Figure \ref{fig:noanom}.  
However, such a configuration is potentially ambiguous because of the cross ``$+$" between the two lines.  
To make sense of the cross, there are two possible resolutions, as shown in Figure~\ref{fig:Lpm}, and each defines an action on the defect Hilbert space ${\cal H}_{\cal L}$.  We denote these two actions on ${\cal H}_{\cal L}$ by
\begin{align}
\widehat{\cal L}^{\pm} :~ {\cal H}_{\cal L} \to {\cal H}_{\cal L}\,.
\end{align}
The two actions are related by a crossing move (Figure~\ref{fig:crossing}):
\begin{align}\label{Lpm}
\widehat{\cal L}^+ = \alpha \widehat{\cal L}^-\,.
\end{align}
Thus, when the $\bZ_2$ is non-anomalous ($\A = +1$), the configuration in Figure~\ref{fig:noanom} is unambiguous and can be interpreted as either $\widehat {\cal L}^+$ or $\widehat{\cal L}^-$. However, when the $\bZ_2$ is anomalous ($\A = -1$), then the two resolutions differ by a sign, and the configuration in Figure~\ref{fig:noanom} is ambiguous. 

\begin{figure}[h!]
\centering
\includegraphics[width=.45\textwidth]{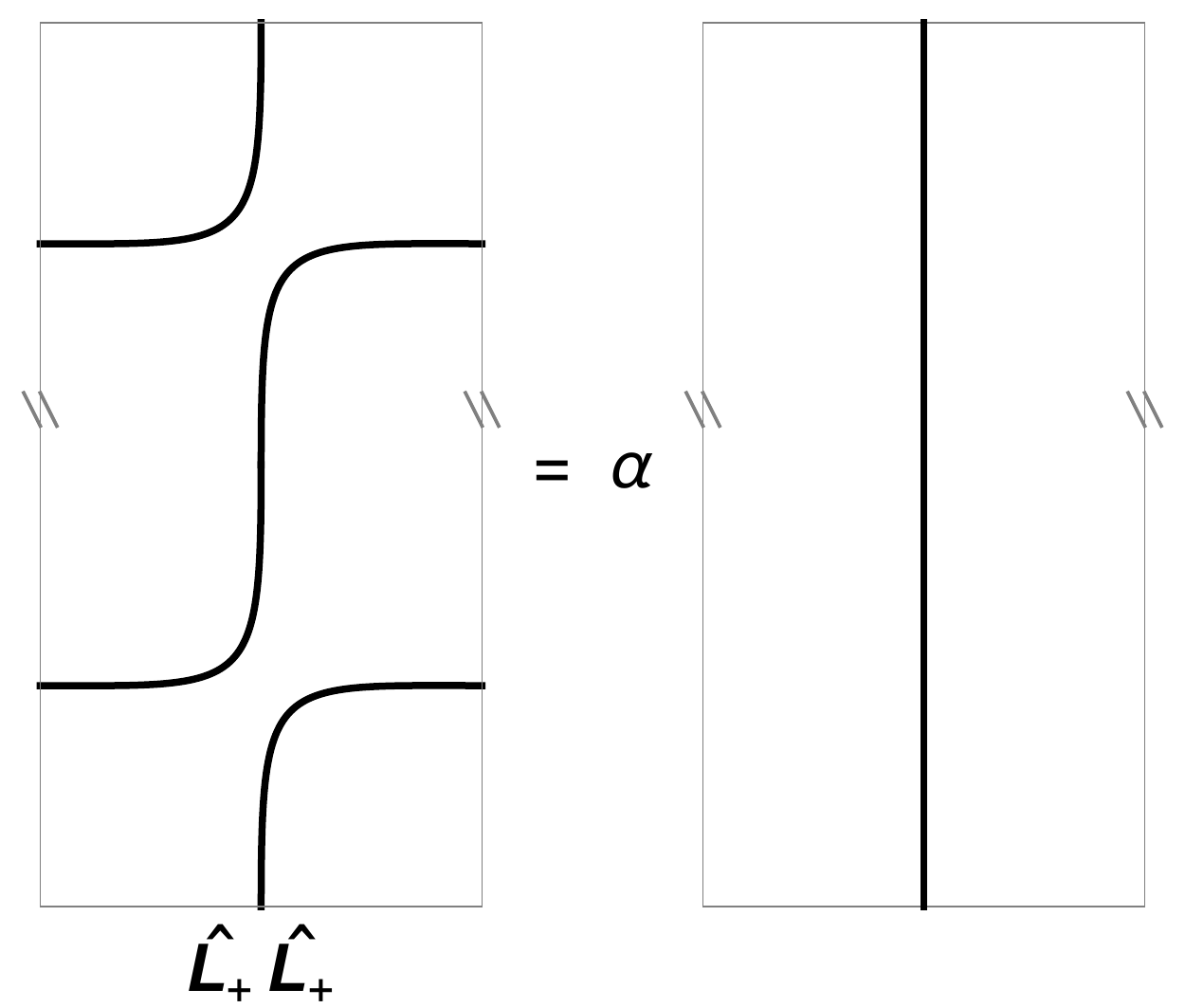}
\caption{The square of $\widehat {\cal L}_+$ can be computed using the crossing relation in Figure \ref{fig:crossing}.
}\label{fig:Lsq}
\end{figure}

\begin{figure}[h!]
\centering
\includegraphics[width=.6\textwidth]{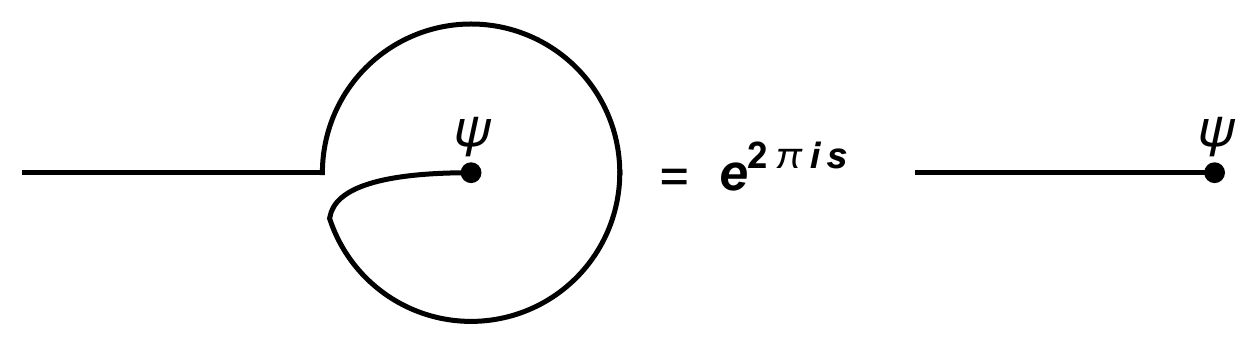}
\caption{Using the operator-state correspondence map from the cylinder $S^1\times \mathbb{R}$ to the plane $\mathbb{R}^2$, we can map the action of  $\widehat{\cal L}_+$ on a state $|\psi\rangle\in {\cal H}_{\cal L}$ to a  ``lassoing" configuration on the plane (left figure).  The $\bZ_2$ line can then be unwound to give the right figure, at a price of a  phase  due to the  fractional spin of the operator $\psi$ living at the end of the defect.
}\label{fig:plane}
\end{figure}

Let us discuss how $\widehat{\cal L}^\pm$ acts on the a state with conformal weight $(h,\bar h)$ in the defect Hilbert space ${\cal H}_{\cal L}$. First, applying the crossing move to the left of Figure~\ref{fig:Lsq} and then unwinding the line, we have
\begin{align}\label{Lsq}
(\widehat{\cal L}^+)^2 =\alpha\,,~~~~(\widehat{\cal L}^-)^2 =\alpha\,.
\end{align}
To determine $\widehat{\cal L}^\pm |h,\bar h\rangle$, we perform an operator-state map from the cylinder to the plane. After this map, we see that the action of $\widehat{\cal L}^\pm $ corresponds to the unwinding of a $\bZ_2$ line, as depicted in Figure \ref{fig:plane}, giving
\begin{align}\label{Laction}
\widehat{\cal L}^\pm  |h,\bar h\rangle  =  e^{\pm 2\pi i (h-\bar h) } |h,\bar h\rangle\,.
\end{align}
This phase in \eqref{Laction} is only consistent with \eqref{Lsq} if the spin $s=  h-\bar h$ obeys
\begin{align}\label{spinselection}
s \in \begin{cases}
& {\bZ\over2}\,,~~~~~~~~~~~(\alpha=+1,~\text{non-anomalous}~~\bZ_2)\,,\\
&\frac 14 + {\bZ\over2}\,,~~~~~~(\alpha=-1,~\text{anomalous}~~\bZ_2)\,,
\end{cases}
\end{align}
for any $|h,\bar h\rangle\in {\cal H}_{\cal L}$.   

Importantly, if the $\bZ_2$ is anomalous, it follows that the scaling dimensions $\Delta=h+\bar h$ of states in the defect Hilbert space ${\cal H}_{\cal L}$ are bounded from below by $1\over4$:
\begin{align}\label{twistedgap}
(\alpha=-1)~~~~\Delta  \ge \frac 14\,,~~~~\forall ~~|h,\bar h\rangle \in {\cal H}_{\cal L}.
\end{align}
This fact will be crucial when we argue for a universal bound on operators in the $\bZ_2$ odd  sector, when the $\bZ_2$ is anomalous (see Section \ref{Sec:Cardy}). 
This lower bound on the defect Hilbert space ground state implies that an anomalous $\bZ_2$ line in a gapped phase -- where all the operators in $\cal H$ and ${\cal H}_{\cal L}$ have vanishing weight -- is unbreakable ({\it i.e.} ${\cal H}_{\cal L}$ is empty). 

Finally, let us comment on an interesting spin-charge relation in the defect Hilbert space ${\cal H}_{\cal L}$. 
Since $\widehat{\cal L}^+ $ differs from $\widehat{{\cal L}}^-$ only by an overall phase $\alpha$, we focus on the former from now on.  
From \eqref{Laction}, we see that the eigenvalue of $\widehat{\cal L}^+$ is determined by the spin $s$ of the state in the defect Hilbert space ${\cal H}_{\cal L}$. 
The spin-charge relation on ${\cal H}_{\cal L}$ is then as follows. When the $\bZ_2$ is non-anomalous, 
\begin{align}\label{spincharge1}
(\alpha=+1)~~~~ \widehat{\cal L}^+=
 \begin{cases}
 &+1\,,~~~~~\text{if}~~s\in \bZ\,,\\
 &-1\,,~~~~~\text{if}~~s\in \frac 12+\bZ\,.
 \end{cases}
 \end{align}
Note that even though the 2d CFT is bosonic, we encounter half-integral spin operators living at the end of $\bZ_2$ lines.  
These states are $\bZ_2$ odd, and are projected out if we were to gauge the $\bZ_2$ symmetry. 
These half-integral spin states are the gapless version of the emerging fermionic excitations from string-like objects in the lattice spin models \cite{PhysRevB.67.245316}.  
On the other hand, when the $\bZ_2$ is anomalous,
 \begin{align}\label{spincharge2}
(\alpha=-1)~~~~ \widehat{\cal L}^+=
 \begin{cases}
 &+i\,,~~~~~\text{if}~~s\in \frac 14 + \bZ\,,\\
 &-i\,,~~~~~\text{if}~~s\in -\frac 14+\bZ\,.
 \end{cases}
 \end{align}
Hence there are anyonic excitations at the endpoints of anomalous $\bZ_2$ lines. 
 
Let ${\cal H}_{\cal L}^+$ denote the subsector of ${\cal H}_{\cal L}$ which has ${\widehat{\cal L}^+}=+1$ in the non-anomalous case, or that with ${\widehat{\cal L}^+}= +i$ in the anomalous case.  
Similarly, let ${\cal H}_{\cal L}^-$ denote the subsector of ${\cal H}_{\cal L}$  in which all states have $\widehat{\cal L}^+=-1$ in the non-anomalous case, or that with $\widehat{\cal L}^+= -i$ in the anomalous case.  
The defect Hilbert space Hilbert space ${\cal H}_{\cal L}$ can be decomposed as
\ie\label{HLsub}
{\cal H}_{\cal L}  ={\cal H}_{\cal L}^+ \oplus {\cal H}_{\cal L}^-\,.
\fe

\subsection{Relation to 3d TQFTs}

If we couple a 2d CFT with $\bZ_2$ symmetry to a 3d SPT and gauge the 2d-3d system, then the operators living at the end of the original $\bZ_2$ line (which no longer exists in the gauged theory) now become the endpoints of the anyons in the 3d TQFT. 
Hence the spin selection rule \eqref{spinselection} in ${\cal H}_{\cal L}$ is related to the spins of the anyons in the 3d TQFT.  We discuss this relation in more detail below.\footnote{We thank Zohar Komargodski  and Pavel Putrov for discussions on this point.}

\paragraph{The Anomalous Case}

We start with a 2d bosonic CFT with an anomalous $\bZ_2$ symmetry. 
The spin selection rule in ${\cal H}_{\cal L}$ is $s =h-\bar h \in \frac 14  +{\bZ\over2}$.  
We couple the 2d CFT to a the 3d SPT  ${2\pi i\over2 }\int_{3d} A\cup A\cup A$, where by anomaly inflow the $\bZ_2$ symmetry of this 2d-3d system can now be gauged.
In the 3d bulk, we obtain the Dijkgraaf-Witten theory \cite{Dijkgraaf:1989pz} associated to the nontrivial element of $H^3(\bZ_2, U(1))=\bZ_2$.   This 3d bosonic TQFT admits a continuum description \cite{Maldacena:2001ss,Banks:2010zn,Kapustin:2014gua}
\ie
S= { 2i \over 2\pi}\int_{3d} bda  +  {2i \over 4\pi } \int_{3d} ada\,,
\fe
where $a$ and $b$ are (continuous) 1-form gauge fields. 
There are four anyons: the trivial line 1, the electric line $e$, the magnetic line $m$, and the dyonic line $d$.  Their  spins are given by\footnote{Recall that the spin of an anyon in 3d TQFT is defined modulo integer, while the spin $s=h-\bar h$ of a 2d CFT operator is a real number.}
\ie
\left.\begin{array}{c|c|c|c|c} \text{Anyon}&~~1~~&~~ e=e^{i \oint a}~~ &~~m=e^{i \oint b}  ~~& ~~d= e^{i \oint a+b}~~\\
\hline \text{Spin}  & 0& 0& -\frac14 & \frac14 \end{array}\right.
\fe
In the gauged 2d-3d system, the $\bZ_2$ even local operators of the ungauged theory are local operators on the 2d boundary.  The $\bZ_2$ odd local operators of the ungauged theory are now the endpoints of the electric line $e$, whose spin is 0.  The operators with spin $\frac 14+ \mathbb{Z}$ in the defect Hilbert space ${\cal H}_{\cal L}$ are the endpoints of the dyonic line $d$, which has spin $\frac 14$. Finally, the rest of the operators with spin $-\frac 14+\bZ$ in ${\cal H}_{\cal L}$ are the endpoints of the magnetic line $m$, which has spin $-\frac14$.
We summarize the above relation as follows:
\ie
\left.\begin{array}{c|c|c|c|c} ~ \text{Ungauged 2d CFT} ~& ~~{\cal H}^+:s\in \bZ ~~&~~ {\cal H}^- :s\in \bZ~~ &~~ {\cal H}_{\cal L}^+ :s\in \frac 14 +\bZ  ~~& ~~ {\cal H}_{\cal L}^-:s\in -  \frac 14 +\bZ ~~\\
\hline \text{Anyon} & 1 & e& d & m\end{array}\right.\notag
\fe

\paragraph{The Non-Anomalous Case}

The spin selection rule in  ${\cal H}_{\cal L}$ for a non-anomalous $\bZ_2$ is $s =h-\bar h \in {\bZ\over2}$. 
We start with a 2d CFT with a non-anomalous $\bZ_2$ coupled to  a trivial 3d $\bZ_2$ SPT, and then make the $\bZ_2$ gauge field dynamical. 
In the bulk, we obtain the 3d $\bZ_2$ gauge theory (without the Dijkgraaf-Witten twist):
\ie
S= { 2i \over 2\pi}\int_{3d} bda  \,,
\fe
where $a$ and $b$ are 1-form continuous gauge fields. 
The anyons in this 3d TQFT are
\ie
\left.\begin{array}{c|c|c|c|c} \text{Anyon}&~~1~~&~~ e=e^{i \oint a}~~ &~~m=e^{i \oint b}  ~~& ~~d= e^{i \oint a+b}~~\\
\hline \text{Spin} & 0 & 0 & 0 & \frac12\end{array}\right.
\fe
In the gauged 2d-3d system,  the 2d operators become the endpoints of different anyons as follows:
\ie
\left.\begin{array}{c|c|c|c|c} ~~ \text{Ungauged 2d CFT} ~~& ~~{\cal H}^+:~s\in \bZ ~~&~~ {\cal H}^- :~s\in \bZ~~ &~~ {\cal H}_{\cal L}^+ :~s\in \bZ   ~~& ~~ {\cal H}_{\cal L}^-:~s\in   \frac 12+\bZ ~~\\
\hline \text{Anyon} & 1 & e& m & d\end{array}\right.\notag
\fe

\subsection{Computation of the Anomaly}\label{Sec:Computation}

The above discussion shows that given a bosonic 2d CFT with a $\bZ_2$ symmetry, there is a simple algorithm to compute its anomaly: one first puts the theory on a cylinder $S^1 \times \bR$ in the presence of a $\bZ_2$ line along the time $\bR$ direction, which modifies the periodic boundary condition (as in Figure \ref{fig:HL}). Then, from the spins of the states in this Hilbert space ${\cal H}_{\cal L}$, we can determine the anomaly from the spin selection rule \eqref{spinselection}.  

The spectrum of the defect Hilbert space ${\cal H}_{\cal L}$ is determined by the $\bZ_2$ action on $\cal H$ via a modular $S$ transformation.  
We first start with the torus partition function with the $\bZ_2$ line wrapped around the spatial $S^1$ at a constant time.  This can be interpreted as a trace over  $\cal H$ with a $\bZ_2$ action inserted:
\begin{align}\label{ZLu}
Z^{\cal L} (\tau ,\bar \tau ) \equiv \Tr_{\cal H}[\widehat {\cal L} \, q^{L_0 -c/24} \bar q^{\bar L_0 -c/24} ] \,,
\end{align}
where $\widehat{\cal L}:{\cal H}\to {\cal H}$ is the $\bZ_2$ symmetry realized on the Hilbert space $\cal H$.  
Here, $q=\exp(2\pi i \tau)$ and $\bar q = \exp(-2\pi i \bar\tau)$. 
The modular $S: \tau \to -1/\tau$ transformation then gives us the partition function over the defect Hilbert space ${\cal H}_{\cal L}$:
\begin{align}\label{ZLd}
S[Z^{\cal L}](\tau,\bar\tau) 
= Z_{\cal L}(\tau ,\bar \tau)& \equiv \Tr_{ {\cal H}_{\cal L}}[ q^{L_0 -c/24} \bar q^{\bar L_0 -c/24} ] \notag,
\end{align}
where $S[f](\tau,\bar \tau ) \equiv f(-1/\tau,-1/\bar \tau)$. 
In this way, we obtain the defect Hilbert space spectrum from the $\bZ_2$ action on the Hilbert space $\cal H$ of local operators.  The spin content of $Z_{\cal L}(\tau,\bar\tau)$ then reveals the anomaly of the $\bZ_2$ symmetry. 

We can further perform a modular $T:\tau \to \tau+1 $ transformation on $Z_{\cal L}(\tau,\bar\tau)$ to obtain the left figure of Figure~\ref{fig:Lpm}:
\begin{align}
T[Z_{\cal L}](\tau,\bar \tau)
 = Z^{{\cal L}^+}_{\cal L}(\tau,\bar\tau) &\equiv \Tr_{ {\cal H}_{\cal L}}[ \widehat{\cal L}^+\,q^{L_0 -c/24} \bar q^{\bar L_0 -c/24} ] \notag,
\end{align}
where $T[f](\tau, \bar \tau) \equiv f(\tau+1,\bar \tau+1)$.
The modular $S$ transformation of $Z^{{\cal L}^+}_{\cal L}$ then depends on the anomaly:
\begin{align}
S[Z^{{\cal L}^+}_{\cal L}] (\tau,\bar \tau)  = Z^{{\cal L}^-}_{\cal L} (\tau,\bar \tau) = \alpha Z^{{\cal L}^+}_{\cal L} (\tau,\bar \tau)\,,
\end{align}
where we have used $\widehat{\cal L}^+  = \alpha\widehat{\cal L}^-$. 
The computation of more general discrete group anomalies from the torus partition function is discussed in \cite{Freed:1987qk} (see also \cite{Felder:1988sd,Sule:2013qla,Numasawa:2017crf}).

These modular  properties can be summarized as follows. When the $\bZ_2$ is non-anomalous, the partition function $Z^{\cal L}(\tau,\bar \tau)$ is invariant under $ST^2S$ and $T$, which generate the congruence subgroup $\Gamma_0(2)$.   When the $\bZ_2$ is anomalous, the partition function $Z^{\cal L}(\tau,\bar \tau)$ is invariant under $ST^4S$ and $T$, which generate the congruence subgroup $\Gamma_0(4)$.

We illustrate this computation in two examples, one with a non-anomalous $\bZ_2$, the $c=1/2$ Ising model, and the other with an anomalous $\bZ_2$, the $c=1$ self-dual free compact boson.

\subsubsection{Non-Anomalous Example: Ising Model}\label{Sec:IsingExample}

The 2d Ising model has  three Virasoro primaries, the vacuum $1$ with $h=\bar h=0$, the energy operator $\epsilon$ with $h=\bar h=\frac 12$, and the spin field $\sigma$ with $h=\bar h={1\over 16}$.  There is a $\bZ_2$ symmetry that flips the sign of the spin field
\begin{align}
\widehat{\cal L}:~~1\to 1\,,~~~~\epsilon\to\epsilon\,,~~~~\sigma\to -\sigma\,.
\end{align}

The torus partition function of the Ising model can be written as the sum of the contributions from the three primaries (and their descendants):
\begin{align}
Z(\tau,\bar \tau) = |\chi_0(\tau)|^2  + |\chi_{\frac12}(\tau)|^2 + |\chi_{1\over 16}(\tau)|^2\,.
\end{align}
Their characters are
\begin{align}\label{IsingCharacter}
\chi_0(\tau) = \frac 12 \left( \sqrt{\theta_3(\tau )\over \eta(\tau) } + \sqrt{\theta_4(\tau )\over \eta(\tau) }\right)\,,~
\chi_{\frac 12}(\tau) = \frac 12 \left( \sqrt{\theta_3(\tau )\over \eta(\tau) } -\sqrt{\theta_4(\tau )\over \eta(\tau) }\right)\,,
~
\chi_{1\over 16}(\tau) = 
\sqrt{\theta_2(\tau )\over 2 \eta(\tau) } \,,
\end{align}
where the $\theta_i$ are the Jacobi theta functions, defined as $\theta_2(\tau) = 2q^{1/8}\prod_{i=1}^\infty (1-q^i)(1+q^i)^2$, $\theta_3(\tau) = \prod_{i=1}^\infty (1-q^i)(1+q^{i-1/2})^2$, $\theta_4(\tau) = \prod_{i=1}^\infty (1-q^i)(1-q^{i-1/2})^2$, and $\eta$ is the Dedekind eta function defined as $\eta(\tau ) = q^{1/24}\prod_{i=1}^\infty (1-q^i)$. 
The torus partition function with the $\bZ_2$ action ($\bZ_2$ line inserted along the spatial direction) is
\begin{align}
Z^{\cal L} (\tau,\bar \tau) =
|\chi_0(\tau)|^2  + |\chi_{\frac12}(\tau)|^2 - |\chi_{1\over 16}(\tau)|^2\,.
\end{align}

To perform the modular $S$ transformation of $Z^{\cal L}(\tau,\bar \tau)$, we note that 
the modular $S$ matrix is
\begin{align}
S=  \frac 12 \left(\begin{array}{ccc}1 & 1 & \sqrt{2} \\1 & 1 & -\sqrt{2} \\\sqrt{2} & -\sqrt{2} & 0\end{array}\right) \,.
\end{align}
It follows that the defect Hilbert space partition function is:
\begin{align}
Z_{\cal L}(\tau,\bar\tau) = S[Z^{\cal L}](\tau,\bar\tau) = \chi_0(\tau)\chi_{1\over2} (\bar \tau)+\chi_{1\over2}(\tau)\chi_{0} (\bar \tau) 
+\chi_{1\over 16}(\tau)\chi_{1\over16} (\bar \tau) \,.
\end{align}
Hence, we see that there are three primaries in the defect Hilbert space ${\cal H}_{\cal L}$, with weights 
\begin{align}
(h,\bar h)=(0,\frac12), ~(\frac 12,0), ~(\frac {1}{16},{1\over 16})\,.
\end{align}
Via the operator-state correspondence, they are mapped to operators living at the end of the $\bZ_2$ line. 
The $(0,\frac 12)$ and $(\frac 12,0)$ states are the free Majorana fermions with half integral spins. 
The scalar $({1\over 16},{1\over 16})$ state in the defect Hilbert space is the disorder operator $\mu(x)$, which is not mutually local with the spin field $\sigma(x)$ because the latter is $\bZ_2$ odd.

The spin spectrum in ${\cal H}_{\cal L}$ corresponds to the non-anomalous ($\alpha=+1)$ spin selection rule in \eqref{spinselection}. We therefore conclude that the $\bZ_2$ symmetry in the Ising model is non-anomalous. Indeed, it is well-known that the Ising CFT is self-dual under the $\bZ_2$ gauging.

We can further do a modular $T$ transformation on $Z_{\cal L}$ to obtain $Z^{{\cal L}^+}_{\cal L}$:
\begin{align}
Z^{{\cal L}^+} _{\cal L} (\tau,\bar\tau) = -\chi_0(\tau)\chi_{1\over2} (\bar \tau)-\chi_{1\over2}(\tau)\chi_{0} (\bar \tau) 
+\chi_{1\over 16}(\tau)\chi_{1\over16} (\bar \tau) \,.
\end{align}
One can easily check that $Z^{{\cal L}^+}_{\cal L}$ is invariant under $S$, hence $Z^{{\cal L}^+}_{\cal L} (\tau,\bar\tau)=Z^{{\cal L}^-}_{\cal L} (\tau,\bar\tau)$, consistent with $\alpha=+1$. From $Z^{{\cal L}^+}_{\cal L}$, we see that the two free fermions are $\bZ_2$ odd and the disorder operator $\mu(x)$ is $\bZ_2$ even.  If we perform the $\bZ_2$ orbifold, then the two free fermions are projected out, while the disorder operator survives in the orbifold theory. On the other hand, the original spin field $\sigma$ is projected out because it is $\bZ_2$ odd.  Therefore, under $\bZ_2$ orbifolding, the order $\sigma$ and disorder $\mu$ operators are exchanged, implementing the Kramers-Wannier duality \cite{PhysRev.60.252}.

\subsubsection{Anomalous Example: $\widehat{\mathfrak{su}(2)}_1$ WZW Model}
\label{Sec:SelfDualExample}

The $c=1$ $\widehat{\mathfrak{su}(2)}_1$ WZW model can be equivalently described by the self-dual free compact boson.  It has two current algebra primaries, the vacuum $|0,0\rangle$ and the spin-$1\over2$ primary  (see Appendix \ref{App:FreeBoson} for our convention)
\ie
|h=\frac 14 , \bar h=\frac14\rangle_{\pm,\pm} = 
\exp\Big[\pm i X_L(0)  \pm i X_R(0)  \Big] |0,0\rangle\,.
\fe
This theory has an anomalous $\bZ_2$ global symmetry which commutes with the $\widehat{\mathfrak{su}(2)}\times \widehat{\mathfrak{su}(2)}$  current algebra, and acts on the primaries by
\begin{align}
\widehat{\cal L}:~~|0,0\rangle  \mapsto |0,0\rangle\,,~~~~|\frac14,\frac14\rangle_{\pm,\pm} \mapsto -| \frac14,\frac14\rangle_{\pm,\pm}\,.
\end{align}
It is well known that the 2d $\mathbb{CP}^1$ model at $\theta=\pi$ flows  to the  $\widehat{\mathfrak{su}(2)}_1$ WZW model in the IR. This $\bZ_2$ anomaly has been analyzed in the UV $\mathbb{CP}^1$ sigma model in \cite{Metlitski:2017fmd,Wan:2018zql}.

 The torus partition function without any line is
\begin{align}
Z(\tau  ,\bar \tau )  = |\chi_0(\tau ) |^2 +  | \chi_{1\over4}  (\tau )|^2\,  ,
\end{align}
where
\begin{align}
\chi_0 (\tau )  =  {\theta_3(2\tau ) \over \eta(\tau)}\,,~~~~\chi_{1\over4} (\tau )  =  {\theta_2(2\tau ) \over \eta(\tau)}\,,
\end{align}
are the  $\widehat{\mathfrak{su}(2)}$ current algebra characters.\footnote{As a slight abuse of notation, in this section, $\chi_h(\tau)$ denotes the current algebra character of a primary with weight $h$, not the Virasoro character \eqref{vcharacter}.} 
The modular $S$ matrix is
\begin{align}
S={1\over \sqrt{2}} \left(\begin{array}{cc}1 & 1 \\1 & -1\end{array}\right)\,.
\end{align}

The torus partition function with the $\bZ_2$ line inserted along the spatial direction is
\begin{align}
Z^{\cal L} (\tau ,\bar \tau ) =  |\chi_0(\tau)|^2 - | \chi_{1\over4}(\tau)|^2\,.
\end{align}
The partition function over the defect Hilbert space is the modular $S$ transformation:
\begin{align}
Z_{\cal L}(\tau ,\bar \tau) = S[Z^{\cal L}](\tau ,\bar \tau) = \chi_0(\tau) \chi_{1\over4}(\bar \tau)   + \chi_{1\over4}(\tau ) \chi_0(\bar \tau)\,.
\end{align}
Hence we learn that the defect Hilbert space ${\cal H}_{\cal L}$ has the following two current algebra primaries:
\begin{align}
 |\frac 14,0\rangle_\pm  \leftrightarrow e^{\pm i X_L(0)}\,,
 ~~&~~ |0,\frac 14 \rangle_\pm\leftrightarrow e^{\pm i X_R(0)}\,.
\end{align}
Note in particular that they have spins $\pm\frac 14$. 
The corresponding operators $e^{\pm i X_{L,R}(0)}$ are not local and are attached to the  end of the $\bZ_2$ line. 
From the spin selection rule  \eqref{spinselection}, we conclude that this $\bZ_2$ in the self-dual free compact boson theory is anomalous. 

Next, we can perform a modular $T$ transformation on $Z_{\cal L}(\tau ,\bar \tau)$ to obtain the left figure of Figure~\ref{fig:Lpm}:
\begin{align}
 Z_{\cal L}^{{\cal L}_+}(\tau ,\bar \tau) = \Tr_{ {\cal H}_{\cal L}}[ \widehat{\cal L}_+\,q^{L_0 -1/24} \bar q^{\bar L_0 -1/24} ]  
 = T[Z_{\cal L}](\tau ,\bar \tau)
 = i  \chi_0(\tau) \chi_{1\over4}(\bar \tau)   -i  \chi_{1\over4}(\tau ) \chi_0(\bar \tau)\,.
\end{align}
It follows that the spin-$1\over4$ states have  $\widehat{\cal L}^+$-charge $+i$, while the spin-$(-{1\over4})$ states have  $\widehat{\cal L}^+$-charge $-i$. 
On the other hand, the right figure of Figure~\ref{fig:Lpm} can be obtained by acting $T^{-1}$ on $Z_{\cal L}(\tau ,\bar \tau)$:
\begin{align}
 Z_{\cal L}^{{\cal L}_-}(\tau ,\bar \tau) = \Tr_{ {\cal H}_{\cal L}}[ \widehat{\cal L}_-\,q^{L_0 -1/24} \bar q^{\bar L_0 -1/24} ]  
 = T^{-1}[Z_{\cal L}](\tau ,\bar \tau)
 = - i  \chi_0(\tau) \chi_{1\over4}(\bar \tau)   +i  \chi_{1\over4}(\tau ) \chi_0(\bar \tau)\,.
\end{align}
Since $  Z_{\cal L}^{{\cal L}_+}(\tau ,\bar \tau)=- Z_{\cal L}^{{\cal L}_-}(\tau ,\bar \tau)$, we have confirmed the $\alpha=-1$ sign in Figure \ref{fig:crossing}.

\section{Modular Bootstrap}\label{Sec:ModularBootstrap}

We now discuss how the torus partition functions with different configurations of topological defect lines are related under modular $S$ transforms, and how these relations together with the Hilbert space definition of these partition functions allow a systematic study of universal constraints.  
 The 2d CFT will be assumed to be compact, unitary, bosonic, and with $c=c_L=c_R>1$.

With the exception of Section \ref{Sec:Analytic}, we will define the {\it gap} in a sector of the spectrum as the scaling dimension of the lightest {\it non-degenerate} Virasoro primary.  
We will be mainly interested in deriving an {\it upper} bound (which depends on the central charge) on the gap in each sector of the spectrum,  and stressing the role of the 't Hooft anomaly.  \footnote{There is no universal lower bound in each sector stronger than the unitarity bound, which is $1\over4$ in ${\cal H}_{\cal L}$ for an anomalous $\bZ_2$, and 0 in every other  case.
For example, in the Hilbert space ${\cal H}$ of local operators, one can achieve an arbitrarily small gap by considering a sigma model CFT with a large target space.  For the defect Hilbert space of an anomalous $\bZ_2$, there is a universal lower bound $1\over4$ on the ground state, but it is saturated by the self-dual boson.  Thus the non-trivial question is whether there is an upper bound on the gap in each sector.}

\subsection{Partition Functions and Characters}

We consider the following four torus partition functions dressed with topological defect lines: no line $Z(\tau, \bar\tau)$, a single line along the spatial direction $Z^{\cal L}(\tau, \bar\tau)$, a single line along the time direction $ Z_{\cal L}(\tau, \bar\tau)$, and $ Z_{\cal L}^{{\cal L}_+}(\tau, \bar\tau) $ as defined in Section~\ref{Sec:Computation}. We remind the readers their definitions in terms of traces over the Hilbert space $\cal H$ and the defect Hilbert space $\cal H_{\cal L}$:
\begin{center}
\begin{tabular}{m{2cm} m{8cm}}
\includegraphics[height=.1\textwidth]{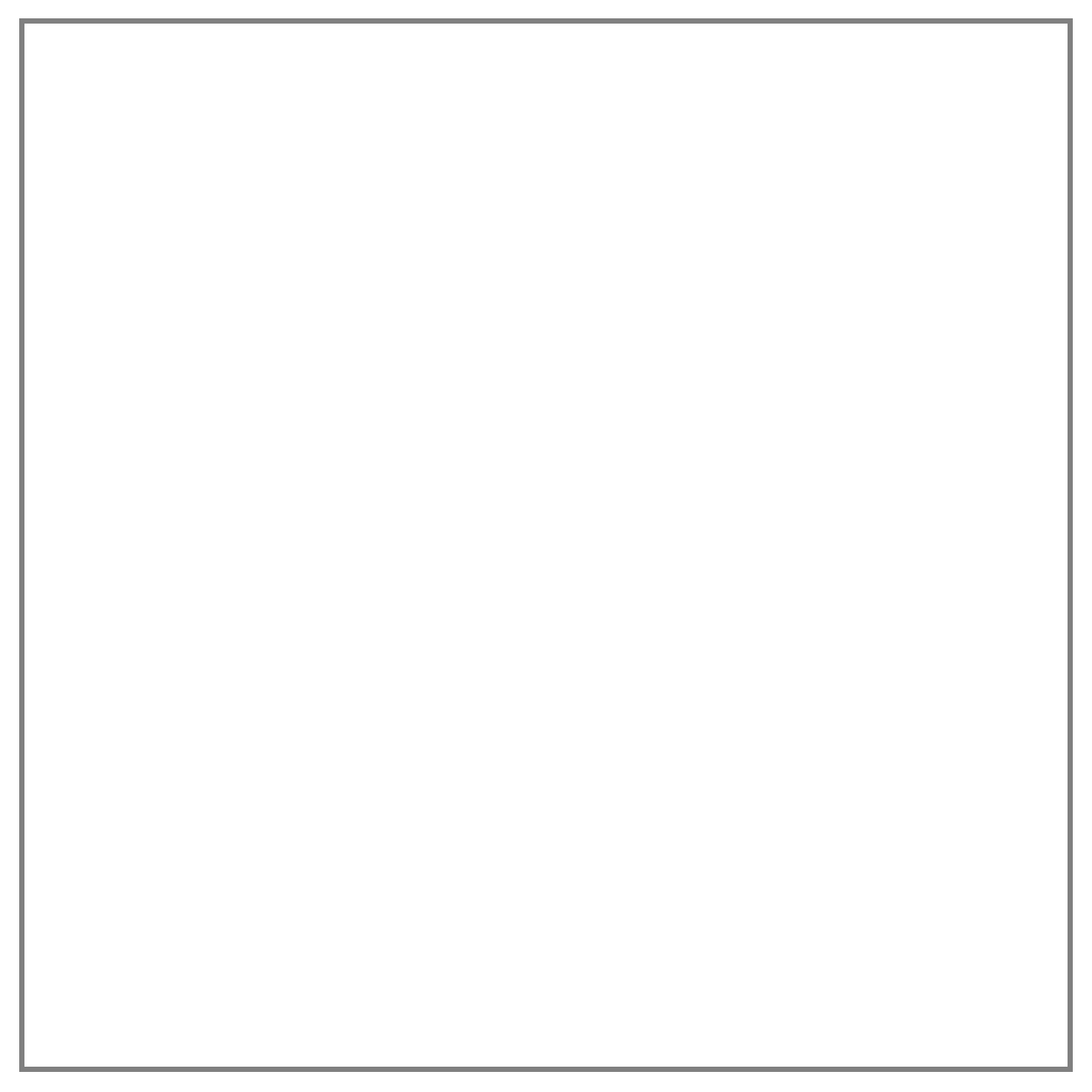} & $Z(\tau, \bar\tau )  = \text{Tr}_{\cal H} [\,q^{L_0-c/24} \bar q^{\bar L_0-c/24}\,]$
\\
\includegraphics[height=.1\textwidth]{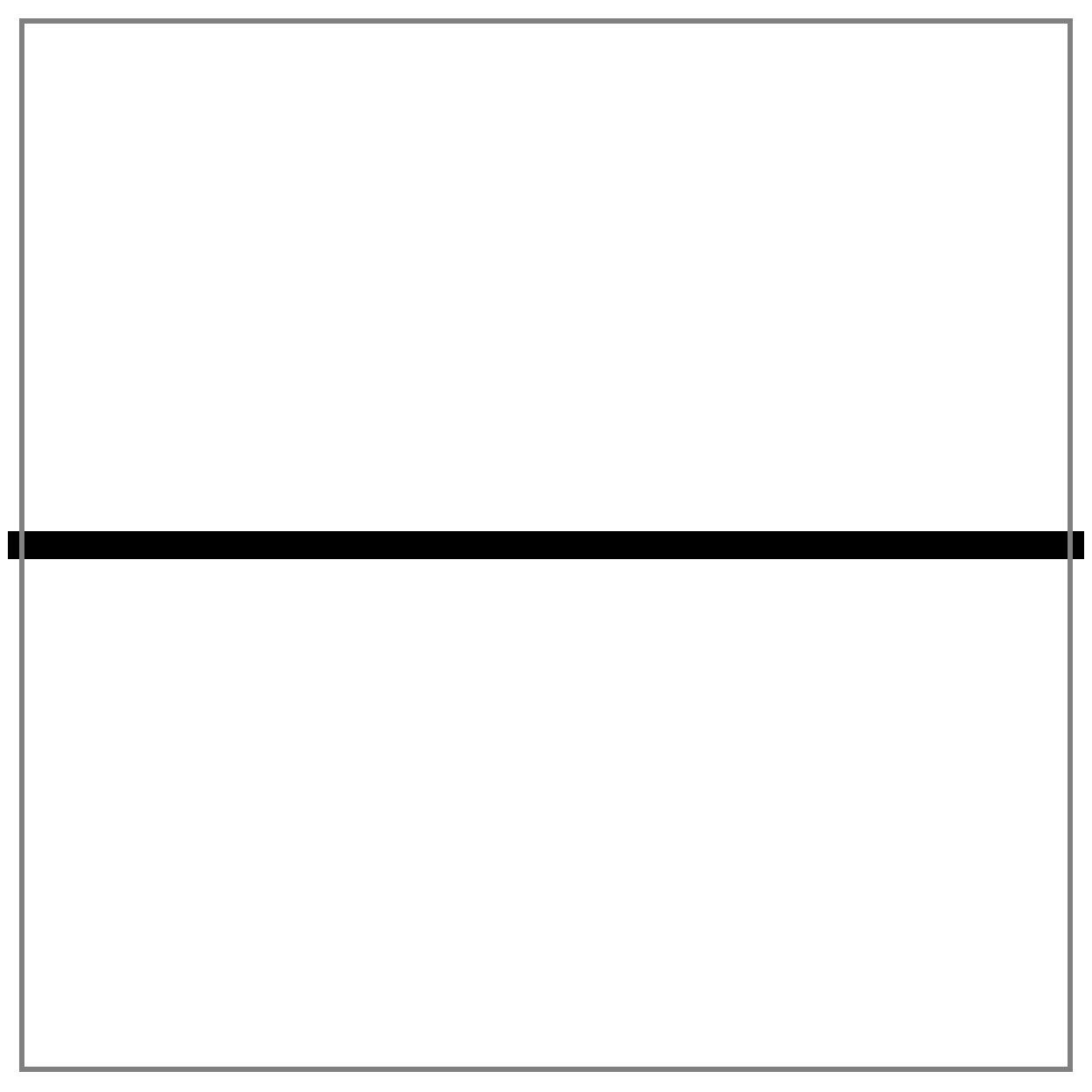} & $Z^{\cal L}(\tau, \bar\tau )  = \text{Tr}_{\cal H} [\, \widehat{\cal L} \, q^{L_0-c/24} \bar q^{\bar L_0-c/24}\,]$
\\
\includegraphics[height=.1\textwidth]{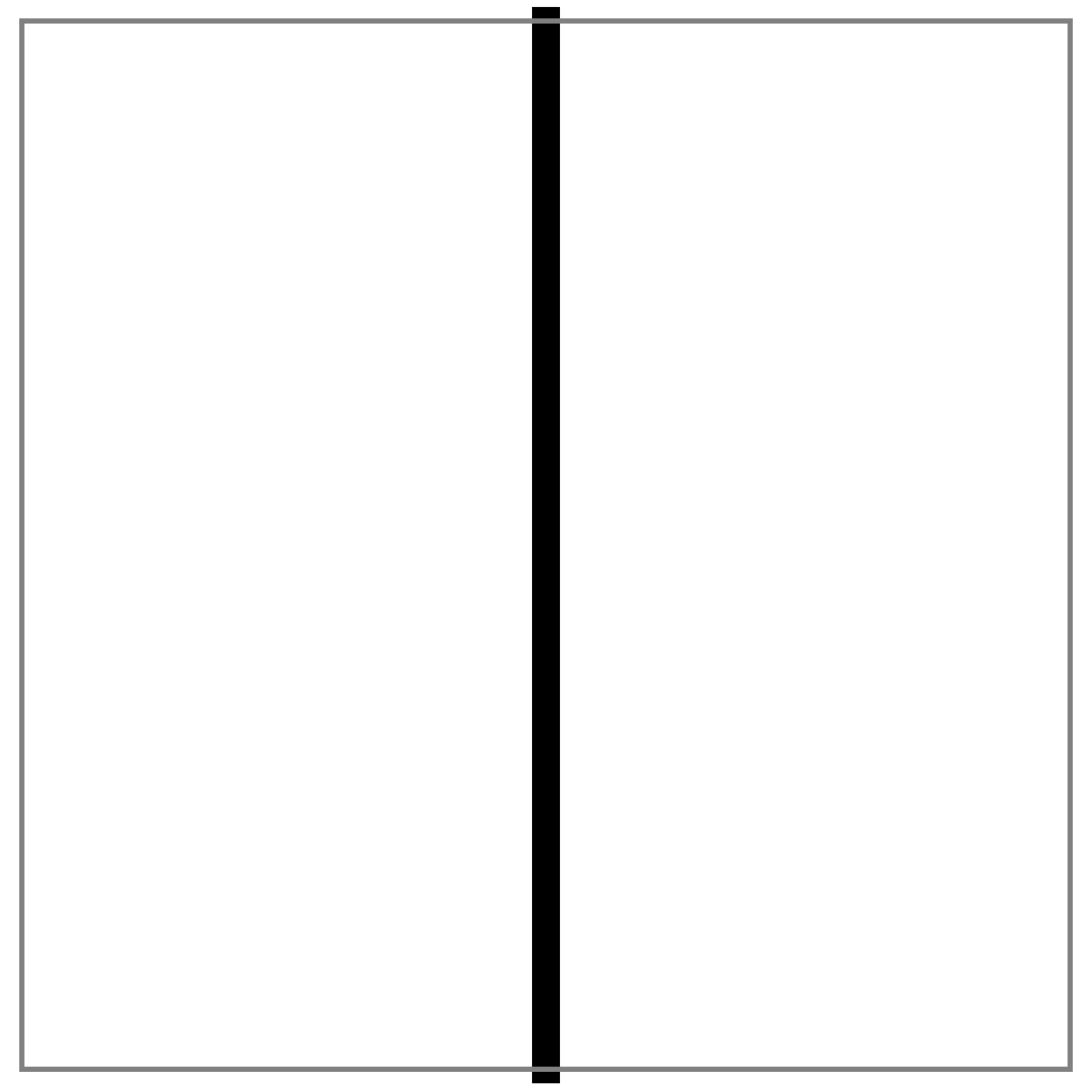} & $Z_{\cal L}(\tau, \bar\tau )  = \text{Tr}_{\cal H_{\cal L}} [\,q^{L_0-c/24} \bar q^{\bar L_0-c/24}\,]$
\\
\includegraphics[height=.1\textwidth]{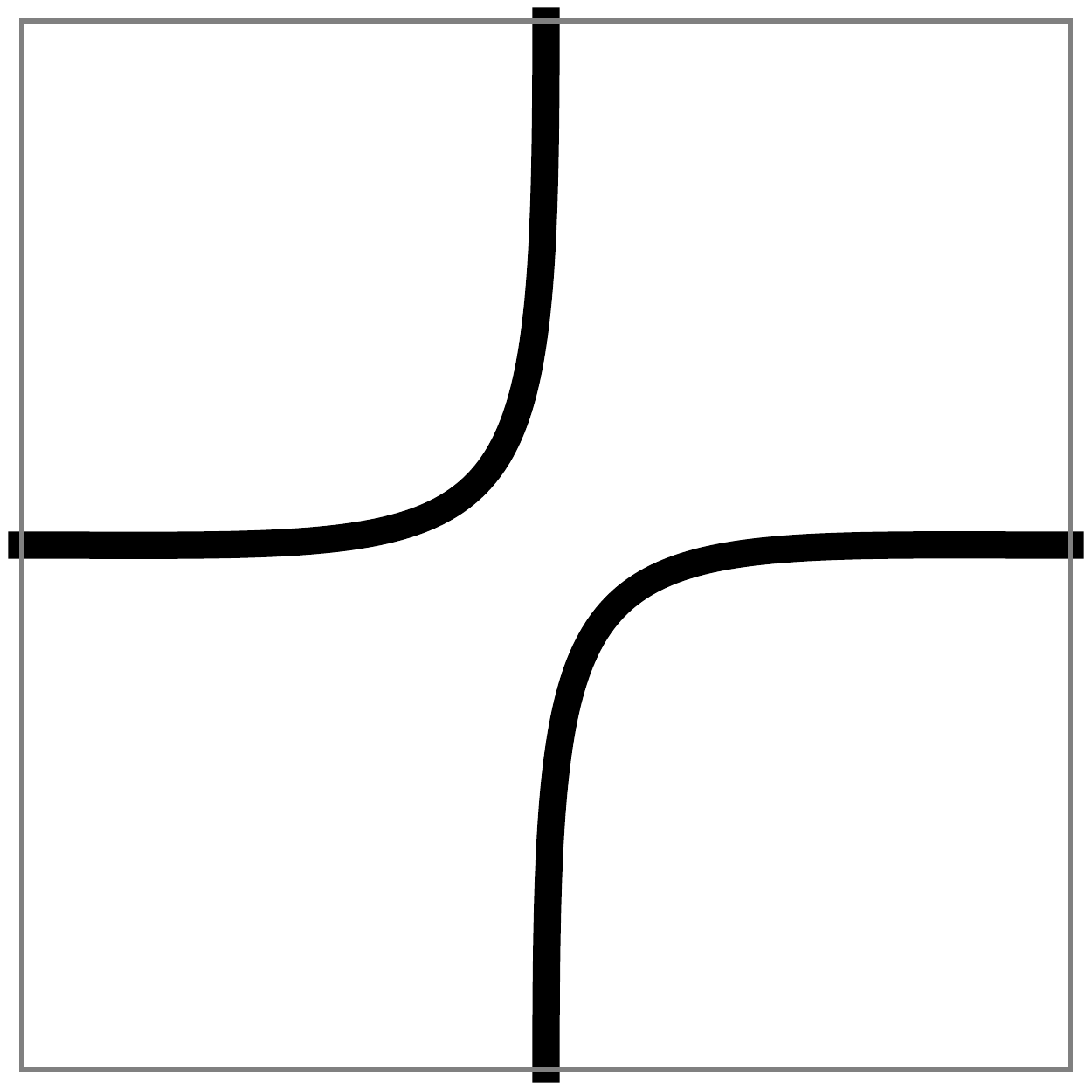} & $Z^{{\cal L}^+}_{\cal L}(\tau, \bar\tau )  = \text{Tr}_{\cal H_{\cal L}} [\,\widehat{\cal L}^+\,q^{L_0-c/24} \bar q^{\bar L_0-c/24}\,]$
\end{tabular}
\end{center}
where $\widehat{\cal L}:{\cal H}\to {\cal H}$ is the $\bZ_2$ symmetry action on $\cal H$  while $\widehat{\cal L}^+$ is an action defined on the defect Hilbert space ${\cal H}_{\cal L}$ (see Figure \ref{fig:Lpm}).  
The consistency of the partition functions under the modular $T$ transform is guaranteed by the spin selection rule derived in Section~\ref{Sec:SpinSelection}. 
In the remainder, we will study the nontrivial constraints imposed by the modular $S$ transform on the partition functions. 

The assumption of $c>1$ (together with the unitarity bound $h,\bar h>0$) simplifies the possible modules of the Virasoro algebra.  There is one degenerate module, {\it i.e.} the vacuum module $h=0$, and a continuous family of non-degenerate modules labeled by a positive conformal weight $h>0$.  
The Virasoro characters are given by
\ie\label{vcharacter}
\chi_0(\tau)=(1-q){q^{-{c-1\over 24}}\over \eta(\tau)},~~~\chi_{h>0}(\tau)={q^{h-{c-1\over 24}}\over \eta(\tau)}\,.
\fe
Combining the left with the right, there are three kinds of Virasoro primaries:
\ie\label{virch}
\text{(vacuum)} \quad & \chi_0(\tau){ \chi}_0(\bar\tau)\,,\\
\text{(conserved current)} \quad &\chi_0(\tau){\chi}_{\bar h>0}(\bar\tau), \quad \chi_{h>0}(\tau){ \chi}_0(\bar\tau)\,,
\\
\text{(non-degenerate)} \quad & \chi_{h> 0}(\tau){ \chi}_{\bar h> 0}(\bar\tau)\,.
\fe

Since the $\bZ_2$ line commutes with the stress tensor up to contact terms, the defect Hilbert space states fall into representations of the Virasoro algebras.  
It follows that the torus partition function $Z(\tau,\bar\tau)$ and $Z_{\cal L}(\tau,\bar \tau)$   are both  given by a non-negative sum over Virasoro characters,
\begin{align}
&Z(\tau,\bar\tau)=\sum_{(h,\bar h)\in {\cal H}} n_{h,\bar h}\, \chi_h(\tau)\chi_h(\bar\tau)\,,\\
&Z_{\cal L}(\tau,\bar\tau)=\sum_{(h,\bar h)\in {\cal H}_{\cal L}} (n_{\cal L})_{h,\bar h}\, \chi_h(\tau) \chi_h(\bar\tau)\,,\label{ZL}
\end{align}
where $n_{h,\bar h}\in \bZ_{\ge0}$ and $(n_{\cal L})_{h,\bar h}\in \bZ_{\ge0}$ are the degeneracies of Virasoro primaries of weight $(h,\bar h)$ in  $\cal H$ and in the defect Hilbert space ${\cal H}_{\cal L}$, respectively.  
In $\cal H$, states have non-negative conformal weights $h,\bar h$ and integer spins:
\begin{align}
(h,\bar h)\in {\cal H}:~~h,\bar h\ge0 \,,~~h-\bar h\in \bZ\,.
\end{align}
On the other hand, the defect Hilbert space states obey a novel spin selection rule \eqref{spinselection} that depends on the anomaly:
\begin{align}
(h,\bar h)\in {\cal H}_{\cal L}:~~h,\bar h\ge0 \,,~~h-\bar h\in {1-\alpha\over8}+{\bZ\over2}\,,
\end{align}
where a non-anomalous $\bZ_2$ has $\alpha=+1$ while an anomalous $\bZ_2$ has $\alpha=-1$.

Recall that  $\cal H$ can be decomposed into the $\bZ_2$ even and odd subsectors ${\cal H} = {\cal H}^+ \oplus {\cal H}^-.$ 
Let the degeneracies of primaries with weight $(h,\bar h)$ in ${\cal H}^\pm$ be $n^\pm_{h,\bar h} \in \bZ_{\ge0}$, respectively.  
By definition, $n_{h,\bar h} = n^+_{h,\bar h} + n^-_{h,\bar h}$, and they are related to $Z(\tau,\bar \tau)$ and  $Z^{\cal L}(\tau,\bar \tau)$ by
\begin{align}
&Z^+(\tau,\bar \tau)\equiv\frac12 \left[ \, Z(\tau,\bar \tau ) + Z^{\cal L} (\tau,\bar \tau) \,\right]  = \sum_{(h,\bar h) \in {\cal H}^+} n_{h,\bar h}^+ \, \chi_h(\tau)\chi_{\bar h}(\bar\tau)\,,\label{Zp}\\
&Z^-(\tau,\bar \tau)\equiv\frac12 \left[ \, Z(\tau,\bar \tau ) - Z^{\cal L} (\tau,\bar \tau) \,\right]  = \sum_{(h,\bar h) \in {\cal H}^-} n_{h,\bar h}^- \, \chi_h(\tau)\chi_{\bar h}(\bar\tau)\,.\label{Zm}
\end{align}

For the defect Hilbert space, recall that ${\cal H}^+_{\cal L}$ is the subsector which has $\widehat{\cal L}_+ = \sqrt{\alpha}$, and    ${\cal H}^-_{\cal L}$ is the subsector which has $\widehat{\cal L}_+ =- \sqrt{\alpha}$. 
Here $\sqrt{-1}$ is taken to be $+i$. 
Let $(n_{\cal L}^\pm)_{h,\bar h}\in \bZ_{\ge0}$ be the degeneracies of Virasoro primaries with weight $(h,\bar h)$ in ${\cal H}_{\cal L}^\pm$, respectively.   
By definition, $(n_{\cal L} )_{h,\bar h}=(n_{\cal L} ^+)_{h,\bar h}+(n_{\cal L} ^-)_{h,\bar h}$, and they are related to $Z_{\cal L}(\tau,\bar\tau)$ and $Z^{{\cal L}^+}_{\cal L}(\tau,\bar \tau)$ by
\begin{align}
&\frac12 \left[ \, Z_{\cal L}(\tau,\bar \tau ) +{1\over \sqrt{\alpha}} Z^{{\cal L}^+}_{\cal L} (\tau,\bar \tau) \,\right]  = \sum_{(h,\bar h) \in {\cal H}_{\cal L}^+} (n^+_{\cal L})_{h,\bar h} \, \chi_h(\tau)\chi_{\bar h}(\bar\tau)\,,\\
&\frac12 \left[ \, Z_{\cal L} (\tau,\bar \tau ) -{1\over \sqrt{\alpha}} Z^{{\cal L}^+}_{\cal L} (\tau,\bar \tau) \,\right]  = \sum_{(h,\bar h) \in {\cal H}_{\cal L}^-} (n^-_{\cal L})_{h,\bar h}\, \chi_h(\tau)\chi_{\bar h}(\bar\tau)\,.
\end{align}

\subsection{Modular Crossing Equation}

From Section \ref{Sec:Computation},  the crossing equations for the  torus partition functions $Z(\tau, \bar\tau)$, $Z^{\cal L}(\tau, \bar\tau)$, $Z_{\cal L}(\tau, \bar\tau)$, $Z^{{\cal L}^+}_{\cal L}(\tau, \bar\tau)$ under the modular $S$ transform are
\begin{align}\label{excrossing}
\begin{pmatrix}
Z(\tau, \bar\tau) \\ Z^{\cal L}(\tau, \bar\tau) \\ Z_{\cal L}(\tau, \bar\tau) \\ Z^{{\cal L}^+}_{\cal L}(\tau, \bar\tau)
\end{pmatrix}
\underset{S}{\longrightarrow}
\begin{pmatrix}
Z(-1/\tau, -1/\bar\tau) \\ Z^{\cal L}(-1/\tau, -1/\bar\tau) \\ Z_{\cal L}(-1/\tau, -1/\bar\tau) \\ Z^{{\cal L}^+}_{\cal L}(-1/\tau, -1/\bar\tau)
\end{pmatrix}
=
\begin{pmatrix}
1~ &  \\  &&1~  \\ &1~ \\ &&&\alpha
\end{pmatrix}
\begin{pmatrix}
Z(\tau, \bar\tau) \\ Z^{\cal L}(\tau, \bar\tau) \\ Z_{\cal L}(\tau, \bar\tau) \\ Z^{{\cal L}^+}_{\cal L}(\tau, \bar\tau)\end{pmatrix}
\end{align}
Note that the anomaly $\alpha=\pm1$ explicitly enters into the crossing equation. 

The dependence on the anomaly in the modular transform of $Z^{{\cal L}^+}_{\cal L}(\tau, \bar\tau)$ can be equivalently implemented via the spin selection rule \eqref{spinselection}. Once the spin is specified, the action of $T$ on a state is determined. 
In this way, we only have to consider the three partition functions, $Z(\tau, \bar\tau)$, $Z^{\cal L}(\tau, \bar\tau)$, $Z_{\cal L}(\tau, \bar\tau)$, while the fourth one $Z^{{\cal L}^+}_{\cal L}(\tau, \bar\tau)$ can be obtained by applying $T$ on $Z_{\cal L}(\tau, \bar\tau)$, {\it i.e.} $Z^{{\cal L}^+}_{\cal L}(\tau, \bar\tau) = T[Z_{\cal L}](\tau, \bar\tau)$.

Define
\ie
{\bf Z}(\tau, \bar\tau) \equiv \begin{pmatrix} Z^+(\tau, \bar\tau) \\  Z^-(\tau, \bar\tau) \\  Z_{\cal L}(\tau, \bar\tau) \end{pmatrix}\,.
\fe
Every component of the vector $\bf Z$  has a non-negative expansion on the Virasoro characters (see \eqref{ZL}, \eqref{Zp}, and \eqref{Zm}). 
Then the crossing equation \eqref{excrossing} under $S$ reduces to 
\ie
\label{SubCrossing}
{\bf Z}(-1/\tau, -1/\bar\tau) = 
 {\bf F} \, {\bf Z}(\tau, \bar\tau)\,,
 \fe
 where the crossing matrix is
 \ie\label{crossingmatrix}
 {\bf F}\equiv
\begin{pmatrix}
{1\over2} & {1\over2} & {1\over2}
\\
{1\over2} & {1\over2} & -{1\over2}
\\
1 & -1 & 0
\end{pmatrix}\,.
\fe
In this new way of writing the crossing equation, while the anomaly $\alpha$ does not explicitly enter into the equation, the allowed spins in the defect Hilbert space ${\cal H}_{\cal L}$ are constrained by the anomaly via \eqref{spinselection}.  
 
We claim that for the purpose of constraining the gaps or scalar gaps in the various sectors, we can assume without loss of generality that in all sectors,
\ie
\label{Parity}
{\bf Z}(\tau, \bar\tau) = {\bf Z}(\bar\tau, \tau),
\fe
which is equivalent to
\ie
\label{nParity}
n^\pm_{h, \bar h} = n^\pm_{\bar h, h}, \quad (n_{\cal L}^\pm)_{h, \bar h} = (n_{\cal L}^\pm)_{\bar h, h}.
\fe
The resulting bounds apply to all partition functions, even the ones that do not satisfy \eqref{Parity}.  
In particular, they apply to CFTs with $c_L = c_R$ but no time-reversal symmetry. The reason is as follows. First, imposing this extra constraint clearly makes the bounds stronger or remain the same. Conversely, for any modular covariant $\bf Z(\tau, \bar\tau)$ that does not necessarily satisfy \eqref{Parity}, its gap and scalar gap are the same as those of ${\bf Z}'(\tau, \bar\tau) \equiv {1\over2}({\bf Z}(\tau, \bar\tau)+{\bf Z}(\bar\tau, \tau))$, which does satisfy \eqref{Parity}.\footnote{In this work, we do not use the fact that the degeneracies are integers, so we are allowed to divide by 2.} Therefore, imposing \eqref{nParity} cannot make the bounds stronger. Hence the claim. By assuming \eqref{nParity}, the partition function in every sector takes the form
\ie
Z(\tau, \bar\tau) = \sum_h n_{h, h} \chi_h(\tau) \chi_h(\bar\tau) + \sum_{h > \bar h} n_{h, \bar h} \left[ \chi_h(\tau) \chi_{\bar h}(\bar\tau) + \chi_{\bar h}(\tau) \chi_h(\bar\tau) \right].
\fe

\subsection{The Linear Functional Method}
\label{Sec:Bootstrap}

The most general putative spectrum $\cal S=\{ {\cal H}^+,{\cal H}^-,{\cal H}_{\cal L}\}$ we will consider contains the following:
\begin{enumerate}
\item Vacuum ($h = \bar h = 0$) {\it only} in the untwisted $\bZ_2$ even sector ${\cal H}^+$.
\item Conserved currents ($h = 0$ or $\bar h = 0$, but not both) in all sectors, including twisted.
\item Non-degenerate primaries ($h, \bar h > 0$) in all sectors.
\end{enumerate} 
Furthermore, the spins in  $\cal H$ are integers while those in the defect Hilbert space ${\cal H}_{\cal L}$ obey the spin selection rule \eqref{spinselection}. 

The first assumption requires some explanations.  Since we assume that there is a unique vacuum in  $\cal H$ that is invariant under the global symmetry, there is no weight-$(0,0)$ operator in the $\bZ_2$ odd sector ${\cal H}^-$. 
In the defect Hilbert space ${\cal H}_{\cal L}$, on the other hand, the existence of a  weight-$(0,0)$ state would have implied that the $\bZ_2$ symmetry commutes with all local operators  (see Section 2.2.5 of \cite{Chang:2018iay}), thus violating the  assumption that global symmetry acts faithfully on local operators. 

Let us write the modular crossing equation \eqref{SubCrossing} in component form as\footnote{For notational convenience, we will allow ourselves to freely raise and lower the $\pm$ index, and identify $Z_\pm  = Z^\pm$, $(n^\pm)_{h,\bar h} = (n_\pm)_{h,\bar h}$, etc. However, the $Z_{\cal L}$ is completely different from $Z^{\cal L}$; the former is the defect Hilbert space partition function, while the latter is the partition function of $\cal H$ weighted by the $\bZ_2$ action $\widehat {\cal L}$.}
\ie
\label{CrossingComponentForm}
Z_i (-1/\tau, -1/\bar\tau) - \sum_{j=\pm,{\cal L}} F_i^{\, j} Z_j (\tau, \bar\tau) = 0,
\fe
where the index $i,j$ runs over $+, - , {\cal L}$, corresponding to the untwisted $\bZ_2$ even, untwisted  $\bZ_2$ odd, and the defect Hilbert spaces, respectively. 
 Next, we define $\tilde Z_i(\tau, \bar\tau) \equiv Z_i (-1/\tau, -1/\bar\tau)$, and introduce the following shorthand for characters
\ie
\chi_{h, \bar h}(\tau, \bar\tau) \equiv \chi_h(\tau) \chi_{\bar h}(\bar\tau), \quad \tilde\chi_{h, \bar h}(\tau, \bar\tau) \equiv \chi_h(-1/\tau) \chi_{\bar h}(-1/\bar\tau).
\fe
The  linear functional method is implemented as follows. Suppose $\A$ is a  linear functional acting on functions of $\tau,\bar \tau$, then
\ie
\label{FunctionalOnCrossing}
0 &= \sum_j ( \D_i^{\,j} \A[\tilde Z_j] - F_i^{\, j} \A[Z_j] ) = \sum_{j, h, \bar h} (n_j)_{h, \bar h} \left( \D_i^{\, j} \A[\tilde\chi_{h, \bar h}] - F_i^{\, j} \A[\chi_{h, \bar h}] \right)\,,
\fe
where  $(n_i)_{h,\bar h}$'s are all non-negative integers. 
A putative spectrum  ${\cal S}=\{{\cal H}^+, {\cal H}^-, {\cal H}_{\cal L}\}$ is ruled out if there exists a functional such that for each $j=+,-,{\cal L}$,
\ie
\label{Non-negativity}
\sum_{i=\pm, {\cal L}}  \left( \D_i^{\, j} \A[\tilde\chi_{h, \bar h}] - F_i^{\, j} \A[\chi_{h, \bar h}] \right) \geq 0, \quad \forall ~~ (h, \bar h) \in {\cal H}_j \,.
\fe
In practice, the functional $\A$ will be taken to be linear combinations of  derivatives $\partial_\tau^{m}\partial_{\bar \tau}^{\bar m}$ evaluated at $\tau = -\bar\tau = i$. 

To give an upper bound on the gap in a particular sector labeled by $j=\pm, {\cal L}$, we assume that the non-degenerate primaries in ${\cal H}_j$ all have scaling dimensions above a certain value $\Delta_\text{gap}^j$, and ask if a functional $\A$ exists that satisfies the non-negativity conditions. If it exists, then the assumption is ruled out, so we lower $\Delta_\text{gap}^j$ and try again; if no such functional $\A$ exists, then we raise $\Delta_\text{gap}^j$ and try again. This process is repeated until we find the smallest $\Delta_\text{gap}^j$ (to within the targeted precision) for which such $\A$ exists. This smallest $\Delta_\text{gap}^j$ is then the best bound on the the gap in the $j$ sector. Note that we can subject $\cal S$ to other assumptions, say the gaps in the other sectors, or the existence or non-existence of certain conserved currents, to get varying bounds on the gap.  

The gaps in the subsectors ${\cal H}_{\cal L}^\pm$ of ${\cal H}_{\cal L}$ can  be distinguished from the spin-charge relations \eqref{spincharge1} and \eqref{spincharge2}.  For the defect Hilbert space, however, we will only study the overall gap in ${\cal H}_{\cal L}$ but not in the individual subsectors in this paper.

We end this section with a technical comment.  In practice, it is easier to work with the reduced partition function $\hat Z(\tau,\bar\tau)$, defined as 
\ie
\hat Z(\tau,\bar\tau)  \equiv |\tau|^{1\over2}|\eta(\tau)|^2 Z(\tau,\bar\tau)\,.
\fe
The reduced partition functions $\hat Z^\pm(\tau,\bar\tau), \hat Z_{\cal L}(\tau,\bar\tau)$ satisfy the same crossing equation \eqref{SubCrossing} as before, {\it i.e.} $\hat{\bf Z} (-1/\tau ,-1/\bar\tau ) = {\bf F}\, \hat{\bf Z}(\tau,\bar\tau)$.  
The advantage of working with the reduced partition functions is that they can be expanded on the reduced Virasoro characters
\ie\label{reducedch}
\hat\chi_{h,\bar h} (\tau,\bar \tau ) \equiv |\tau|^{1\over2}|q^{-{c-1\over24}} (1-q)|^2\,,~~\hat \chi_{h,\bar h}(\tau,\bar\tau )\equiv q^{h-{c-1\over24}}\bar q^{\bar h-{c-1\over24}}\,,
\fe
which are simpler functions than the Virasoro characters.

\section{Analytic Bound on the $\bZ_2$ Odd Operators}\label{Sec:Analytic}

We begin with an analytic study of the crossing equation, to serve as a conceptual guideline and a warmup for the systematic study of the modular bootstrap in Section~\ref{Sec:General}.  
We focus on the upper bound on the lightest  primary in each sector and how it depends on  the 't Hooft anomaly.  
In this section, we define the gap in each sector to be the lightest (nontrivial) primary, which can either be a conserved current or a non-degenerate primary.  In Section~\ref{Sec:General}, the gap will be defined as the lightest {\it non-degenerate} Virasoro primary in the sector of interest.

We pay special attention to the bound on the lightest $\bZ_2$ odd primary in ${\cal H}^-$, denoted by ${\bf \Delta}_{\rm gap}^-$. In particular, we will find that 
\begin{itemize}
\item There is a universal upper bound on the lightest $\bZ_2$ odd primary if the $\bZ_2$ is anomalous, but not otherwise.
\end{itemize}
In Section~\ref{Sec:OddBound}, we derive such a bound for an anomalous $\bZ_2$ for $0\le c\le3$, while a stronger numerical bound for larger values of $c$ will be presented in later sections.

\subsection{Free Compact Boson Example}\label{Sec:CompactBoson}

Let us consider the lightest $\bZ_2$ odd operator in the Hilbert space $\cal H$ of local operators.  Since the $\bZ_2$ is a global symmetry, this operator must necessarily be a Virasoro primary.  
When do we expect there to be an upper bound on the scaling dimension for this operator? 

A motivating example is the $c=1$ free compact boson theory with radius $R$, {\it i.e.} $X(z,\bar z)\sim X(z,\bar z)+2\pi R$. 
We review the theory and analyze its $\bZ_2$ symmetries and anomalies in Appendix \ref{App:FreeBoson}.  
At every radius $R$, there are two non-anomalous $U(1)$ global symmetries, the momentum $U(1)_n$ and the winding $U(1)_w$.  
We consider the $\bZ_2$ subgroups of these $U(1)_n$ and $U(1)_w$, and denote them by $\bZ_2^{(1,0)}$ and $\bZ_2^{(0,1)}$, respectively. 
While the $\bZ_2^{(1,0)}$ and $\bZ_2^{(0,1)}$ are separately non-anomalous, there is a mixed anomaly between the two. 
Consequently, the diagonal $\bZ^{(1,1)}_2$ subgroup is anomalous. 
The anomalies can be computed, for example, by comparing the spin content in the defect Hilbert space \eqref{c=1twgs} to the spin selection rule \eqref{spinselection}.  
Each current algebra primary ${\cal O}_{n,w}(z,\bar z)= \exp[i({n\over R}+wR)X_L(z) +i ({n\over R} -wR)X_R(\bar z)]$ has the following $\bZ_2$ charges:
\ie
\left.\begin{array}{c|c|c|c}
&~~ \bZ_2^{(1,0)} ~~&~~\bZ _2^{(0,1)} ~~&~~ \bZ _2^{(1,1)} ~~\\\hline
~~{\cal O}_{n,w}(z,\bar z)~~& e^{i\pi n} & e^{i\pi w} & e^{i\pi (n+w)} \\ 
\end{array}\right.
\fe

Let us examine the lightest $\bZ_2$ odd primary in the $c=1$ free compact boson theory for each of the above $\bZ_2$ symmetries.  
The lightest (non-anomalous) $\bZ_2^{(0,1)}$ odd primary is the minimal winding exponential operator ${\cal O}_{0,1}$, which has scaling dimension ${R^2\over 2}$ (see \eqref{expweight}).  
This minimal winding state becomes arbitrarily heavy as we take the radius $R$ to be large. Hence, for $c=1$, there is no upper bound on the lightest $\bZ_2^{(0,1)}$ odd primary. Similarly, the lightest (non-anomalous) $\bZ_2^{(1,0)}$ odd primary is the minimal momentum operator ${\cal O}_{1,0}$, whose scaling dimension is $1\over 2R^2$, so there is no bound either.

The above $c=1$ example can be extended to larger values of $c$ by considering the tensor product with any other CFT to produce a theory with $\bZ_2$ symmetry whose lightest odd primary is not bounded from above.  Indeed, in Section~ \ref{Sec:c>1}, the numerical bootstrap finds no bound for $1 \le c \le 25$.  It is therefore reasonable to expect that \textit{for all $c \ge 1$ CFTs, there is no upper bound for the lightest $\bZ_2$ odd primary if the $\bZ_2$ is non-anomalous, $\A=+1$.} 

By contrast, the lightest (anomalous) $\bZ_2^{(1,1)}$ odd primary is {\it either} the minimal momentum operator ${\cal O}_{1,0}$ {\it or} the minimal winding operator ${\cal O}_{0,1}$.  It is impossible to make {\it both} of them heavy as we vary $R$, so we do have an upper bound:  the lightest $\bZ_2^{(1,1)}$ odd primary is bounded from above by $1\over2$ on the moduli space of the free compact boson.  In Section \ref{Sec:OddBound} (and more generally in Section \ref{Sec:c>1}), we will show that for general $c \ge 1$, there is a bound on the lightest $\bZ_2$ odd primary if the symmetry is anomalous $\A=-1$.  

We summarize the above discussions in the following table:
\ie
\left.\begin{array}{c|c|c|c}
&~~ \bZ_2^{(1,0)} ~~&~~\bZ _2^{(0,1)} ~~&~~ \bZ _2^{(1,1)} ~~\\
\hline
\text{anomaly}~~\alpha~~&+1&+1&-1 \\
\hline
~~\text{lightest odd op.}~~& {\cal O}_{1,0} & {\cal O}_{0,1}&  ~{\cal O}_{1,0}~~\text{or}~~{\cal O}_{0,1} \\ 
\hline
{\bf \Delta}_{\rm gap}^-&{1\over 2R^2}& {R^2\over2}& ~\text{Min}({1\over2R^2}, {R^2\over2})
\end{array}\right.
\fe

\subsection{The Cardy Limit}
\label{Sec:Cardy}

In the free compact boson example, we saw that there is a bound in ${\cal H}^-$ if the $\bZ_2$ is  anomalous. 
In this subsection, we argue that this is true more generally by taking the high/low temperature (Cardy) limit of the modular $S$ transformation of the torus partition function.  The lower bound \eqref{twistedgap} on the scaling dimension in the defect Hilbert space will turn out to be the key.

Let us review the standard Cardy argument for a modular invariant partition function $Z(\tau, \bar\tau)$. For simplicity, we set $\tau=-\bar \tau = it$, with $t > 0$.  We have,
\ie
Z(t )  = Z(1/t)\,.
\fe
Now we take $t\to0$, so that the righthand side is dominated by the vacuum. The equation becomes
\ie\label{Cardy}
\int d\Delta\, \rho(\Delta)\, e^{-2\pi t(\Delta-{c\over 12})} \sim e^{{ \pi\over 6t} c}\,,
\fe
where $\rho(\Delta)$ is the density of states with respect to the scaling dimension $\Delta$.  
  The Casimir energy $c \over 12$ results in a divergence as we take $t\to0$, which must be reproduced on the lefthand side by the exponential growth of the high energy states:
\ie\label{rho}
\log \rho(\Delta )\sim 2\pi \sqrt{ { c\over 6}  (\Delta-{c\over 12})}\,.
\fe

With the $\bZ_2$, we consider the $t\to0$ limit of
\ie\label{CardyZL}
Z^{\cal L}(t) = Z_{\cal L}(1/t)\,.
\fe
The righthand side is dominated by the defect Hilbert space ground state with scaling dimension $\Delta_{\cal L}$:
\ie\label{CardyZ2}
\int d\Delta\, \left[\rho^+(\Delta)- \rho^-(\Delta)\right]\, e^{-2\pi t(\Delta-{c\over 12})} \sim e^{{ \pi\over 6t} (c-12\Delta_{\cal L})}\,,
\fe
where $\rho^\pm(\Delta)$ are the densities of $\bZ_2$ even/odd states of scaling dimension $\Delta$, respectively.  

In the anomalous case \eqref{twistedgap}, $\Delta_{\cal L}$ of the defect Hilbert space ground state is bounded from below by $1\over4$, so the divergence on the righthand side of \eqref{CardyZ2} is smaller than that of \eqref{Cardy} as $t\to0$. This means that the weighted density of states $\rho^+(\Delta)-\rho^-(\Delta)$ has, if any, a {\it slower} exponential growth
\ie
\log(\rho^+(\Delta)-\rho^-(\Delta)) \sim 2\pi \sqrt{ {c-12\Delta_{\cal L} \over 6} (\Delta - {c\over12}) }
\fe
than that of the total density of states $\rho(\Delta)=\rho^+(\Delta)+\rho^-(\Delta)$ in \eqref{rho}.  
This means that there must be huge cancellations between the $\bZ_2$ even and odd heavy states when the $\bZ_2$ is anomalous. 
Hence, the $\bZ_2$ odd states cannot be pushed to be arbitrarily heavy for an anomalous $\bZ_2$. 

By contrast, $\Delta_{\cal L}$ for a non-anomalous $\bZ_2$ can be arbitrarily close to 0, so the $\bZ_2$ odd states can be arbitrarily heavy, consistent with the analysis in Section~\ref{Sec:CompactBoson}.

The  common weakness in the above arguments is that by taking the Cardy limit alone, we only arrive at {\it asymptotic} formulae for the relation between the defect Hilbert space ground state and the heavy $\bZ_2$ odd states, but we do not have quantitative control over the regime of validity of the formulae.  
The quantitative bound will be derived in Section \ref{Sec:OddBound} (and more generally in Section \ref{Sec:c>1}), by considering the medium temperature expansion of the modular crossing equation.

\subsection{Analytic Bound on the  $\bZ_2$ Odd Operators}\label{Sec:OddBound}

In this subsection, we derive an analytic bound for the lightest $\bZ_2$ odd primary under an anomalous $\bZ_2$, for $0\le c\le3$.  
For simplicity, we only use the dilitation symmetry, instead of the full Virasoro symmetry.
We also ignore the dependence on the spin $h-\bar h$, and only keep track of the scaling dimension $\Delta=h+\bar h$ of the operators. 
We set $\tau=-\bar \tau = it$, and expand the torus partition function in each sector as
\ie
&Z^\pm(t) \equiv\frac12 \left[ Z(t) \pm Z^{\cal L}(t)\right]= \sum_{\Delta\in {\cal H}^\pm}(N_\pm)_\Delta g_\Delta (t)\,,\\
&Z_{\cal L}(t) =\sum_{\Delta\in {\cal H}_{\cal L} } (N_{\cal L})_\Delta g_\Delta(t)\,,
\fe
where $(N^\pm)_ \Delta\in \bZ_{\ge 0}$ and $(N_{\cal L})_\Delta\in \bZ_{\ge0}$ are the degeneracies of {\it all states} (not necessarily primaries) with scaling dimension $\Delta$, in ${\cal H}^\pm$ and in ${\cal H}_{\cal L}$, respectively.  
Here, $g_\Delta(t)$ is the {\it scaling character} that counts the contribution from a single operator of dimension $\Delta$:
\ie\label{scalingch}
g_\Delta(t)= e^{-2\pi t (\Delta-{c\over12}) }\,.
\fe
The only dependence on the anomaly $\A$ is that the scaling dimensions in the defect Hilbert space ${\cal H}_{\cal L}$ is bounded from below by (see  \eqref{twistedgap}):
\ie 
\Delta\ge {1-\A\over8}\,,~~~~~\forall~~\Delta\in {\cal H}_{\cal L}\,.
\fe
This difference turns out to be crucial for the existence of  a bound in ${\cal H}^-$ in the anomalous case.

Let us write the  modular crossing equation ${\bf Z}(1/t)  - {\bf F} \, {\bf Z}(t) = 0$ \eqref{SubCrossing} as a vector equation with contributions from the three sectors ${\cal H}^\pm$ and ${\cal H}_{\cal L}$:
\ie\label{crossing2}
\sum_{\Delta\in {\cal H}^+} (N^+)_\Delta \,  M_{i}^{\,+} (t) +\sum_{\Delta\in {\cal H}^-} (N^-)_\Delta \, M_{i}^{\,-} (t)
 +\sum_{\Delta\in {\cal H}_{\cal L}} (N_{\cal L})_\Delta  \,M_{i}^{\,{\cal L}} (t)  = 0 \,,~~~i=\pm, {\cal L}
\fe
where  the matrix $M_{i}^{\,j}(\Delta,t)$ is defined as
\ie\label{M}
M_{i}^{\,j}(\Delta,t) \equiv \delta_i^{\,j} g_\Delta(1/t) - F_i^{\,j} g_\Delta (t)\,.
\fe
For a fixed $j$, we write ${\bf M}^j(\Delta,t)$ to denote a vector-valued function of $\Delta$ and $t$ with components $M_i^{\,j}(\Delta,t)$.

To bound the lightest $\bZ_2$ odd primary, we look for a real linear functional $\A$ that acts on vector-valued functions of $t$ (with 3 components). If for some ${\bf \Delta}_{\rm gap}^->0$, we can find a functional $\A$ such that
\ie\label{functional}
&\A[{\bf M}^{+}(\Delta, t)]\ge0 \,,~~~~\forall~~\Delta\ge0\,,\\
&\A[{\bf M}^- (\Delta,t)]\ge 0 \,,~~~~\forall~~\Delta \ge{\bf \Delta}_{\rm gap}^-\,,\\
&\A[{\bf M}^{\cal L} (\Delta,t) ]\ge0\,,~~~~\forall ~~\Delta\ge {1-\A\over 8}\,,
\fe
 then by applying $\A$ on \eqref{crossing2}, we reach a contradiction unless there is a $\bZ_2$ odd operator in ${\cal H}$ below ${\bf \Delta}_{\rm gap}^-$.   In this way, we obtain an upper bound ${\bf \Delta}_{\rm gap}^-$ on the lightest $\bZ_2$ odd primary. 

Consider a derivative basis for linear functionals acting on vector-valued functions ${\bf V}$ of $t$:\footnote{This is a linear functional in the sense that it acts on functions of $t$ over the field of functions of $\Delta$. The normalization and exponential factor are chosen for later convenience.
}
\ie
\label{alphabasis}
\A_{n,i} [ {\bf V}(t) ] \equiv  e^{2\pi (\Delta-{c\over12})}  \left( {6 t \over \pi {c}} {d \over dt} \right)^n  V_i(t) \Big|_{t = 1}.
\fe
We will expand our functional on this basis.  
When acting linear functionals on $\bf  M$, there are linear relations among $\A_{n, i}$, and the number of independent functionals is determined by the matrix rank of ${\bf I} - (-)^n {\bf F}$. For the $\bf F$ considered here \eqref{crossingmatrix}, there is one independent functional at each even derivative order, and two at each odd order. Thus the most general cubic linear functional takes the form
\ie\label{alphaexpand}
\A = \sum_{n\text{ even}} \gamma^{n, 1} \A_{n, 1} + \sum_{n\text{ odd}} \sum_{i=1}^2 \gamma^{n, i} \A_{n, i} \,.
\fe
As a proof of principle, we restrict our functionals to derivative order 3, {\it i.e.} $0\le n\le3$.  Stronger numerical bounds using more general functionals will be presented in Section \ref{Sec:c>1}.

\begin{figure}[H]
\centering
\subfloat{
\includegraphics[height=.2\textwidth]{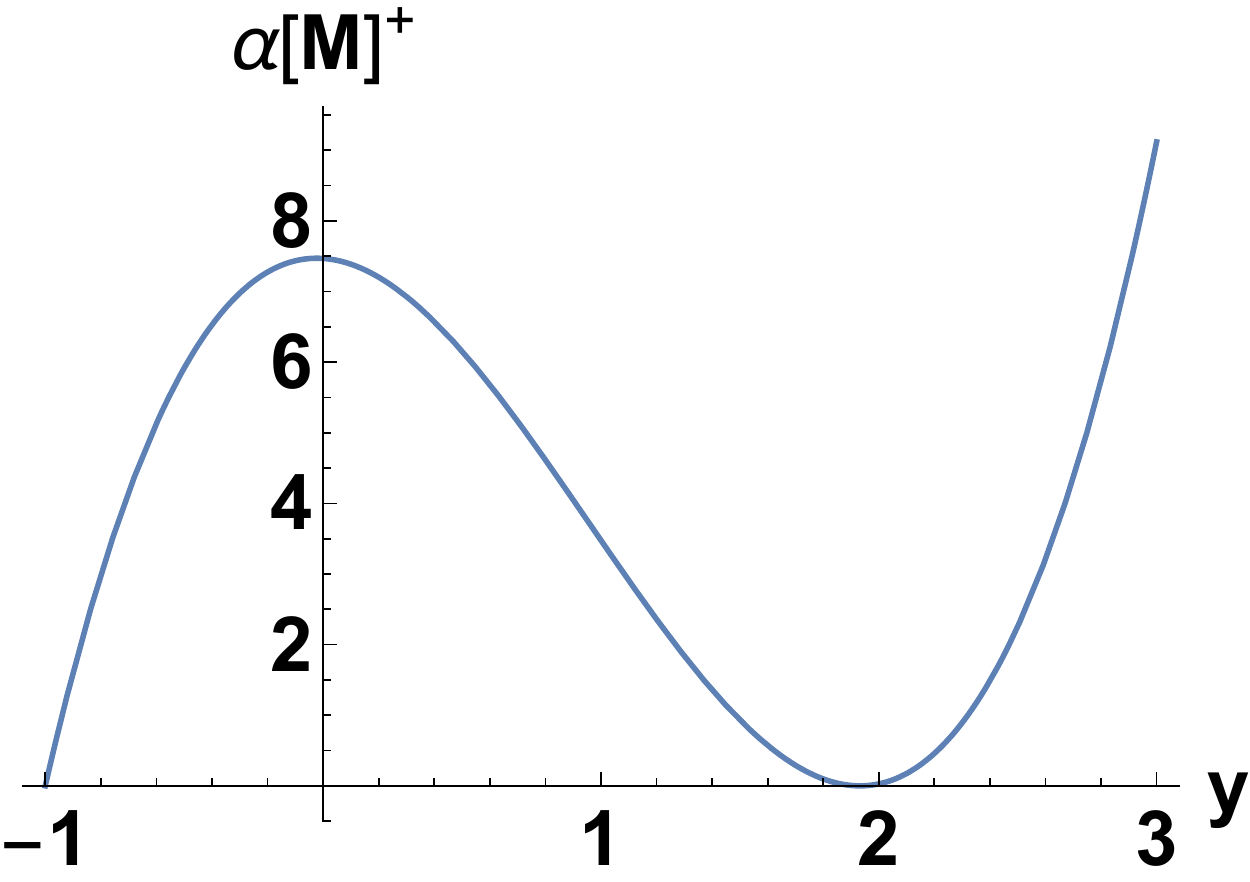}
}
\subfloat{
\includegraphics[height=.2\textwidth]{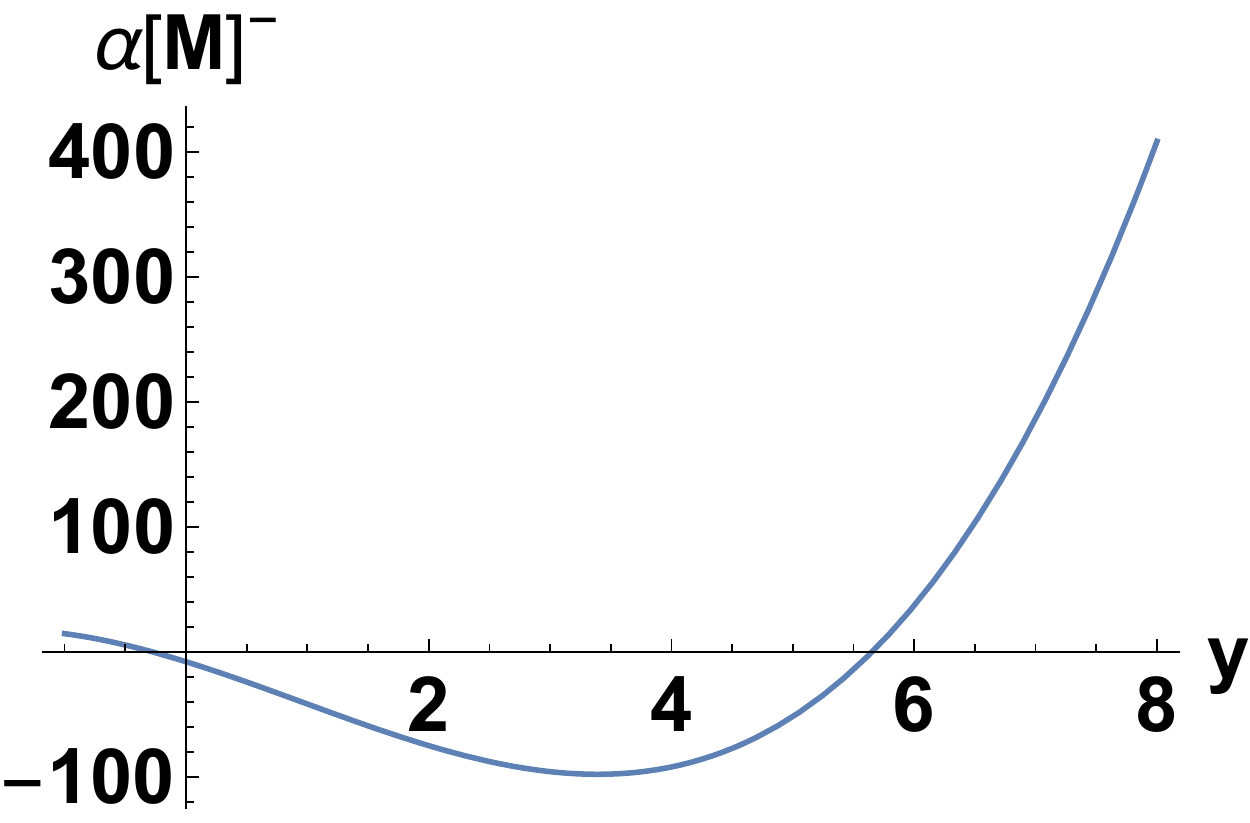}
}
\subfloat{
\includegraphics[height=.2\textwidth]{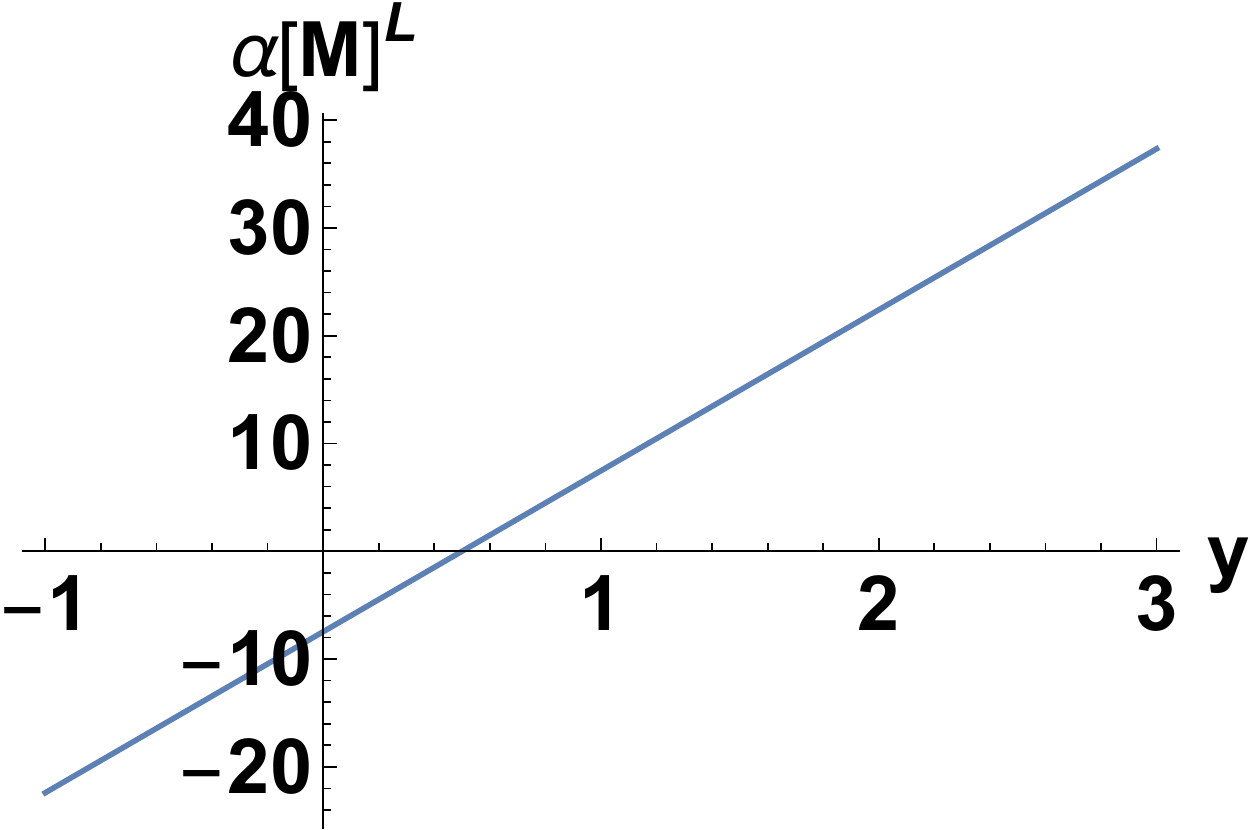}
}
\caption{Linear functional \eqref{ChargedFunctional} acted on ${\bf M}^j(\Delta,t)$ for $c = 2$.}
\label{Fig:ChargedFunctional}
\end{figure}

\begin{figure}
\centering
\includegraphics[height=.33\textwidth]{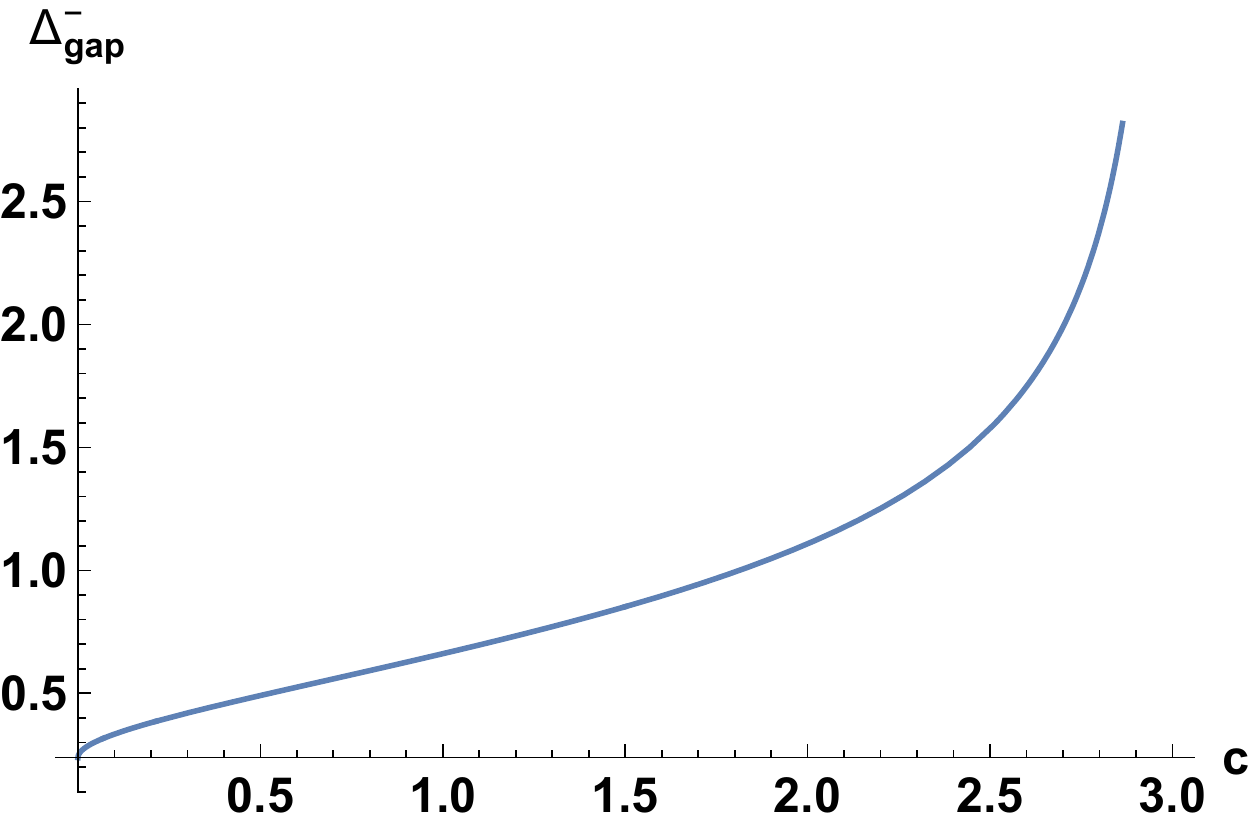}
\caption{Upper bound  on the lightest  $\bZ_2$ odd primary in $\cal H$ as a function of the central charge $c$.  The region below the curve is allowed.   A stronger bound is presented in the right figure of Figure \ref{Fig:AnomalousGeneral}.}\label{Fig:Charged}
\end{figure}

In Sections~\ref{Sec:CompactBoson} and~\ref{Sec:Cardy}, we argued that a universal bound on the lightest $\bZ_2$ odd primary should only exist if there is 't Hooft anomaly. 
We find that, indeed, there is no cubic linear functional satisfying \eqref{functional} if the $\bZ_2$ is non-anomalous ($\A=+1$). 
However, if we only need to maintain positivity for $\Delta \ge {1\over4}$ in the defect Hilbert space, such as in the anomalous case, then we are able to construct linear functionals for $c$ sufficiently small. 
In particular, the following choice of linear functional satisfies \eqref{functional} for $0 < c < 3$:
\footnote{At $c=3$, we can alternatively choose the following linear functional  $\gamma^{1,1} = -\frac{9}{16}-\frac{32}{\pi ^2}-\frac{9}{\pi }, 
\gamma^{1,2} = \frac{27}{16}+\frac{72}{\pi ^2}+\frac{21}{\pi },
\gamma^{2,1} = \frac{3(6+\pi )}{\pi },
\gamma^{0,1} = \gamma^{3,2} = 0,  \gamma^{3,1} = -1.$ 
This gives ${\bf\Delta}_{\rm gap}^- =  \frac{24+5 \pi +\sqrt{3 \left(384+112 \pi +9 \pi ^2\right)}}{8 \pi } \approx 3.559.$
}
\ie
\label{ChargedFunctional}
& \gamma^{0,1} = \frac{(\pi  c+18)^2}{\pi ^2 c^2}, \quad \gamma^{1,1} = \frac{4 \pi ^2 c^3-9 (\pi -8) \pi  c^2-36 (3 \pi -4) c+540}{4 \pi ^2 (c-3) c^2}, 
\\
& \gamma^{1,2} = -\frac{9 \left(\pi  (8+\pi ) c^2+4 (32+3 \pi ) c-60\right)}{4 \pi ^2 (c-3) c^2}, \quad \gamma^{2,1} = 0, \quad \gamma^{3,1} = \gamma^{3,2} = -1.
\fe
Its actions on ${\bf M}^j(\Delta,t)$ give (see Figure~\ref{Fig:ChargedFunctional})
\ie\label{ChargedAct}
&\A[{\bf M}^+(\Delta,t) ] = {1\over 2c^2\pi^2} (y+1)(2\pi c y - c\pi -18  )^2\,,\\
&\A[{\bf M}^-(\Delta,t) ] = { 4 \pi ^2 c^2 y^3-72 \pi  c y^2-(\pi  c+18)^2\over 2c^2\pi^2} 
- {(\pi  c+18) \left(\pi  c^2-9 (6+\pi ) c+54\right) y \over2c^2\pi^2(c-3)}\,,\\
&\A[{\bf M}^{\cal L}(\Delta,t) ] = - {(\pi  c+18)^2 (c y+c-3)  \over2c^2\pi^2(c-3)}\,,
\fe
where $y\equiv {12\Delta\over c }-1$.  
The resulting upper bound on the lightest $\bZ_2$ odd primary in the anomalous case is given by
\ie
\boxed{\,
{\bf \Delta}_\text{gap}^- \le (\widehat y+1) {c \over 12}\,}\,,
\fe
where $\widehat y$ is the largest root of $\A[{\bf M}^-(\Delta,t)]$.  See Figure \ref{Fig:Charged}.

\subsection{Tensor Product with TQFTs}\label{Sec:TQFT}

In any spacetime dimension $d$, given a bosonic anomaly $\A$ for a discrete, internal, global symmetry $G$, it was shown that there exists a $d$-dimensional TQFT with symmetry $G$ and anomaly $\A$ \cite{Wang:2017loc,Tachikawa:2017gyf}.
Therefore, for any bosonic QFT with the symmetry $G$ and anomaly $\A'$, we can take the tensor product theory QFT $\otimes$ TQFT, and consider the diagonal symmetry $G$, to realize the anomaly $\A\A'$.  

In 2d, such a TQFT with a nontrivial anomaly always has a degenerate vacua, {\it i.e.} it describes the spontaneously broken phase.  
For example, the TQFT with an anomalous $\bZ_2$ has two degenerate vacua, one $\bZ_2$ even and the other $\bZ_2$ odd.  
Now, if we take the tensor product of a CFT with a non-anomalous $\bZ_2$ symmetry, together with this anomalous $\bZ_2$ TQFT, then the tensor product theory has a diagonal anomalous $\bZ_2$ symmetry whose lightest odd operator is the weight-$(0,0)$ vacuum  that trivially satisfies our bound derived in Section \ref{Sec:OddBound}. 

However, in higher than two spacetime dimensions, the TQFT mentioned above always has a unique zero-weight local operator.  
By taking the tensor product, it follows that we can modify the anomaly of  a discrete, internal, global symmetry $G$ of a bosonic QFT without changing the local operator spectrum. Hence, there can never be a bound on the scaling dimensions and degeneracies of $G$-charged local operators that depends on such anomaly.\footnote{We thank David Simmons-Duffin for pointing out this argument to us.}

\section{General Bootstrap Bounds}\label{Sec:General}
\label{Sec:c=1}

In this section, we perform a numerical study to obtain bounds that harness the full power of Virasoro symmetry and the modular covariance of partition functions, for small to moderate values of the central charge $c$. Allowing $\tau$ to take general values in the upper half plane also lets us distinguish spin, and in particular makes possible the derivation of bounds on the scalar gap.

In Section \ref{Sec:BoundCondition}, we discuss the general expectations for whether particular bounds should exist. 
Section~\ref{Sec:c=1} is a precision test at $c = 1$, where we find our bounds to be saturated by the free compact boson. In Section~\ref{Sec:c>1}, we study the bounds on the lightest even/odd operators for $c \ge 1$, and by saturation by a number of WZW models. 
In Section~\ref{Sec:OrderDisorder}, we present a ``order-disorder" bound for a non-anomalous $\bZ_2$.  
In Section~\ref{Sec:RG}, we derive bounds on the scalar primaries, and discusses their implications on renormalization group flows.
In Section \ref{Sec:Largec}, we present an analytic derivation of the large $c$ asymptotics of certain bounds.

\subsection{When is there a Bound?}\label{Sec:BoundCondition}

We will be interested in the upper bound on the lightest {\it non-degenerate} primary in each sector.\footnote{Note that the notion of the gap in Section \ref{Sec:OddBound} is different from here.  In Section \ref{Sec:OddBound}, the gap ${\bf \Delta}^-_{\rm gap}$ is defined as the lightest $\bZ_2$ odd primary in ${\cal H}$, which can either be a conserved current or a non-degenerate primary. 
By contrast, in this section, $\Delta_{\rm gap}^j$ are defined as the lightest non-degenerate Virasoro primary in each sector.}
We start with the following question: In which sector do we expect a bound, and how does it depend on the anomaly?

As we have already seen in Section \ref{Sec:CompactBoson}, the $c=1$ free compact boson theory is an illuminating example.  By examining the different sectors in the free compact boson theory, and exploring its conformal moduli space, we find that ($\checkmark$ means there is a bound, while $\text{\sffamily x}$ means there is no bound):
\ie\label{yesno}
\left.\begin{array}{|c|c|c|c|c|}\hline  & 
~~{\cal H}^+~~ & ~~{\cal H}^-~~  & ~~{\cal H}_{\cal L}~~ &~~ {\cal H}^- \oplus {\cal H}_{\cal L} ~~\\
\hline ~~\text{Non-Anomalous}~~ & \checkmark & \text{\sffamily x} & \text{\sffamily x} & \checkmark \\
\hline \text{Anomalous} &  \checkmark & \checkmark &  \text{\sffamily x} & \checkmark \\\hline \end{array}\right.
\fe
If there is no bound in either ${\cal H}^-$ or ${\cal H}_{\cal L}$ of the free compact boson theory, then by considering the tensor product theory with any other CFT, we can produce examples of $c>1$ CFT whose gap can be arbitrarily large in that sector.  
By contrast, if there is a bound in a given sector of the $c=1$ free compact boson theory, then it does not immediately follow that such a bound persists to higher $c$.  

We will show that the same conclusion \eqref{yesno} is true for CFTs with larger values of the central charge. 
We show explicitly by  analytic and numerical bootstrap that there are universal bounds in ${\cal H}^+$ and ${\cal H}^-\oplus {\cal H}_{\cal L}$ for all $c>1$ CFTs, to arbitrary large $c$, in both the non-anomalous and the anomalous cases (see Section \ref{Sec:Largec}). 
For the bound in ${\cal H}^-$ in the anomalous case, we cannot find an analytic bound that is valid to arbitrarily large $c$. 
Nevertheless, we find numerical bounds at least up to $c \le 25$, and expect the bounds to continue to exist for $c > 25$.  
We will denote the bound in ${\cal H}^\pm$ as $\Delta_{\rm gap}^\pm$, and in ${\cal H}^-\oplus {\cal H}_{\cal L}$ as $\Delta_{\rm gap}^{\rm ord/dis}$. 
The meaning of the superscript ord/dis will be explained in Section~\ref{Sec:OrderDisorder}.

\subsection{Bootstrap Bounds for $c=1$}
\label{Sec:c=1}

We begin with a precision study of the bounds for $c = 1$, and match with the free compact boson. 
This analysis serves as a consistency check for our numerical method. 
Even though there are additional degenerate Virasoro modules at $c=1$, the free compact boson partition function for any $R$ can always be expanded on the vacuum and non-degenerate characters \eqref{vcharacter} with non-negative coefficients.

The spectra, symmetries, and anomalies in the free compact boson are discussed in Section \ref{Sec:CompactBoson} and in Appendix~\ref{App:FreeBoson}.  
We recall that at generic radii, the $\bZ_2$ symmetries include a non-anomalous momentum $\bZ_2^{(1,0)}$, a non-anomalous winding $\bZ_2^{(0,1)}$, and their diagonal subgroup $\bZ_2^{(1,1)}$ which is anomalous.

\begin{figure}[H]
\centering
\subfloat{
\includegraphics[height=.3\textwidth]{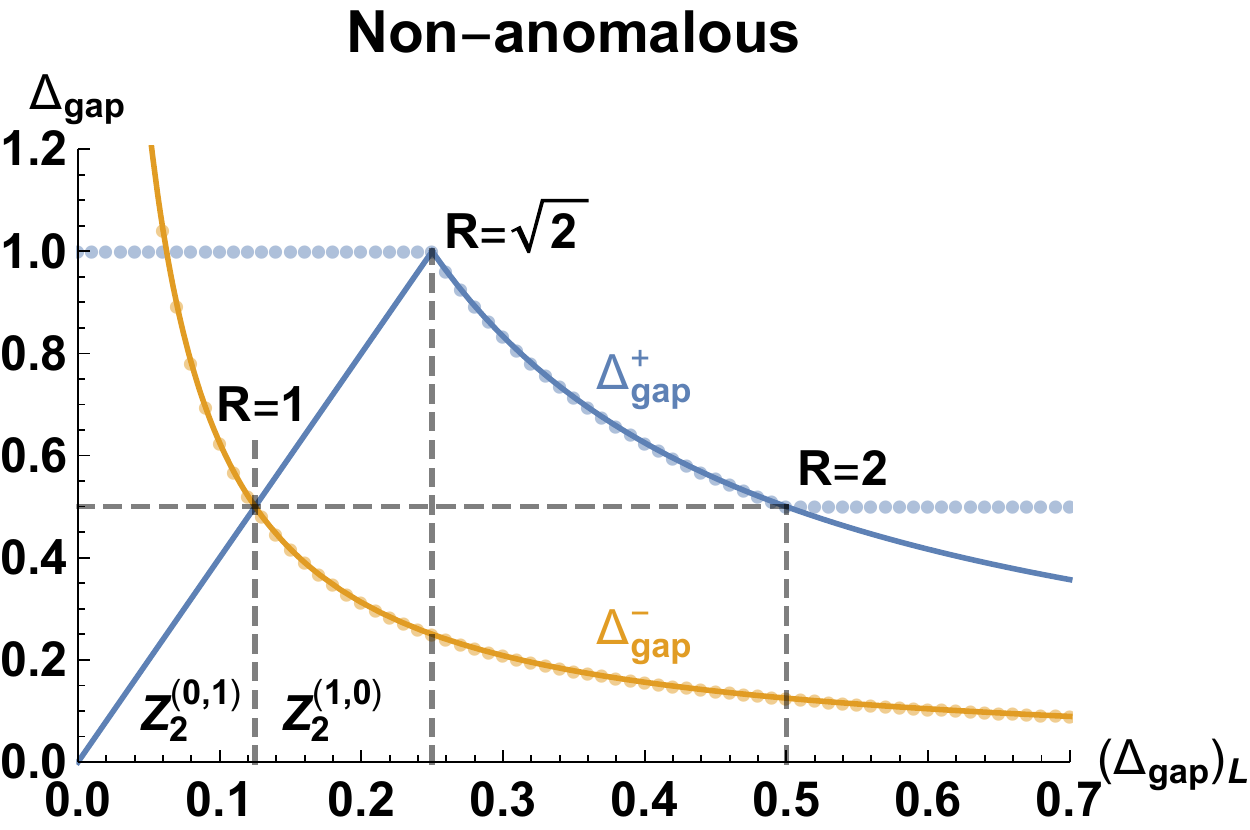}
}
~
\subfloat{
\includegraphics[height=.3\textwidth]{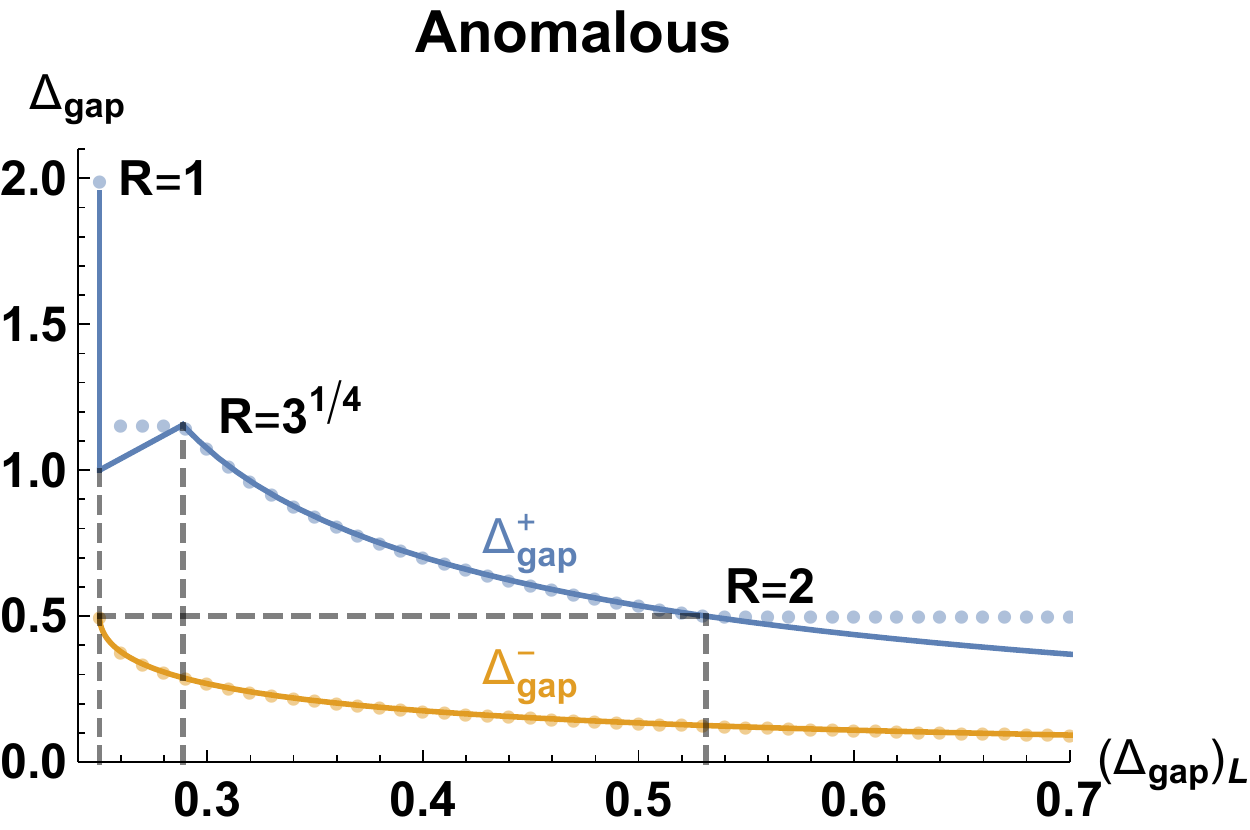}
}
\caption{{\bf Dots:} Upper bounds on the lightest $\bZ_2$ even and odd non-degenerate primaries, varied over the scaling dimension of the lightest primary in the defect Hilbert space ${\cal H}_{\cal L}$, for $c = 1$, at derivative order 19. In the anomalous case, we do not assume the existence of a conserved current with spin $\pm{1\over4}$ in the defect Hilbert space. {\bf Solid lines:} Free compact boson.
}
\label{Fig:anom-s-c=1}
\end{figure}

Let us study the bounds on the lightest $\bZ_2$ even and odd non-degenerate primaries, as functions of the scaling dimension of the lightest primary in the defect Hilbert space ${\cal H}_{\cal L}$, denoted by $(\Delta_{\rm gap})_{\cal L}$, is varied.  See Figure~\ref{Fig:anom-s-c=1}. The bounds are saturated by the free compact boson in continuous ranges of the radii $R$.\footnote{In the anomalous case, we do not assume the existence of a spin-${1\over4}$ conserved current in the defect Hilbert space (but we do allow the existence of all other conserved currents). Otherwise, the bounds away from $(\Delta_\text{gap})_{\cal L} = {1\over4}$ are weaker and are not saturated by the free compact boson. However, though we do not assume the existence of a spin-${1\over4}$ conserved current, this does not rule out the free compact boson at self-dual radius $R = 1$, because the weight-$({1\over4},0)$ conserved current module and one of the weight-$({1\over4},1)$ non-degenerate modules combine via the recombination rule to mimic a weight-$({1\over4},0)$ non-degenerate module, which is allowed when the gap in the defect Hilbert space is  $1\over4$.
} 
See Appendix \ref{App:FreeBoson} for the gaps in the $c=1$ free compact boson theory with respect to various $\bZ_2$ symmetries as functions of the radius $R$.

Some comments on the numerical bounds shown in Figure~\ref{Fig:anom-s-c=1} are in order:
\begin{itemize}
\item The absolute upper bound on the scaling dimension for the  non-degenerate Virasoro primary is the maximal value of the plot as we vary $(\Delta_{\rm gap})_{\cal L}$. 
Note that there is no bound on the lightest $\bZ_2$ odd primary in the non-anomalous case, which is consistent with our analysis in Sections~\ref{Sec:CompactBoson} and~\ref{Sec:Cardy}.  
\item In the non-anomalous case, the gap $(\Delta_{\rm gap})_{\cal L}= {1\over8}$ in the defect Hilbert space is realized at $R = 1$ simultaneously by the momentum $\bZ_2^{(1,0)}$ and winding $\bZ_2^{(0,1)}$. To the right, $(\Delta_{\rm gap})_{\cal L}> {1\over8}$, the shown gaps (solid lines) correspond to the momentum $\bZ_2^{(1,0)}$; to the left, $(\Delta_{\rm gap})_{\cal L} < {1\over8}$, they correspond to the winding $\bZ_2^{(0,1)}$.
\item In both the non-anomalous and anomalous cases, the bootstrap bounds on the lightest $\bZ_2$ odd non-degenerate primaries are saturated by the entire moduli space of the free compact boson, from $R = 1$ to $R = \infty$.
\item In the non-anomalous case, the bootstrap bounds on the lightest $\bZ_2$ even non-degenerate primary are saturated by the free compact boson with radii between $R = \sqrt2$ and $R = 2$. To the left of $R = \sqrt2$, the bounds become flat, because a gap of $(\Delta_{\rm gap})_{\cal L} = {1\over4}$ can also be interpreted as a gap of any smaller value.\footnote{In other words, the bootstrap bounds on $\Delta_\text{gap}^j$ must be monotonically decreasing with increasing $(\Delta_\text{gap})_{\cal L}$.
}
To the right of $R = 2$, the bounds become flat and unsaturated. 
\item In the anomalous case, the situation for the bounds on the lightest $\bZ_2$ even non-degenerate primary is similar to that described in the previous point. 
The jump in the bound at $R=1$ is because the lightest $\bZ_2$ even primary becomes a spin-one conserved current, and is thus excluded from our definition of the gap in Section~\ref{Sec:BoundCondition}.  See Appendix \ref{App:FreeBoson} for more details. 
\end{itemize}

\subsection{Bootstrap Bounds on the $\bZ_2$ Even/Odd Operators}
\label{Sec:c>1}

\begin{figure}[H]
\centering
\subfloat{
\includegraphics[height=.33\textwidth]{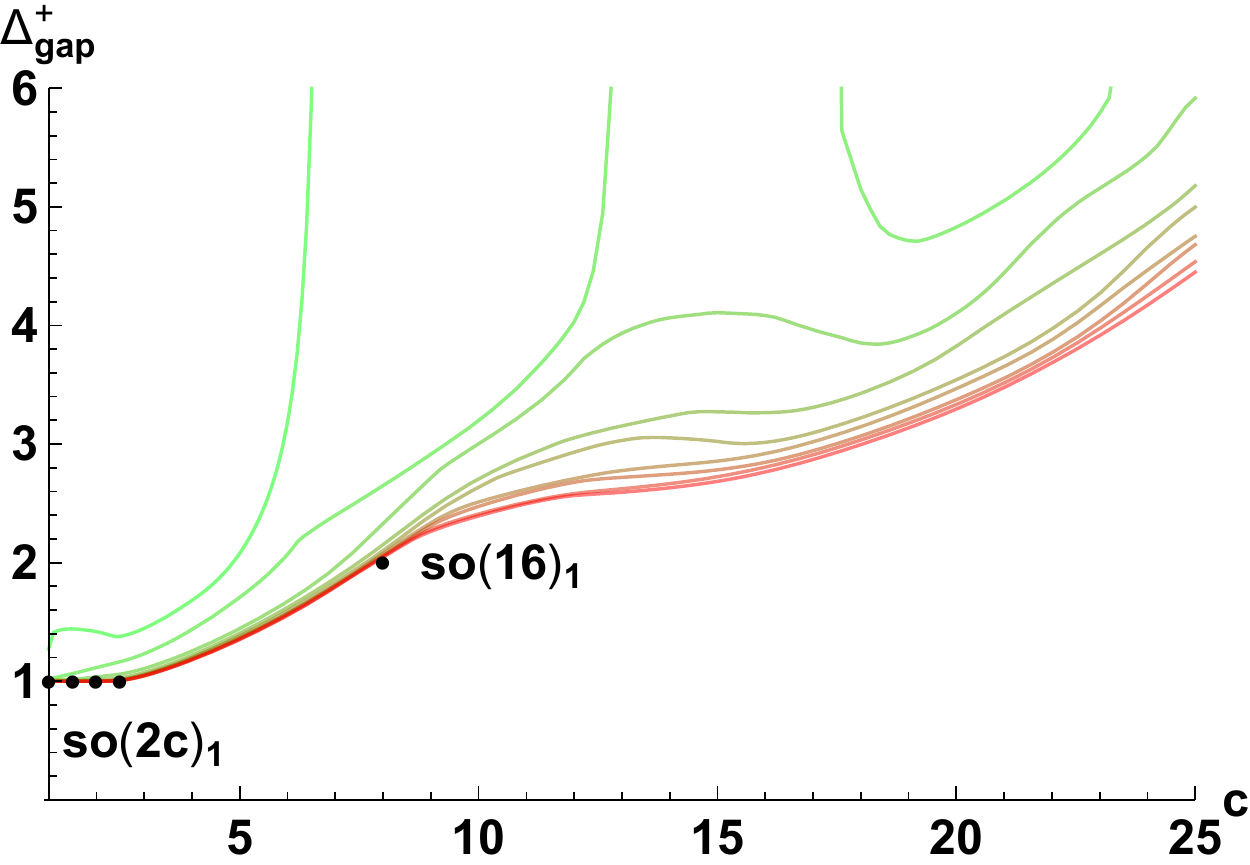}
}
\caption{{\bf Non-anomalous $\bZ_2$}: Upper bound on the lightest $\bZ_2$ even non-degenerate primary.
The green-to-red curves are the bounds at increasing derivative orders, from 3 to 19. The black dots mark the WZW models that (almost) saturate the bound.
}
\label{Fig:NonanomalousGeneral}
\end{figure}

\begin{figure}[H]
\centering
\subfloat{
\includegraphics[height=.33\textwidth]{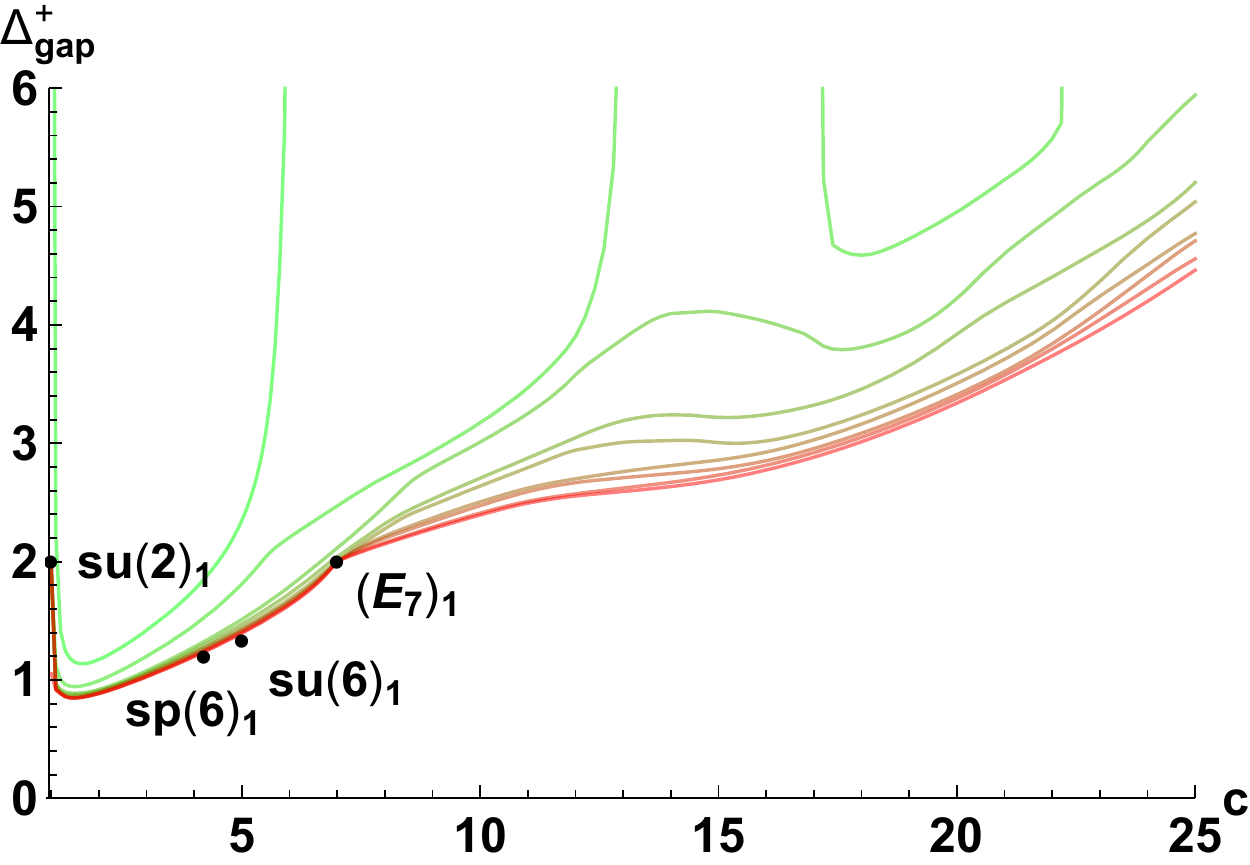}
}
~
\subfloat{
\includegraphics[height=.33\textwidth]{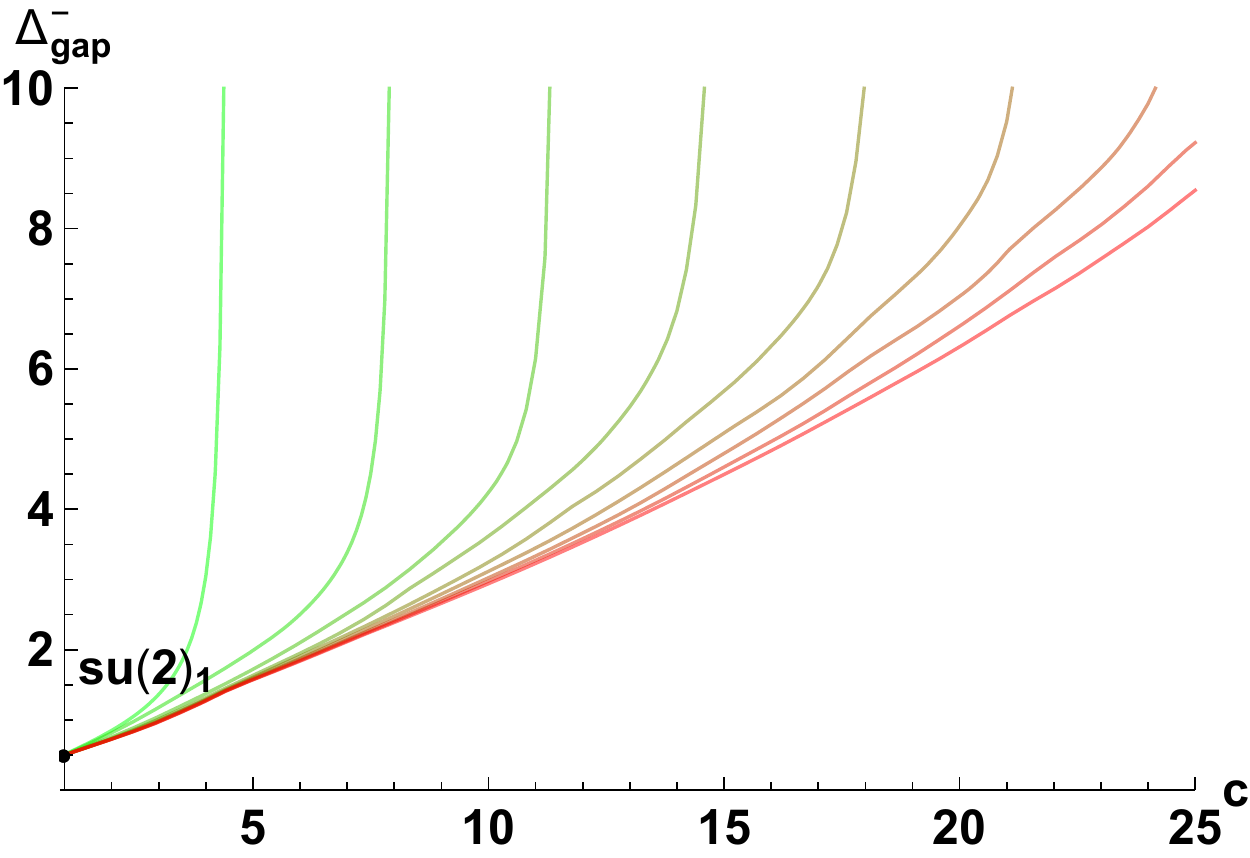}
}
\caption{{\bf Anomalous $\bZ_2$}: Upper bounds on the lightest $\bZ_2$ even and odd non-degenerate primaries. 
The green-to-red curves are the bounds at increasing derivative orders, from 3 to 19. The black dots mark the WZW models that (almost) saturate the bound.
}
\label{Fig:AnomalousGeneral}
\end{figure}

\begin{figure}[H]
\centering
\subfloat{
\includegraphics[height=.33\textwidth]{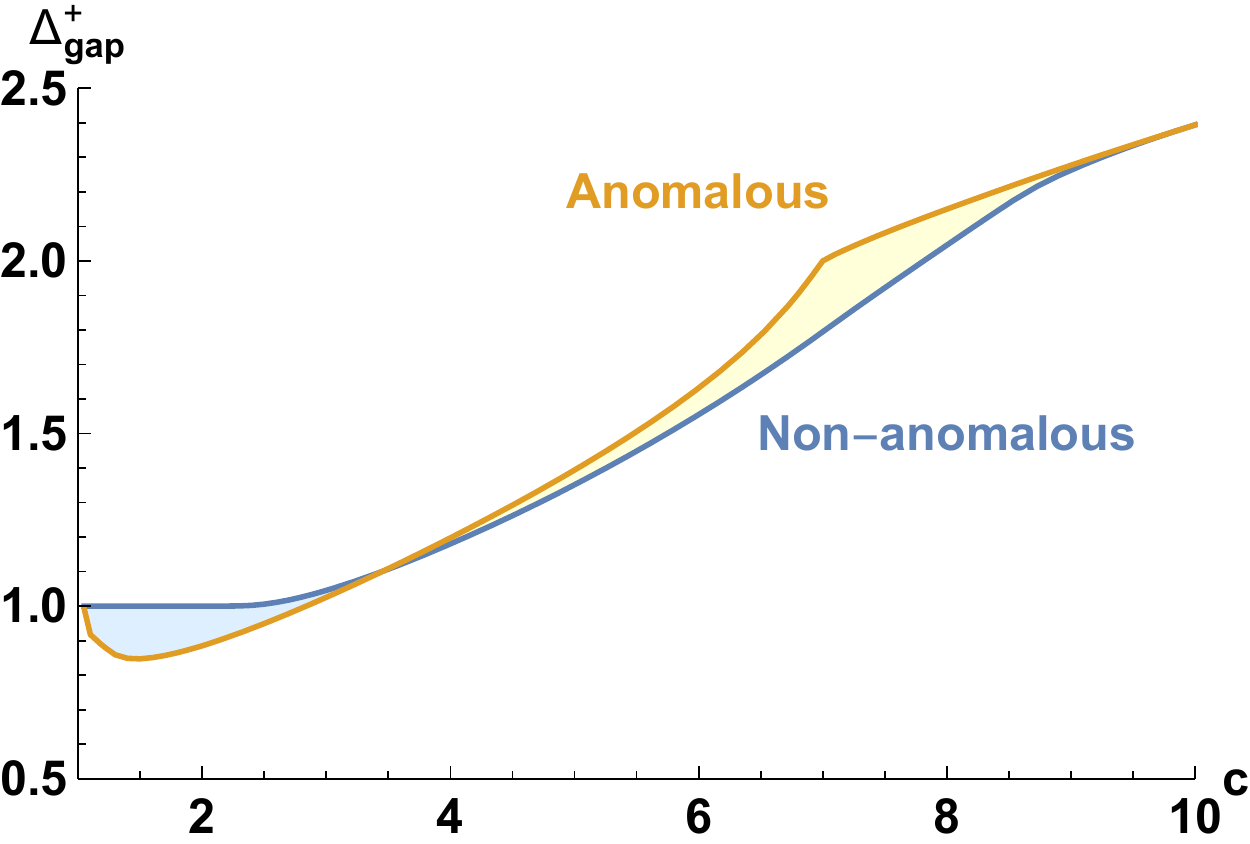}
}
\caption{Juxtaposition of non-anomalous and anomalous 
bounds on the lightest $\bZ_2$ even non-degenerate primary at derivative order 19.  
We do not display the comparison for the 
bounds on the lightest $\bZ_2$ odd primary, because there is no bound when the $\bZ_2$ is non-anomalous.
}
\label{Fig:General}
\end{figure}

Next we extend the bootstrap analysis to $c \ge 1$. The upper bounds on the lightest $\bZ_2$ even/odd non-degenerate primaries are shown in Figures~\ref{Fig:NonanomalousGeneral} and \ref{Fig:AnomalousGeneral}. The green-to-red curves are the bounds at increasing derivative orders, from 3 to 19. To stress the importance of the 't Hooft anomaly, Figure~\ref{Fig:General} juxtaposes the bounds with and without it.

The convergence of the bounds with increasing derivative order becomes slower at larger values of the central charge. In particular, at a fixed derivative order $d$, the bound on $\Delta_\text{gap}^-$ in the presence of 't Hooft anomaly only exists up to a finite value of the central charge $c_\text{max}(d)$, which increases with $d$.

Our bounds are saturated or almost saturated by various WZW models at level 1.\footnote{Recall that $\widehat{\mathfrak{so}(2)}_1=\widehat{\mathfrak{u}(1)}_2$ is the tensor product of two Ising CFTs, and $\widehat{\mathfrak{so}(3)}_1=\widehat{\mathfrak{su}(2)}_2$.}  
We list these cases in Appendix \ref{App:Saturate}.  
When the $\bZ_2$ is non-anomalous, we find that the bound on $\Delta^+_{\rm gap}$ is a plateau at $\Delta^+_{\rm gap}=1$ between $1\le c\le \frac 52$. At integral and half integral values of $c$ on this plateau, the bound is saturated by the $\widehat{\mathfrak{so}(2c)}_1$ WZW model, which can be described as the theory of $n$ free Majorana fermions summed  over the spin structures.\footnote{At $c={5\over2}$, the $\widehat{\mathfrak{so}(5)}_1$ WZW model with $\Delta^+_{\rm gap}=1$ almost saturates the numerical bound, which is $\Delta^+_{\rm gap}\simeq 1.0057$ at derivative order 19.  To know whether or not this is an example of saturation requires numerical data at higher derivative orders.}
The $\bZ_2$ symmetry is a center symmetry that commutes with the current algebra.

When the $\bZ_2$ is anomalous, we find two theories saturating our bounds.  At $c=1$, the bounds $\Delta_{\rm gap}^+ \leq 2$ and $\Delta_{\rm gap}^- \leq 0.5$ are saturated by the self-dual boson discussed in Section~\ref{Sec:SelfDualExample}.  At $c=7$, the bound $\Delta_{\rm gap}^+\le 2$ is saturated by the $(E_7)_1$ WZW model.\footnote{We also find a kink near $c = 1.30$ and $\Delta_\text{gap}^+ = 1$, but the extremal functional method reads off degeneracies that are not integer valued. Hence, we do not expect it to be saturated by a physical theory.}

We summarize our findings below:
\begin{itemize}
\item There is a universal bound on the lightest $\bZ_2$ odd non-degenerate primary if the $\bZ_2$ is anomalous, but not otherwise. 
\item There is a universal bound on the the lightest $\bZ_2$ even non-degenerate primary, with or without 't Hooft anomaly.  They are compared in Figure \ref{Fig:General}.
\item Suppose we have a CFT whose lightest $\bZ_2$ even non-degenerate primary is larger than $\Delta^+_{\rm gap}$ for $\A=+1$ but smaller than $\Delta^+_{\rm gap}$ for $\A=-1$, such as in the blue region in Figure \ref{Fig:General}, then we can conclude that the $\bZ_2$ symmetry must be anomalous, and vice versa if the primary is in the yellow region. 
\end{itemize}

\subsection{Order versus Disorder}\label{Sec:OrderDisorder}

In the non-anomalous case, there is no bound in either the odd sector ${\cal H}^-$ or the defect Hilbert space ${\cal H}_{\cal L}$.  
However, as we will see, there is a bound on the lightest non-degenerate primary in the union of ${\cal H}^- \oplus {\cal H}_{\cal L}$. 
In other words, {\it the lightest $\bZ_2$ odd primary and the defect Hilbert space ground state cannot both be too heavy relative to $c$.} We denote the scaling dimension of this lightest operator by $\Delta_{\rm gap}^{\rm ord/dis}$, for reasons we explain below. 
Note that in the anomalous case, since there is already a bound in ${\cal H}^-$ alone, we are guaranteed to have a bound in ${\cal H}^- \oplus {\cal H}_{\cal L}$.

Consider two  phases separated by a  second order phase transition, where on one side the symmetry is spontaneously broken and the other side unbroken. The two phases can be obtained by perturbing the CFT  at the critical point with a relevant operator, whose sign determines which phase we reach after the flow. Whereas the symmetry-breaking phase can be probed by the non-vanishing two-point function of an order operator at asymptotically large separation, 
 the symmetry-preserving phase can be probed by that of a non-local disorder operator. 
We expect the order and disorder operators to be light operators in ${\cal H}^-$ and ${\cal H}_{\cal L}$, respectively, at the critical CFT point. For example, in the critical Ising CFT, the lightest $\bZ_2$ odd primary is the spin field $\sigma(x)$, which is the order operator for $\bZ_2$. The defect Hilbert space ground state, on the other hand, is the disorder operator $\mu(x)$.  

In the {\it gapped} phase, 
it has recently been shown in \cite{Levin:2019ifu} that in an 
Ising symmetric spin chain, the order and disorder parameters cannot both be nonzero.\footnote{We thank Michael Levin for a discussion on this point and sharing a draft on related questions.} In the {\it gapless} phase, an analogous statement may be that the order and the disorder operators of a CFT with $\bZ_2$ symmetry cannot both be too heavy. Although generally the order and disorder operators in a CFT may not be the lightest operator in their respective spectra, our bound on $\Delta^{\rm ord/dis}_{\rm gap}$ has a similar flavor as the one proven in \cite{Levin:2019ifu} for spin chains in the gapped phase.  
Motivated by this analogy, we will denote the gap in ${\cal H}^-\oplus {\cal H}_{\cal L}$ as $\Delta^{\rm order/dis}_{\rm gap}$.

\begin{figure}[H]
\centering
\subfloat{
\includegraphics[height=.33\textwidth]{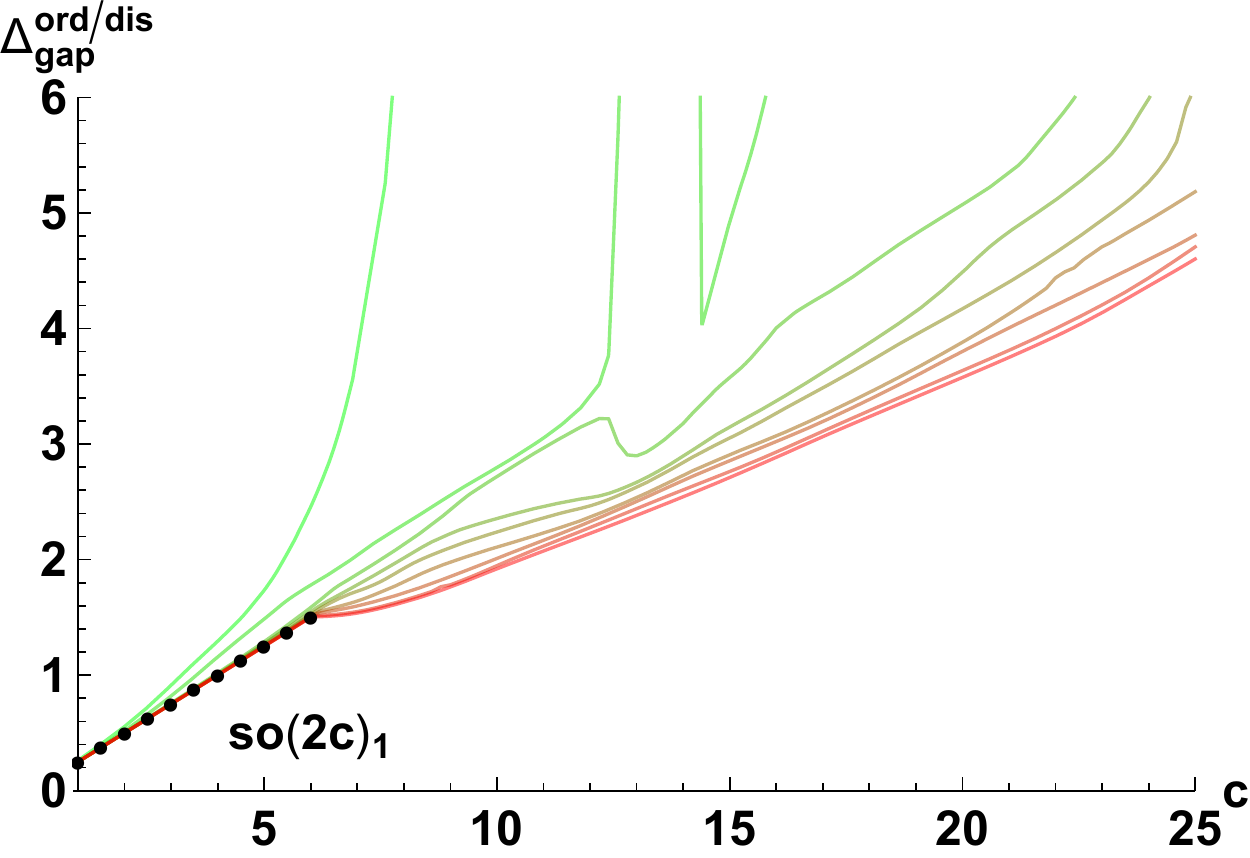}
}
\caption{{\bf Non-anomalous $\bZ_2$}: Upper bound on $\Delta_{\rm gap}^{\rm ord/dis}$ in ${\cal H}^-\oplus {\cal H}_{\cal L}$. 
The green-to-red curves are the bounds at increasing derivative orders, from 3 to 19. The black dots mark the $\widehat{\mathfrak{so}(2c)}_1$ WZW models that saturate the bound at $c \in {\bZ\over2}$ for $1 \le c \le 6$.
}
\label{Fig:NonanomalousGeneral-ord-dis}
\end{figure}

The bootstrap bound on $\Delta^{\rm ord/dis}_{\rm gap}$ is presented in Figure~\ref{Fig:NonanomalousGeneral-ord-dis}. 
We find that the bounds between $c = 1$ and $c = 6$ form a straight line given by $\Delta_{\rm gap}^{\rm ord/dis} \leq {c \over 4}$, where at integral and half-integral values of $c$, the bound is saturated by the $\widehat{\mathfrak{so}(2c)}_1$ WZW model.

\subsection{Scalar Bounds and Renormalization Group Flows}
\label{Sec:RG}

Another quantity of physical interest is the bound on the $\bZ_2$ even/odd {\it scalar} primaries in a CFT, which are related to relevant deformations if $\Delta<2$.  
Consider an RG flow preserving a $\bZ_2$ symmetry in the UV.  
 As the flow approaches a candidate IR fixed point, if there is  a symmetry-preserving relevant operator at a fixed point, then a generic flow would be driven away.  
This implies that without fine-tuning, the flow would miss the fixed point. 
See, for example, \cite{Affleck:1988zj,Affleck:1988wz,Ohmori:2018qza} for applications of this idea. 
A bound on relevant deformations that preserve various global symmetries would provide strong constraints on RG flows. 

\begin{figure}[H]
\centering
\subfloat{
\includegraphics[height=.33\textwidth]{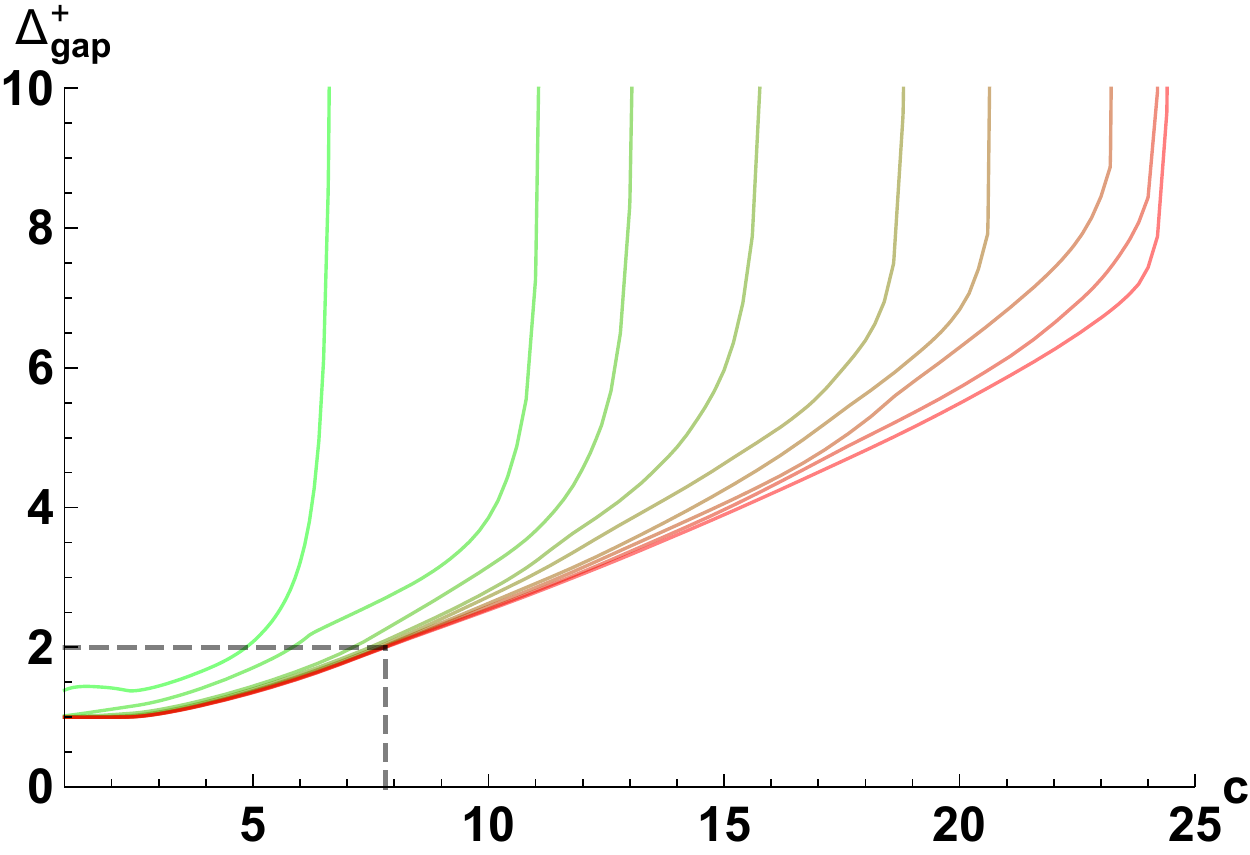}
}
\caption{{\bf Non-anomalous $\bZ_2$}: Upper bounds on the lightest $\bZ_2$ even scalar primary. The green-to-red curves are the bounds at increasing derivative orders, from 3 to 19. The red curve crosses $\Delta_\text{gap}^+ = 2$ at $c \approx 7.81$.}
\label{Fig:ScalarNonanomalous}
\end{figure}

\begin{figure}[H]
\centering
\subfloat{
\includegraphics[height=.33\textwidth]{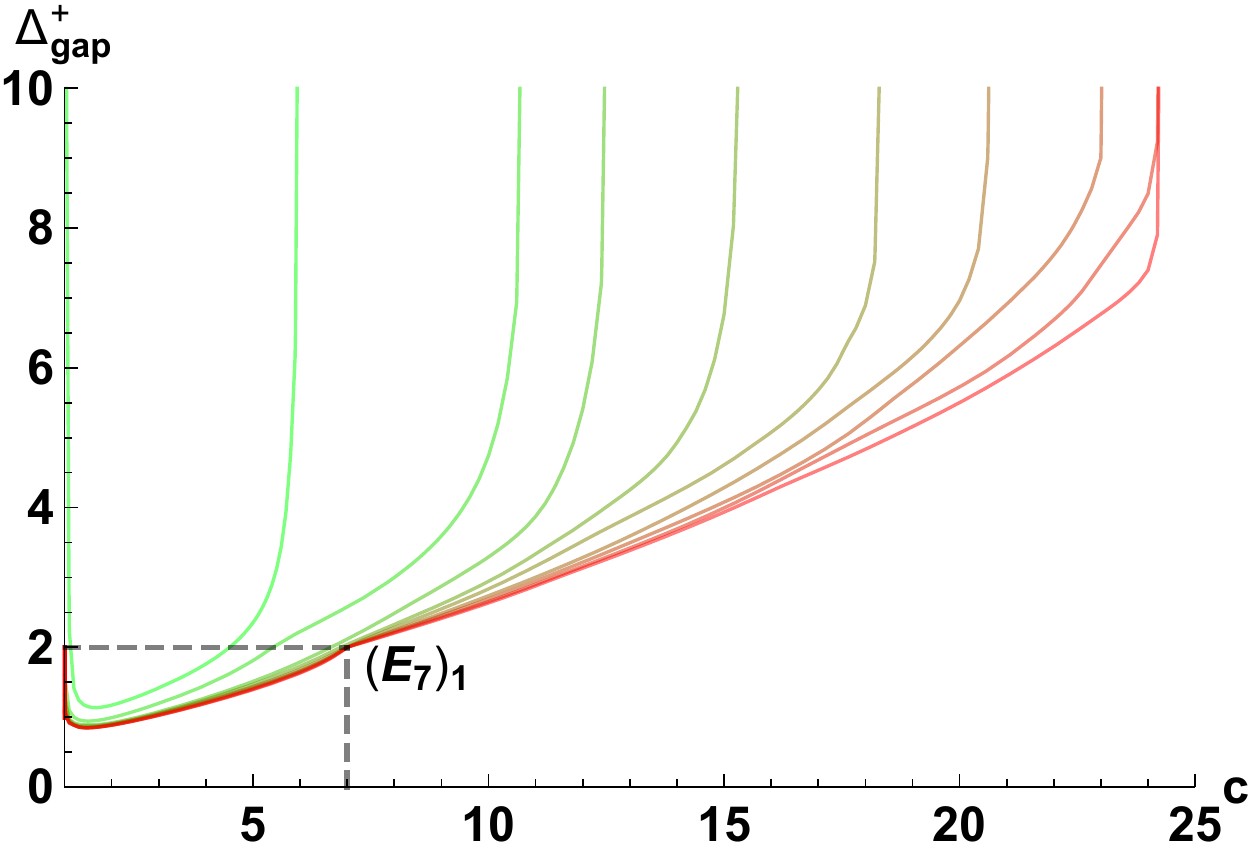}
}
~
\subfloat{
\includegraphics[height=.33\textwidth]{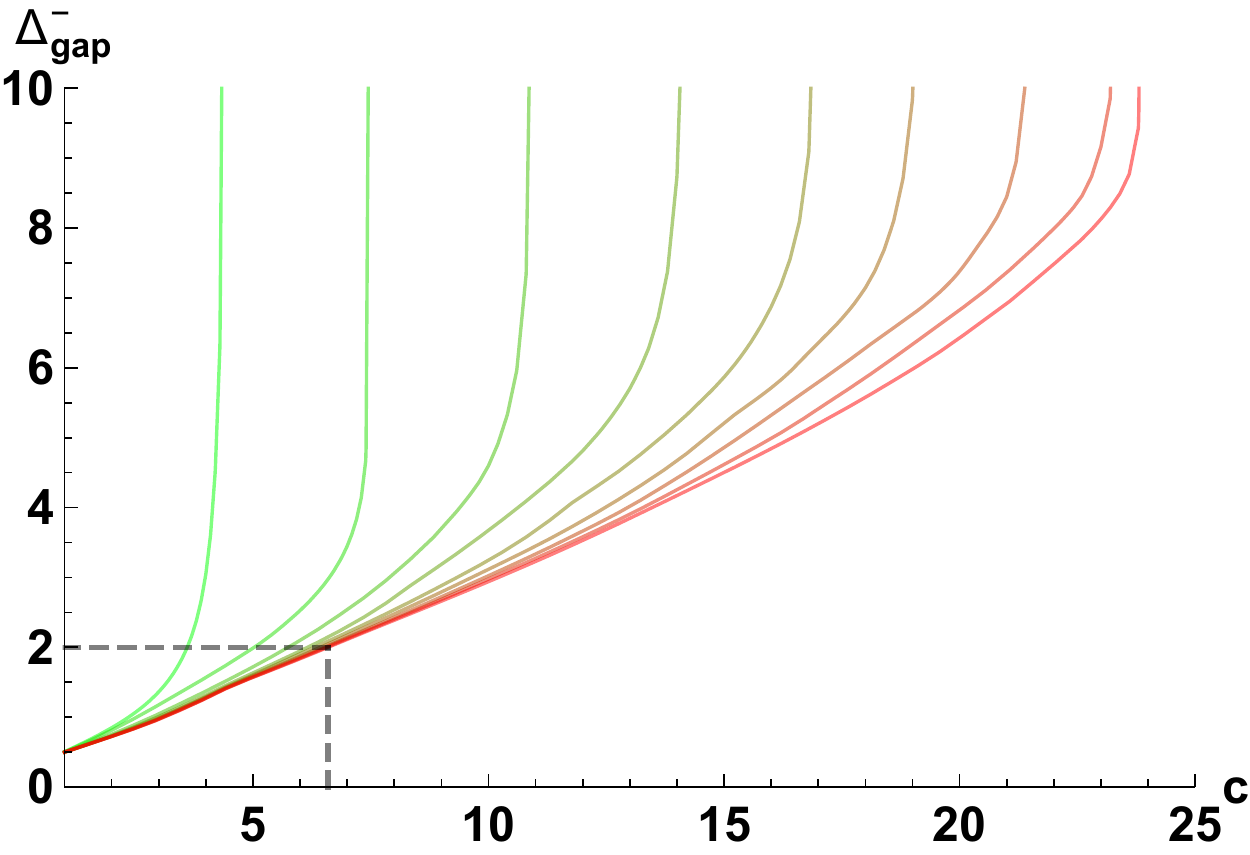}
}
\caption{{\bf Anomalous $\bZ_2$}: Upper bounds on the lightest $\bZ_2$ even and odd scalar primaries. The green-to-red curves are the bounds at increasing derivative orders, from 3 to 19. The red curves crosse $\Delta_\text{gap}^\pm = 2$ at $c = 7$ and $c \approx 6.59$, respectively.}
\label{Fig:ScalarAnomalous}
\end{figure}

Using the modular bootstrap techniques, we obtain bounds on the lightest $\bZ_2$ even non-degenerate scalar primaries in both the anomalous and non-anomalous cases. We also present a bound on the $\bZ_2$ odd scalar primary in the anomalous case.   
See Figures~\ref{Fig:ScalarNonanomalous} and~\ref{Fig:ScalarAnomalous}.\footnote{In the ordinary modular bootstrap, it was observed that no bound on the lightest scalar primary exists for $c \geq 25$ \cite{Collier:2016cls}. The explanation there is the existence of a modular invariant partition function with continuous spectrum, no vacuum, and only non-scalar primaries at $c=25$.  
The same partition function can also be applied to solve the modular bootstrap equation with a $\bZ_2$ symmetry, and hence we do not have a bound when $c>25$.
} 

We summarize our findings below:
\begin{itemize}
\item In the absence of 't Hooft anomaly, a $\bZ_2$-preserving relevant deformation always exists for $1 < c < 7.81$.
\item In the presence of 't Hooft anomaly, a $\bZ_2$-preserving relevant deformation always exists for $1 < c < 7$, and a $\bZ_2$-breaking relevant deformation always exists for $1 < c < 6.59$.  
At $c=7$, the $(E_7)_1$ WZW model has a $\bZ_2$-preserving marginal operator which is a current bilinear $J^a (z) \bar J^b(\bar z)$.
\item The statements above imply that RG flows preserving \textit{only} a $\bZ_2$ symmetry generically do not end at fixed points with $1 < c < 7$ without fine-tuning. 
Put differently, in this range of the central charge, a gapless phase that is only protected by  $\bZ_2$ symmetry is not stable under perturbations. 
If the $\bZ_2$ is non-anomalous, then the range is further extended to $1 < c < 7.81$.\footnote{This statement is true even if the the $\bZ_2$ acts trivially in the IR. In that case there is no symmetry forbidding the relevant operators near the IR fixed point. By the result of \cite{Collier:2016cls}, there is always a relevant operator if $1<c<8$, hence the flow will generically miss such a fixed point without fine-tuning.}
\end{itemize}

\subsection{Large $c$}
\label{Sec:Largec}

In this subsection, we analytically derive bounds that are valid to arbitrary large $c$ on the lightest primaries in ${\cal H}^+$ and in ${\cal H}^-\oplus {\cal H}_{\cal L}$. 
We only use functionals up to 3 derivative order with $\tau = -\bar\tau = it$, so the bounds are far from the strongest possible. 
With finite derivative order, we cannot find a functional that gives a bound in ${\cal H}^-$ for arbitrarily large $c$ (see Section \ref{Sec:c>1}). 
The large $c$ bounds in ${\cal H}^+$ and in ${\cal H}^-\oplus {\cal H}_{\cal L}$ we obtain in this subsection apply to both the anomalous and non-anomalous cases.

We follow the strategy in Section \ref{Sec:OddBound} but now utilizing the full Virasoro symmetry. 
Specifically, we use the reduced Virasoro characters \eqref{scalingch} in place of the scaling character $g_\Delta(t)$:
\ie
\hat\chi_0(t) = { t^{1\over2} e^{2\pi  {c-1\over12} t} } (1-e^{-2\pi t})^2\,,~~~\hat\chi_{\Delta>0}(t) = { t^{1\over2} e^{-2\pi (\Delta-{c-1\over12}) t} }\,.
\fe
Here we make an additional assumption to simplify the analysis. We assume that the partition functions can be expanded in the vacuum and non-degenerate characters alone, without the need for conserved currents. 
In other words, our bounds in this subsection apply to CFTs without conserved currents.\footnote{More generally, the bounds in this subsection apply to what are called {\it generic-type} CFTs in \cite{Collier:2016cls}.  In generic-type CFTs, each holomorphic conserved current of spin $s \geq 1$ is accompanied by a primary of weight $(s, 1)$, and similarly for the anti-holomorphic currents, so that the recombination rule $\chi_h(\tau) \stackrel{h \to 0}{\longrightarrow} \chi_0(\tau) + \chi_1(\tau)$ 
disguises the combination of modules as a non-degenerate module at the unitarity bound. This way, the partition function admits an expansion in the vacuum and non-degenerate characters alone.
}

We now define the matrix \eqref{M} in terms of the reduced Virasoro characters,
\ie
M_{i}^{\,j}(\Delta,t) \equiv \delta_i^{\,j} \hat \chi_\Delta(1/t) -F_i^{\,j} \hat \chi_\Delta (t), \quad y={12\Delta\over c}-1.
\fe  
 As in Section \ref{Sec:OddBound}, we will expand our functional in the derivative basis as ({\it c.f.} \eqref{alphabasis} and \eqref{alphaexpand}) 
\ie
& \A = \sum_{n\text{ even}} \gamma^{n, 1} \A_{n, 1} + \sum_{n\text{ odd}} \sum_{i=1}^2 \gamma^{n, i} \A_{n, i} \,,
\\
& \A_{n,i} [ {\bf V}(t) ] \equiv  e^{2\pi (\Delta-{c-1\over12})}  \left( {6 t \over \pi {c}} {d \over dt} \right)^n  V_i(t) \Big|_{t = 1}.
\fe

While we manage to find analytic functionals to put bounds on the lightest primaries, the exact expressions are not illuminating to present. Therefore, we will present them in $1/c$ expansions.

\paragraph{$\bZ_2$ even sector ${\cal H}^+$} 
We consider the ansatz
\ie
\gamma^{1,2} = \gamma^{3,2} = 0,
\fe
such that the action of $\A$ is identical in ${\cal H}^-$ and ${\cal H}_{\cal L}$, and demand that the corresponding $\A[{\bf M}^j(\Delta,t)]$ has a double zero at $y=1$  for $j=-,{\cal L}$. Together with the requirement that $\A$ annihilates the vacuum, the linear functional is completely fixed (up to overall rescaling).  The resulting linear functional satisfies the positivity conditions (analogous to \eqref{functional}) for $c > 2$. 
To first order in $1/c$, the expansion coefficients of the functional $\A$ are given by
\ie
& \gamma^{0,1} = -1 
+ {1 \over c} \left( 12-\frac{12}{\pi }-12 \coth (\pi ) \right)
+ {\cal O}(c^{-2}),
\\
& \gamma^{1,1} = 1 
+ {6 \over c} \left(-4+\frac{5}{\pi }+4 \coth (\pi )\right)
+ {\cal O}(c^{-2}),
\\
& \gamma^{2,1} = 1 
+ {6 \over c} \left(2-\frac{5}{\pi }-2 \coth (\pi )\right)
+ {\cal O}(c^{-2}),
\\
& \gamma^{3,1} = -1.
\fe
Its action on ${\bf M}^j(\Delta,t)$ gives
\ie
\A[ {\bf M}^j(\Delta,t) ] &=
{1\over2}
\begin{pmatrix}
(-1 + y) (1 + y) (1 + 3 y)
\\
(-1 + y)^2 (1 + y)
\\
(-1 + y)^2 (1 + y)
\end{pmatrix}
\\
&
\hspace{.25in} +
{6 \over \pi c}
\begin{pmatrix}
 -(y+1) (7 y+1)-\pi  (y (y+6)+1) (\coth (\pi )-1) \\
 (y-1)^2 (1+\pi  (\coth (\pi )-1)) \\
 (y-1)^2 (1+\pi  (\coth (\pi )-1)) \\
\end{pmatrix}
 + {\cal O}(c^{-2}).
\fe
The resulting bound on the lightest $\bZ_2$ even primary is
\ie
\Delta_\text{gap}^+ \le {c \over 6} -1+\frac{2}{\pi }+\coth (\pi ) + {\cal O}(c^{-1}) \approx {c \over 6} + 0.6404 + {\cal O}(c^{-1})\,,
\fe
which applies to both the anomalous and the non-anomalous cases.

\paragraph{Order-Disorder ${\cal H}^-\oplus {\cal H}_{\cal L}$} We consider the ansatz
\ie
\gamma^{0,1} = \gamma^{1,2} = 0,
\fe
such that the action of $\A$ is identical in ${\cal H}^-$ and ${\cal H}_{\cal L}$, and demand that $\A[{\bf M}^j(\Delta,t)]$ has a double zero at $y=1$ for $j=-,{\cal L}$. Together with the requirement that $\A$ annihilates the vacuum, the linear functional is completely fixed (up to overall rescaling). The resulting linear functional satisfies the positivity conditions (analogous to \eqref{functional}) for all $c > 1$. 
To first order in $1/c$, the expansion coefficients are given by
\ie
& \gamma^{0,1} = 3 
+ {36 \over c} (\coth (\pi )-1)
+ {\cal O}(c^{-2}),
\\
& \gamma^{1,1} = 1 
+ {6 \over c} \left(-4+\frac{1}{\pi }+4 \coth (\pi )\right),
\\
& \gamma^{2,1} = -3 
+ {1 \over c} \left(\frac{54}{\pi }+36 (\coth (\pi )-1)\right)
+ {\cal O}(c^{-2}),
\\
& \gamma^{3,1} = -1.
\fe
Its action on ${\bf M}^j(\Delta,t)$ gives
\ie
\A[ {\bf M}^j(\Delta,t) ] &=
{1\over2}
\begin{pmatrix}
3 (y-1)^2 (y+1) \\
 (y-1) (y+1) (y+3) \\
 (y-1) (y+1) (y+3) \\
\end{pmatrix}
\\
&
\hspace{.5in} +
{6 \over \pi c}
\begin{pmatrix}
  3 \pi  (y-1)^2 (\coth (\pi )-1) \\
 -2 y (3 y+1)-\pi  (y (3 y+2)+3) (\coth (\pi )-1) \\
 -2 y (3 y+1)-\pi  (y (3 y+2)+3) (\coth (\pi )-1) \\
\end{pmatrix}
 + {\cal O}(c^{-2}).
\fe
The resulting upper bound on the lightest primary in ${\cal H}^-\oplus {\cal H}_{\cal L}$ is
\ie
\Delta_{\rm gap}^{\rm ord/dis} \le {c \over 6} -1+\frac{1}{\pi }+\coth (\pi ) + {\cal O}(c^{-1}) \approx {c \over 6} + 0.3221 + {\cal O}(c^{-1})\,,
\fe
which applies to both the anomalous and the non-anomalous cases.

\section{$U(1)$ Symmetry and the Weak Gravity Conjecture}
\label{Sec:U1}

In this section, we consider bosonic 2d CFTs with $U(1)$ global symmetry.\footnote{We thank Nathan Benjamin for discussions.}
We will see that the existence of a universal upper bound on the lightest $U(1)$ charged operator again depends on the 't Hooft anomaly.

\subsection{$U(1)$ Symmetry and its Anomaly} 

Consider the global symmetry generated by a conserved spin-one current, $J_\mu(z,\bar z)$ (with $\partial^\mu J_{\mu}=0$), in a bosonic 2d CFT.
We require that
\begin{itemize}
\item The symmetry is globally $U(1)$, not $\mathbb{R}$.
\item The $U(1)$ global symmetry acts faithfully on the local operators.
\end{itemize}

Let $J\equiv J_z$ and $\bar J\equiv J_{\bar z}$. 
  In any compact unitary 2d CFT, unitarity implies $\partial \bar J= 0$ and $\bar\partial J=0$, so each of them is separately  a $\mathfrak{u}(1)$ Lie algebra generator. Globally, however, the holomorphic current $J(z)$ may not generate a $U(1)$ group, but rather an $\mathbb{R}$. The same is true for the anti-holomorphic current $\bar J$.  
  We denote their zero modes by $J_0 \equiv \oint {dz\over 2\pi i} J(z)$ and $\bar J_0  \equiv -\oint {d\bar z\over 2\pi i } \bar J(\bar z)$. 

The topological line implementing a $U(1)$ rotation by $ \eta$ is 
\begin{align}
U_\eta = \, \exp \left[ 2\pi i \eta\left(  \oint_{\cal L} dz J(z) - \oint _{\cal L} d\bar z \bar J(\bar z) \right)\right] \,.
\end{align}
The assumption that the symmetry is $U(1)$ instead of $\mathbb{R}$ implies that $\eta$ is circle valued, {\it i.e.} $U_\eta = U_{\eta+1}$, so we may take $\eta\in [0,1)$. 
The faithfulness assumption requires that $U_\eta$ is not an identity operator unless $\eta$ is an integer.  
Furthermore, the $U(1)$ charge of a local operator in the Hilbert space $\cal H$ is always an integer, 
\ie
Q= J_0+\bar J_0 \in \bZ\,.
\fe

The OPEs of $J$ and $\bar J$ are
\begin{align}
&J(z) J(0) \sim {k \over z^2}\,,~~~~~\bar J(\bar z) \bar J(0) \sim {\bar k\over \bar z^2}\,.
\end{align}
Note that the levels $k$ and $\bar k$ are physically meaningful and cannot be scaled away if we assume that our symmetry is globally a $U(1)$ acting faithfully on all local operators.\footnote{For example, had we rescaled both $J \to 2J$ and $\bar J \to 2\bar J$, the new topological line operator $U_\eta$ with $\eta={1\over2}$ would act trivially on the Hilbert space, violating the faithfulness condition.}

When $\bar k=0$ ($k=0$), this $U(1)$ global symmetry is generated by a holomorphic (anti-holomorphic) current. But more generally, the current associated to a $U(1)$ global symmetry can be neither purely holomorphic nor anti-holomorphic, even though its holomorphic and anti-holomorphic components each separately generates a {\it different} $\mathbb{R}$ or $U(1)$ global symmetry.

Using the topological line $U_\eta$ of a $U(1)$ global symmetry, we can similarly consider the defect Hilbert space defined in Figure \ref{fig:HL}. 
The defect Hilbert space  ${\cal H}_\eta$  of the topological line $U_\eta$ is related to the bulk Hilbert space $\cal H$ by a simultaneous spectral flow \cite{Schwimmer:1986mf} on both the left and the right of opposite amounts (see, for example, \cite{Heidenreich:2016aqi}):
\begin{align}\label{sf}
\begin{split}
&h^\eta = h -\eta J_0  + {k\eta^2\over2}\,,~~~J_0^\eta = J_0 -\eta k \,,\\
&\bar h^\eta = \bar h+ \eta \bar J_0  + {\bar k\eta^2\over2}\,,~~~\bar J_0^\eta =\bar J_0 +\eta \bar k \,.
\end{split}
\end{align}
For non-integer $\eta$, these are the quantum numbers of a non-local operator living at the end of the topological line $U_\eta$.  
However, for $\eta\in \bZ$, the topological line $U_\eta$ is trivial, and \eqref{sf} are the quantum numbers of {\it local} operators.  
In other words, starting with an operator in $\cal H$, the $U(1)$ global symmetry (instead of $\bR$) implies the existence of infinitely many spectral partners by applying \eqref{sf} with $\eta\in \bZ$. 

In particular, integer quantization of the spin $h^{\eta=1} -\bar h^{\eta=1}$  demands that
\begin{align}
{k-\bar k\over2} \in \bZ\,,
\end{align}
where we have used the fact that the $U(1)$ charges $Q=J_0+\bar J_0$ are  integers. 
The integer ${k-\bar k\over2}$ is the 't Hooft anomaly of a $U(1)$ global symmetry in a bosonic 2d quantum field theory.  
Indeed, if ${k-\bar k\over2}\neq 0$, the states in the defect Hilbert space ${\cal H}_\eta$  have $U(1)$ charges $Q^\eta = J_0^\eta + \bar J_0^\eta = Q- \eta(k-\bar k)$, which are not integers for generic $\eta$. 
This implies that the theory is not invariant under the $U(1)$ transformation in the presence of the $U(1)$ topological line defect, which is the hallmark of 't Hooft anomaly.\footnote{This is similar to the $\bZ_2$ case, where the $\bZ_2$ eigenvalues \eqref{spincharge2} are $\pm i$ in the defect Hilbert space ${\cal H}_{\cal L}$ when the $\bZ_2$ is anomalous.}  

Some comments are in order:
\begin{enumerate}
\item  The $\bZ_2$ subgroup of a $U(1)$ is anomalous if ${k-\bar k\over2}$ is odd, and non-anomalous if ${k-\bar k\over2}$ is even. 
\item A  holomorphic $U(1)$ has $k\neq0$ and $\bar k=0$ (so $\bar J$ is a trivial operator), which  is always anomalous. Similarly, an anti-holomorphic $U(1)$ is also always anomalous.
\item The integral spectral flow \eqref{sf} of the identity operator is given by
\ie
h=  {k\eta^2\over2}\,,~~\bar h = {\bar k\eta^2\over2}\,,~~Q=  -\eta(k-\bar k)\,,~~~\eta\in\bZ\,.
\fe
In particular, when the $U(1)$ is anomalous, {\it i.e.} when $k\neq\bar k$, there are always spectral flow partners of the identity that are charged (see, for example, \cite{Heidenreich:2016aqi,Montero:2016tif}). 
Hence an anomalous $U(1)$ global symmetry guarantees the existence of charged operators via spectral flow, while this is not true for an $\bR$ global symmetry.  
However, when the $U(1)$ is non-anomalous, {\it i.e.} $k=\bar k$, the spectral flows of the identity  are all charge neutral.
\end{enumerate}

\paragraph{Free compact boson example}
Let us illustrate the above discussion with the $c=1$ free compact boson. We normalize the OPE to be
\begin{align}
X(z,\bar z) X(0,0)\sim -\frac 12 \log |z|^2\,.
\end{align}
Hence $\partial X(z) \partial X(0)\sim -{1\over 2z^2}$ and $\bar \partial X(\bar z) \bar\partial X(0)\sim -{1\over 2\bar z^2}$.  The current algebra primaries are the exponential operators ${\cal O}_{n,w}(z,\bar z) = \exp\left[  i\left( {n\over R} +wR\right) X_L(z)  +   i\left( {n\over R} -wR\right) X_R(\bar z)  \right]$ labeled by the momentum $n\in \bZ$ and the winding number $w\in\bZ$. We have the OPE
\begin{align}
i\partial X(z) {\cal O}_{n,w}(0) \sim { \left( \frac nR +wR\right)\over2 z} {\cal O}_{n,w}(0)\,,~~~
i\bar\partial X(\bar z) {\cal O}_{n,w}(0) \sim { \left( \frac nR -wR\right)\over2\bar z} {\cal O}_{n,w}(0)\,.
\end{align}
For generic $R$, the charges of $i\partial X(z)$ and $i\bar\partial X(\bar z)$ are irrational. 
Hence,  the holomorphic current $i\partial X$ and the anti-holomorphic current $i\bar \partial X$ generate two $\mathbb{R}$ symmetries, not  $U(1)$. 

On the other hand, there are two $U(1)$ global symmetries, the momentum $U(1)_n$ and the winding $U(1)_w$ for all radii, under which ${\cal O}_{n,w}$ has charges $n$ and $w$, respectively. 
The currents of the momentum and winding $U(1)$'s are  combinations of $\partial X(z)$ and $\bar \partial X(\bar z)$:
\begin{align}
&U(1)_{n} :~~J(z) =  iR  \partial X(z)\,,~~~~~\bar J(\bar z) = i R \bar\partial X(\bar z)\,,\\
&U(1)_{w} :~~J(z) =  {i\over R}  \partial X(z)\,,~~~~~\bar J(\bar z) =- {i\over R} \bar\partial X(\bar z)\,,
\end{align}
under which ${\cal O}_{n,w}$ has integer charges $n$ and $w$, respectively.  We find that
\begin{align}
&U(1)_{n} :~~ k =\bar k= { R^2\over2} \,,\\
&U(1)_{w} :~~ k =\bar k= { 1\over2R^2} \,.
\end{align}
In particular, they both obey $k=\bar k$, which means that they are non-anomalous. Note that while $k$ and $\bar k$ are not separately quantized in general,  $k-\bar k\over2$ is always an integer.  
Any combination of the momentum $U(1)_n$ and winding $U(1)_w$ with an integer coefficient is also a $U(1)$ symmetry. When $R^2$ is rational, there exists a $U(1)$ that is generated by a holomorphic current, and another $U(1)$ generated by an anti-holomorphic current. The chiral algebra is hence enhanced at rational $R^2$.

The  spectral flow \eqref{sf} for $U(1)_{n}$ by one unit $\eta=1$ takes the exponential operator ${\cal O}_{n,w}$ to ${\cal O}_{n,w-1}$, and similarly for the spectral flow of $U(1)_{w}$.

\subsection{Bounds on the $U(1)$ Charged Operator}

We start with a simple statement for all $c \ge 1$ bosonic 2d CFTs:
\begin{itemize}
\item There is {\it no} universal upper bound on the lightest $U(1)$ charged operator if the $U(1)$ is {\it non-anomalous} ({\it i.e.} $k=\bar k$).
 \end{itemize}
Indeed, in the $c=1$ free compact boson theory, the lightest charged operator under the non-anomalous winding $U(1)_w$ is the minimal winding operator ${\cal O}_{0,1}$, which has scaling dimension $R^2 \over 2$.  By taking the radius to be arbitrarily large, the minimal winding operator is arbitrarily heavy, and hence there is {\it no} upper bound on the lightest charged operator for this non-anomalous $U(1)_w$.  
To arrive at the same conclusion for larger values of $c$, we tensor product the free compact boson theory with any other CFT, and consider the $U(1)_w$ of the former.  
Again, the lightest charged operator of this $c>1$ CFT can be made arbitrarily heavy by taking the radius of the boson to be large.  While this tensor product construction does not rigorously cover the entire range of $c \ge 1$, we expect the statement to be true for all $c \ge 1$.

In \cite{Benjamin:2016fhe,Montero:2016tif,Bae:2018qym}, a bound on the lightest charged operator is derived for a {\it holomorphic} $U(1)$, which always has an 't Hooft anomaly.   
Hence, it is consistent with the above statement. 
More generally, we argue that, just as in the $\bZ_2$ case, the existence of a bound on the lightest charged operator is not directly related to holomorphicity, but to the 't Hooft anomaly.\footnote{Incidentally, the authors of \cite{Montero:2016tif} considered the   $\bZ_N$ subgroup of a holomorphic $U(1)$ with $k=2N$ and $\bar k=0$.  This   $\bZ_N$ is a non-anomalous subgroup of  an anomalous $U(1)$. Indeed, while they obtained a universal bound on charged operators for the anomalous $U(1)$, they did not find a bound for the non-anomalous $\bZ_N$ subgroup.   This is consistent with our general observation that there is a bound only when the symmetry is anomalous.}  

Indeed, consider an anomalous  $U(1)$ global symmetry with 
\ie\label{oddanomaly}
{k-\bar k\over2} \in 2\bZ+1\,.
\fe 
The $\bZ_2$ subgroup of this $U(1)$ is anomalous, $\A=-1$. 
From Section~\ref{Sec:OddBound} and Section~\ref{Sec:c>1}, we know that there is a bound on the lightest $\bZ_2$ odd operator, which is also a bound on the lightest $U(1)$ charged operator when \eqref{oddanomaly} is true. In particular, both $k$ and $\bar k$ can be nonzero, and the $U(1)$  can be neither holomorphic nor anti-holomorphic. 

This leads us to argue that {\it there is a bound on the lightest $U(1)$ charged operator if the symmetry is anomalous, but not otherwise.}  
We will leave the study for bounds on the lightest charged operator for a general $U(1)$ global symmetry for the future. 

Let us finally comment on the interpretation of our bounds from the weak gravity conjecture in  AdS$_3$/CFT$_2$ \cite{Montero:2016tif}.\footnote{We thank Clay Cordova  and Kantaro Ohmori for discussions on this point.}   
 The 't Hooft anomaly of a $U(1)$ current in an even-dimensional holographic CFT is captured by the level of the Chern-Simons term for the dual gauge field $a_\mu$ in AdS \cite{Witten:1998qj} (see also \cite{Kraus:2006wn} for the specific context of AdS$_3$/CFT$_2$).  
In the presence of a 3d Chern-Simons term, the gauge field acquires a mass, and there is no electric charge confinement.  In this case, there are finite-energy charged particles in the bulk, and the weak gravity conjecture applies.  
Indeed, we expect there to be a bound on the lightest charged operator in the CFT when the $U(1)$ is anomalous.  
The importance of the Chern-Simons term was already emphasized by \cite{Montero:2016tif} in the formulation of the weak gravity conjecture in AdS$_3$/CFT$_2$. 

In the absence of the Chern-Simons term, on the other hand, charged particles are confined.  
Consequently, the argument for the weak gravity conjecture does not apply to such a $U(1)$.  
This is perhaps consistent with our observation in the boundary CFT that there is no bound on the lightest charged $U(1)$ operator if the $U(1)$ is non-anomalous. 
However, this argument is not complete because there could be mixed Chern-Simons terms with other gauge fields in the bulk, rendering the photon massive.\footnote{In AdS$_3$/CFT$_2$, the most natural boundary condition for a Chern-Simons gauge theory is to hold $a_z$ fixed if the level is positive, and $a_{\bar z}$ fixed if the level is negative \cite{Kraus:2006wn}, giving rise to holomorphic and anti-holomorphic currents on the boundary, respectively.  
To couple to a non-holomorphic $U(1)$ symmetry, we can, for example, start with two $U(1)$ gauge fields in AdS$_3$ with a mixed Chern-Simons term $ {i N \over 4\pi }\int_{\text{AdS}_3}( adb+bda)$, 
and choose to hold both $a_z+\gamma b_z$ and $ a_{\bar z}-\gamma b_{\bar z}$ fixed on the boundary, with $\gamma$ any real number. For a generic irrational $\gamma$, the  boundary $U(1)$ symmetry is neither holomorphic nor anti-holomorphic.} 
We leave a more complete treatment for the future.

\section{Outlook}

Continuous global symmetries are typically associated to conserved currents, and some of their 't Hooft anomalies enter into the current correlation functions, and constrain the local operator data. 
Rather surprisingly, our investigation shows even  discrete 't Hooft anomalies place strong constraints on the light charged operator spectrum in 2d CFTs.  
There are several interesting open avenues for future study:
\begin{itemize}
\item Extend the analysis to general discrete and continuous internal global symmetries in two dimensions.
\item Even more generally, one can incorporate the non-invertible (``non-symmetry") topological defect lines \cite{Frohlich:2004ef,Frohlich:2006ch,Chang:2018iay} into the modular crossing equation, and study how the bounds on local operators depend on the associated fusion category. 
\item Generalize to spacetime symmetries  such as the time-reversal symmetry. 
\item Extend to fermionic theories which depend on the choice of the spin structure. 
 For example, the 't Hooft anomaly for an internal $\bZ_2$ symmetry in a 2d fermionic CFT is classified by $\bZ_8$.
\item Higher dimensional generalizations.
\end{itemize}

Let us ask whether our story generalizes to spacetime dimensions greater than two.  
The answer is negative for a  discrete, internal, global symmetry $G$ in a bosonic QFT.
As discussed in Section \ref{Sec:TQFT}, given such a $G$ and its anomaly $\A$ in $d>2$ spacetime dimensions, there is always a $d$-dimensional TQFT with a {\it unique vacuum} that carries this symmetry and anomaly \cite{Wang:2017loc,Tachikawa:2017gyf}.  Therefore, we can modify the anomaly of a $d$-dimensional QFT by taking its tensor product with the above TQFT, without changing the local operator data.  Hence, in $d>2$, there cannot possibly be an anomaly-dependent bound on the local operator spectrum for  discrete, internal, bosonic symmetries.

On the other hand, there are symmetries and  anomalies that cannot be carried by TQFTs.  For example, every continuous global symmetry is associated to a conserved current, which cannot exist in a TQFT.  It is therefore interesting to ask whether such an anomaly has non-trivial implications on the charged local operator spectrum.

\section*{Acknowledgements}

We thank N.\ Arkani-Hamed, N.\ Benjamin, C.\ Cordova, P.-S.\ Hsin,  Z.\ Komargodski, M. Levin,  K.\ Ohmori, P.\ Putrov, T.\ Rudelius, N.\ Seiberg, D\ Simmons-Duffin, J.\ Wang, E.\ Witten, and F.\ Yan for interesting discussions. YL is supported by the Sherman Fairchild Foundation, and by the U.S. Department of Energy, Office of Science, Office of High Energy Physics, under Award Number DE-SC0011632.
The work of  S.H.S.\ is supported  by the National Science Foundation grant PHY-1606531 and by the Roger Dashen Membership.

\appendix

\section{$c = 1$ Free Compact Boson}
\label{App:FreeBoson}

The moduli space of $c=1$ CFTs consists of two branches, the $S^1$ branch and the $S^1/\bZ_2$ branch, together with three isolated theories. This appendix reviews the $S^1$ branch, which has the description of the free compact boson $X(z, \bar z) = X_L(z) + X_R(\bar z)$ at radii $R \in \bR_{>0}$ (with $R$ and $1 \over R$ identified via T-duality). 
The free boson field is normalized such that $X(z,\bar z) X(0,0 )\sim -\frac 12 \log|z|^2$.  
In $\cal H$, the exponential operators are
\ie\label{expop}
{\cal O}_{n,w}(z,\bar z) =e^{i p_L X_L (z)+ i p_R X_R(\bar z)},
\fe
which are labeled by two integers, the momentum number $n$ and the winding number $w$:
\ie
 p_L = {n \over R} + wR, \quad p_R = {n \over R} - wR, \quad n, w \in \bZ \,.
\fe 
The conformal weights of ${\cal O}_{n,w}$ are
\ie\label{expweight}
(h, \bar h) = ( {p_L^2 \over 4}, {p_R^2 \over 4} ).
\fe

The global symmetry at a generic radius contains $( U(1)_n \times U(1)_w ) \rtimes \bZ_2$, where the last $\bZ_2$ acts as $X \to -X$. 
The $U(1)_n$ and $U(1)_w$ correspond to momentum and winding, which act by phases $e^{in\theta}$ and $e^{iw\theta}$ on the exponential operator \eqref{expop}, respectively.  
We will focus on the  $( \bZ_2^{(1,0)} \times \bZ_2 ^{(0,1)}) \times \bZ_2$  subgroup. 
More explicitly, the two $\bZ_2$ symmetries that are subgroups of the $U(1)$'s are simultaneous shifts in $X_L$ and $X_R$ such that the exponential operators have signs $e^{i\pi n}$ and $e^{i\pi w}$, respectively.  If we parameterize such a shift by
\ie
(X_L, X_R) \to (X_L, X_R) + (\ell_L, \ell_R),
\fe
then the condition for every exponential operator to have charge $\pm 1$ under this shift is ${\ell_L + \ell_R \over R}, \ (\ell_L - \ell_R) R \in \pi \bZ$. 
For generic $R$, this condition can be achieved if $(\ell_L, \ell_R)$ belongs to the lattice spanned by
\ie
v_1 = {\pi \over 2} ( R, R ), \quad v_2 = {\pi \over 2} ( {1 \over R}, - {1 \over R} ).
\fe
Let us denote the $\bZ_2$ generated by $m_1 v_1 + m_2 v_2$ as $\bZ_2^{(m_1,m_2)}$, with  $m_i=0,1$.  In particular, $\bZ_2^{(1,0)}$ and $\bZ_2^{(0,1)}$ are the momentum and winding $\bZ_2$, respectively. The topological line for $\bZ_2^{(m_1,m_2)}$ is\footnote{The extra sign in front of $\bar\partial X_R$ comes from $\oint ds^\mu j_\mu  = \oint dz j_z - \oint d\bar z \bar j_{\bar z}$.}
\ie
\exp\left[
  {i\over \pi} (m_1 v_1 + m_2 v_2)\cdot\left( \oint dz \partial X_L \, , \,
  - \oint d\bar z\bar \partial X_R\right)
\right]\,.
\fe

There is no 't Hooft anomaly for the momentum $\bZ_2^{(1,0)}$, nor for the winding $\bZ_2^{(0,1)}$ alone, but there is a mixed anomaly between the momentum $\bZ_2^{(1,0)}$ and the winding $\bZ_2^{(0,1)}$.  
This means that the defect Hilbert spaces of $\bZ_2^{(1,0)}$ have $\pm1$  charges under $\bZ_2^{(1,0)}$, but $\pm i$ charges under $\bZ_2^{(0,1)}$, and vice versa.

\paragraph{Defect Hilbert space ${\cal H}_{\cal L}$} 
The defect Hilbert space  states of $\bZ_2^{(m_1,m_2)}$ $(m_i=0,1)$ are given by
\ie\label{twexp}
\exp \left[ {i \over \pi} (s_1 v_1 + s_2 v_2) \cdot (X_L,- X_R) \right]
=\exp \left[ {i \over 2} \left(
(s_1R+ {s_2\over R} ) X_L 
-(s_1R- {s_2\over R} ) X_R 
\right) \right]\,,
\fe
with $s_1 = m_1 ~\text{mod}~2$ and $s_2 = m_2 ~\text{mod}~2$.  In particular, the defect Hilbert space ground states correspond to $s_1=m_1$ and $s_2=m_2$. 
The conformal weights of \eqref{twexp} is
\ie\label{twweight}
h = \frac {1}{16}  \left( s_1R+{s_2\over R} \right)^2\,,
\quad
\bar h = \frac {1}{16}  \left(s_1R-{s_2\over R} \right)^2\,.
\fe
Note that the spin is
\ie\label{c=1twgs}
h-\bar h  = {s_1s_2\over 4} \,.
\fe
This is consistent with the spin selection rule and the mixed anomaly between $\bZ_2^{(1,0)}$ and $\bZ_2^{(0,1)}$. The momentum $\bZ_2^{(1,0)}$ and winding $\bZ_2^{(0,1)}$ charges of the operator \eqref{twexp}  are
\begin{align}
\bZ_2^{(1,0)}:~~e^{\pm{ i \pi s_2\over2} }\,,
\quad
\bZ_2^{(0,1)}:~~e^{\pm{ i \pi s_1\over2} }\,.
\end{align}

To be completely explicit:
\begin{itemize}
\item The $\bZ_2^{(1,0)}$ defect Hilbert space ground states are
\ie
\exp\left[ \pm {i R \over 2} \left( X_L - X_R \right) \right],
\quad
(h, \bar h) = ( {R^2 \over 16}, {R^2 \over 16} ).
\fe
whose spin corroborates with the absence of an anomaly. Note that this defect Hilbert space ground state has $+1$ charge under $\bZ_2^{(1,0)}$, but $\pm i$ charges under $\bZ^{(0,1)}$, implying the mixed anomaly between the two $\bZ_2$ symmetries.
\item Similarly, the $\bZ_2^{(0,1)}$ defect Hilbert space ground states are
\ie
\exp\left[ \pm {i \over 2R} \left( X_L + X_R \right) \right],
\quad
(h, \bar h) = ( {1 \over 16R^2}, {1 \over 16R^2} ).
\fe
Note that this defect Hilbert space ground state has $+1$ charge under $\bZ_2^{(0,1)}$, but $\pm i$ charges under $\bZ^{(1,0)}$, implying the mixed anomaly between the two $\bZ_2$ symmetries.
\item The $\bZ_2^{(1,1)}$ defect Hilbert space ground states are
\ie
& \exp\left[ \pm \left( {i \over 2} (R + {1 \over R} ) X_L - {i \over 2} (R - {1 \over R} ) X_R \right) \right],
\quad
(h, \bar h) = ( { (R+1/R)^2 \over 16}, { (R-1/R)^2 \over 16} ),
\\
& \exp\left[ \pm \left( {i \over 2} (R - {1 \over R} ) X_L - {i \over 2} (R + {1 \over R} ) X_R \right) \right],
\quad
(h, \bar h) = ( { (R-1/R)^2 \over 16}, { (R+1/R)^2 \over 16} ),
\fe
satisfying the anomalous spin selection rule $s \in {\bZ \over 2} + {1\over4}$. 
\item When $R^2 \in \mathbb{N}$, there exist holomorphic currents in one of the four Hilbert spaces ${\cal H}, {\cal H}_{(1,0)}, {\cal H}_{(0,1)}, {\cal H}_{(1,1)}$. 
For example, there are currents with $(s_1, s_2) = \pm (1, R^2)$, giving rise to a pair of defect Hilbert space states with weight $({R^2 \over 4}, 0)$. 
These states are in ${\cal H}_{(1,0)}$ if $R^2$ is even, and in ${\cal H}_{(1,1)}$ if $R^2$ is odd. 
The above can be generalized to when $R^2$ is rational. 
\item Finally, for the $\bZ_2$ that acts as $X \to -X$, the defect Hilbert space ground states are two-fold degenerate, corresponding to the two fixed points, and have conformal weights $({1\over 16},{1\over 16})$.
\end{itemize}

\paragraph{Hilbert space of local operators ${\cal H}$}

We list the lightest $\bZ_2$ even/odd states in $\cal H$ with respect to various $\bZ_2$ symmetries below, to compare with the bootstrap bounds in Section \ref{Sec:c=1}. 
Without loss of generality, we assume $R \geq 1$.  
\begin{itemize}
\item With respect to the non-anomalous momentum $\bZ_2^{(1,0)}$, the $(n, w) = (\pm 2, 0), (0, \pm 1)$ states are even, while the $(n, w) = (\pm 1, 0), (\pm 1, \pm 1)$ states are odd. The maximal gap in the even sector is 1, realized at $R = \sqrt2$.
\item With respect to the non-anomalous winding $\bZ_2^{(0,1)}$, the $(n, w) = (\pm 2, 0), (\pm 1, 0)$ states are even, while the $(n, w) = (0, \pm 1), (\pm 1, \pm 1)$ states are odd. The maximal gap in the even sector is $1\over2$, realized at $R = 1$.
\item With respect to the anomalous $\bZ_2^{(1,1)}$, the $(n, w) = (\pm 2, 0), (\pm 1, \pm 1)$ states are even, while the $(n, w) = (\pm 1, 0), (0, \pm 1)$ states are odd. The maximal gap in the even sector is 2, realized at $R = 1$, where the $(n, w) = (\pm 1, \pm 1)$ states become conserved currents and are thus excluded from the definition of the gap. For $1 < R < 3^{1\over4}$, the states with the lowest scaling dimensions are the $(n, w) = (\pm 1, \pm 1)$ operators with scaling dimension ${ R^2 + 1/R^2 \over 2}$, while for $R > 3^{1\over4}$, they are the $(\pm 2, 0)$ scalars with scaling dimension $2 \over R^2$.
\end{itemize}

\section{WZW Models}

Let us consider the center symmetries in WZW models for simple Lie groups.  Their anomalies have been computed in \cite{Furuya:2015coa,Numasawa:2017crf}.  
We will consider the WZW models whose center contains a $\bZ_2$ subgroup. 

Let us start with some basics of the WZW models. 
We follow the conventions in \cite{DiFrancesco:1997nk}.
Let $\mathfrak{g}$ be  a simple Lie algebra and $r=\text{rank}(\mathfrak{g})$.  
The central charge of the $\mathfrak{g}$ WZW model at level $k$ is
\begin{align}
c= {k \dim \mathfrak{g}\over k+h^\vee}\,,
\end{align}
where $h^\vee$ is the dual Coxeter number.  
Given a positive integer level $k$, the  current algebra module is labeled by a weight $\lambda$  satisfying
\begin{align}
0 \le \sum_{i=1}^r a_i^\vee \lambda_i \le k\,,
\end{align}
where $a_i^\vee$ are the comarks.  Here $\lambda_i\in \bZ_{\ge 0}$ are the Dynkin labels of $\lambda$, and we write $\lambda= (\lambda_1,\lambda_2,\cdots,\lambda_r)$. 
The conformal weight of a current algebra primary of weight $\lambda$ is
\begin{align}
h_\lambda = {(\lambda, \lambda+2\rho )\over 2(k+h^\vee)}\,,
\end{align}
where $\rho= (1,1,\cdots,1)$ is the Weyl vector.  
We will only consider diagonal WZW models where the left and the right modules are identical for every  primary.  
The current algebra primary will be denoted by $|h_\lambda, h_\lambda\rangle$. 

We write the Dynkin labels of an affine weight $\hat \lambda$ as $\hat \lambda=  [\lambda_0 , \lambda_1 ,\cdots, \lambda_r]$. 
The affine fundamental weights are denoted by $\hat \omega_0= [ 1,0,\cdots ,0]$, $\hat \omega_1= [0 ,1,\cdots, 0]$, ..., $\hat \omega_r= [ 0,0,\cdots ,1]$,

\subsection{$\bZ_2$ Center Symmetry and its Anomaly}

We focus on the $\bZ_2$ center symmetry (if exists) of the diagonal WZW model for a simple Lie algebra $\mathfrak{g}$.  
The advantage of considering the $\bZ_2$ center symmetry is that it commutes with the left and the right current algebras, so many calculations can be done with the help of the current algebra $\hat{\mathfrak{g}}$.  The nontrivial centers of simple Lie algebras are
\ie
\begin{array}{c|c|c|c|c|c|c|c}
 \mathfrak{g} & A_{r} & B_r & C_r & D_{2n} & D_{2n+1} &    E_6 &  E_7 \\
 \hline
 \text{Center}& \bZ_{r+1}  &  \bZ_2 &  \bZ_2 &  \bZ_2\times \bZ_2& \bZ_4&\bZ_3&\bZ_2    
  \end{array}
\fe
while $G_2,F_4,E_8$ have no center.  Among the above, the centers of $A_{2n-1}, B_r,C_r , D_{r}, E_7$ have $\bZ_2$ subgroups.

Each element of the center group is associated to an outer automorphism of the affine Lie algebra $\hat {\mathfrak{g}}$.  
We list the outer automorphism $A$ associated to the generator of each $\bZ_2$ center subgroup in Table \ref{Table:WZWZ2}.  
The $\bZ_2$ acts on the current algebra primary $|h_\lambda,h_\lambda\rangle$ as
\ie
\bZ_2 :~ |h_\lambda ,h_\lambda \rangle \to e^{2\pi  i (A\hat\omega_0 ,\lambda)} |h_\lambda ,h_\lambda \rangle \,.
\fe

\begin{table}
\centering
\begin{tabular}{c|c|c|c}
$\mathfrak{g}$ & \text{Action of the $\bZ_2$ Outer Automorphism}  & $A\hat\omega_0$ & $\A$ \\
\hline
$A_{2n-1}$ &  $A[\lambda_0 ,\lambda_1,\cdots, \lambda_{2n-1} ] =  [\lambda_n , \lambda_{n+1}, \cdots, \lambda_{2n-1},\lambda_0,\lambda_1,\cdots,\lambda_{n-1}] $
 & $\hat\omega_{n} $  &  $e^{i \pi  k n }$\\
\hline
$B_r$ & $ A[\lambda_0 ,\lambda_1,\cdots, \lambda_{r} ]  = [\lambda_1, \lambda_0 ,\cdots,\lambda_{r-1},\lambda_r] $
& $\hat \omega_1$ & 1 \\
\hline
$C_r$ & $ A[\lambda_0 ,\lambda_1,\cdots, \lambda_{r} ]  = [\lambda_r,\lambda_{r-1} , \cdots,\lambda_1,\lambda_0]$ & $\hat\omega_r$ & $e^{i\pi k r}$ \\
\hline
$D_{r=2n\ge 4}$ & $A[\lambda_0 ,\lambda_1,\cdots, \lambda_{r} ]  = [\lambda_1, \lambda_0,\lambda_2 , \cdots,\lambda_r, \lambda_{r-1}]  $
& $\hat\omega_ 1 $ & 1\\
& $\widetilde A[\lambda_0 ,\lambda_1,\cdots, \lambda_{r} ] =  [\lambda_r ,\lambda_{r-1} ,\lambda_{r-2} ,\cdots,\lambda_1,\lambda_0]$
& $\hat\omega_r$   &  $e^{i\pi k n} $\\
& $A\widetilde A[\lambda_0 ,\lambda_1,\cdots, \lambda_{r} ]  = [\lambda_{2n-1}  ,\lambda_{2n} , \lambda_{2n-2},\cdots, \lambda_2,\lambda_0,\lambda_1]$
& $\hat\omega_{r-1}$  & $ e^{i\pi k n}$\\
\hline
$D_{r=2n+1}$ & $A[\lambda_0 ,\lambda_1,\cdots, \lambda_{r} ]  = [\lambda_1, \lambda_0 , \lambda_2,\cdots, \lambda_{r-2} ,\lambda_r ,\lambda_{r-1}] $ & $\hat\omega_1 $ &1\\
\hline
$E_7$ & $A[\lambda_0 ,\lambda_1,\cdots, \lambda_{7} ]  = [\lambda_6, \lambda_5,\lambda_4, \lambda_3,\lambda_2,\lambda_1,\lambda_0,\lambda_7]$ & $\hat\omega_6$ &$e^{i \pi k}$
\end{tabular}
\caption{The outer automorphisms associated to the $\bZ_2$ centers in simple affine Lie algebras. Note that the center $\bZ_2\times \bZ_2$ of $D_{2n+2}$ has three $\bZ_2$ subgroups, whose generators are $A,\widetilde A, A\widetilde A $.  We also list the 't Hooft anomaly for each $\bZ_2$ in the level $k$ WZW model.}\label{Table:WZWZ2}
\end{table}
 
The 't Hooft anomaly $\alpha$ of the $\bZ_2$ center subgroup in the level $k$ WZW model is computed in \cite{Numasawa:2017crf} as
\ie
\alpha =  e^{ 2\pi i k |A \hat\omega_0|^2} \,,
\fe
where $\hat\omega_0=[1,0,\cdots,0]$ is the 0-th affine fundamental weight.  We list the anomaly for each $\bZ_2$ in the rightmost column in Table \ref{Table:WZWZ2}.

\subsection{$A$ Series}

\paragraph{$\boldsymbol{\widehat{\mathfrak{su}(2)}_k}$}

As a warm-up, let us start with the diagonal $\widehat{\mathfrak{su}(2)}_k$ WZW model of central charge $c={3k\over k+2}$.   The current algebra primary $|j,j\rangle$ is labeled by the spin $j = {\lambda_1\over2} \in  {\bZ\over2}$, with $0\le j \le {k\over2}$.  
The scaling dimension of $|j,j\rangle$ is:
\begin{align}
\Delta_j  =  { 2j ( j+1) \over k+2}\,. 
\end{align}
The center $\bZ_2$ charge of $|j,j\rangle$ acts as
\begin{align}
\bZ_2:~|j,j\rangle \to (-1)^{2j}  |j,j\rangle\,.
\end{align}
The center $\bZ_2$ is anomalous if $k \in 2\mathbb{N}-1$, and non-anomalous otherwise.

We summarize the lightest $\bZ_2$ even/odd non-degenerate Virasoro primaries in each $\widehat{\mathfrak{su}(2)}_k$ WZW model.    
Degenerate primaries such as $J^a_{-1}|0,0\rangle$ are excluded  in accordance with our definition of the gap in Section \ref{Sec:General}.
\begin{itemize}
\item $\widehat{\mathfrak{su}(2)}_{k=1}~~(\A=-1)$
\ie
&\bZ_2 ~\text{even}:~~J_{-1}^a \bar J^b_{-1}|0,0\rangle\,,~~~~\Delta^+_{\rm gap}=2\,,\\
&\bZ_2~\text{odd}:~~ |j=\frac 12 , j=\frac 12\rangle\,,~~~~\Delta^-_{\rm gap } = \frac12\,.
\fe
This is the self-dual free compact boson discussed in Section \ref{Sec:SelfDualExample}. 
\item $\widehat{\mathfrak{su}(2)}_{k>1}~~(\A=  e^{i\pi k})$
\ie
&\bZ_2~\text{even}:~~ |j=1,j=1\rangle\,,~~~~\Delta^+_{\rm gap} = {4\over k+2}\,,\\
&\bZ_2~\text{odd}:~~ |j=\frac 12,j=\frac 12\rangle\,,~~~~\Delta^-_{\rm gap} = {3\over 2(k+2)}\,.
\fe
\end{itemize}

\paragraph{$\boldsymbol{\widehat{\mathfrak{su}(2n)}_k}$}

Let us move on to the $\widehat{\mathfrak{su}(2n)}_k$ WZW models with $n>1$. 
The central charge is  $c=  {(4n^2-1) k \over k+2n}$. 
The center of $SU(2n)$ is $\bZ_{2n}$.  
The generator of  $\bZ_{2n}$ acts on the current algebra primary $|h_\lambda,h_\lambda\rangle$ by the phase $\exp[{2\pi i\over 2n}\sum_{j=1}^{2n-1} j \lambda_j]$.  
It follows that the $\bZ_2$ subgroup acts by a phase
\ie
\bZ_2 :~|h_\lambda,h_\lambda \rangle \to e^{ i \pi \sum_{j=1}^{2n-1}j \lambda_j}|h_\lambda,h_\lambda\rangle\,.
\fe
From Table \ref{Table:WZWZ2}, the $\bZ_2$ is anomalous  if and only if both $k$ and $n$ are odd. 
For any $k$, the lightest $\bZ_2$ even and odd current algebra primaries are, respectively,
\ie\label{11}
(\A=e^{i\pi kn})~~~~
&\bZ_2~\text{even}:~~\lambda= (0,1,0,\cdots, 0) = \Lambda^2 \square\,,~~~
\Delta^+_{\rm gap}= {2(n-1)(2n+1)\over n(k+2n)}\,,\\
&\bZ_2~\text{odd}:~~~\lambda =(1,0,0,\cdots,0) = \square\,,~~~\Delta^-_{\rm gap} = { 4n^2-1\over 2n(k+2n)}\,.
\fe
Note that the scaling dimension $\Delta$ of the current algebra primary $\Lambda^2\square$ is always lighter than 2 (of $J_{-1}^a \bar J_{-1}^b |0,0\rangle$), so it is also the lightest $\bZ_2$ even Virasoro primary. 
Also, the lightest $\bZ_2$ odd current algebra primary is trivially the lightest odd Virasoro primary, because the generators of the current algebra are $\bZ_2$ even. The same applies to all the other WZW models in the later subsections.

\subsection{$B$ Series}

The $B_{r}=\widehat{\mathfrak{so}(2r+1)}$ WZW model at level $k$ has central charge $c={ k (2r^2+r)\over k+2r-1}$.  The $\bZ_2$ center is always non-anomalous.  The lightest even/odd current algebra primaries are
\ie
(\A=+1)~~~~
&\bZ_2~\text{even}:~~\lambda= (1,0,\cdots,0)\,,~~~\Delta^+_{\rm gap}= {2r\over k+2r-1} \,,\\
&\bZ_2~\text{odd}:~~\lambda = (0,\cdots,0,1)\,,~~~\Delta^-_{\rm gap}= {2r^2+r \over 4(k+2r-1)}\,.
\fe

\paragraph{The free fermions}

The $\widehat{\mathfrak{so}(2r+1)}$ WZW model  at level 1 with $c={2r+1\over2}$ can be described as $2r+1$ free Majorana fermions summed over spin structures, so that we end up with a bosonic theory. 
There are three current algebra primaries labeled by the affine weights $\hat\omega_0$, $\hat\omega_1$, and $\hat\omega_r$. Their conformal weights are  $h_{\hat\omega_0}=0,~~h_{\hat\omega_1}= \frac12,~~ h_{\hat\omega_r}={2r+1\over 16}$.  
Their characters are
\ie
&\chi_{\hat \omega_0}   = \frac 12 \left( {\theta_3^{r+1/2} + \theta_4^{r+1/2}\over\eta^{r+1/2}} \right)\,,~~~\chi_{\hat \omega_1} =  \frac 12 \left( {\theta_3^{r+1/2} - \theta_4^{r+1/2}\over\eta^{r+1/2}} \right)\,,~~~\chi_{\hat \omega_{r}} =  {1\over\sqrt{2}}  {\theta_2^{r+1/2} \over\eta^{r+1/2}} \,,
\fe
where $\theta_i = \theta_i(0|\tau)$.  When $r=0$, we recover the Ising characters \eqref{IsingCharacter}.  The $r=1$ theory is the $\widehat{\mathfrak{su}(2)}_2$ WZW model. 
Under the modular $S$ transformation,  $\theta_3/\eta \to \theta_3/\eta$, $\theta_2/\eta \leftrightarrow \theta_4/\eta$.

The even and odd current algebra primaries under the center $\bZ_2$ are
\ie
&{\cal H}^+:~~|0,0\rangle\,,~~|\frac12, \frac 12\rangle\,,~~~~{\cal H}^- :~~|{2r+1\over 16} , {2r+1\over 16}\rangle\,.
\fe
The torus partition function  with the insertion of a spatial $\bZ_2$ line is
\ie
Z^{\cal L} (\tau ,\bar\tau ) = |\chi_{\hat\omega_0}(\tau)|^2 +|\chi_{\hat\omega_1}(\tau)|^2 -|\chi_{\hat\omega_r}(\tau)|^2 \,.
\fe
Applying $S$, we derive the defect Hilbert space partition function
\ie
Z_{\cal L}(\tau,\bar\tau) = \chi_{\hat\omega_0}(\tau) \chi_{\hat\omega_1}(\bar\tau)+ \chi_{\hat\omega_1}(\tau) \chi_{\hat\omega_0}(\bar\tau)+| \chi_{\hat\omega_r}(\tau) |^2\,.
\fe
Hence, the current algebra primaries in the defect Hilbert space are
\ie
{\cal H}_{\cal L}:~~|0,\frac 12\rangle\,,~~|\frac 12,0\rangle\,,~~|{2r+1\over 16},{2r+1\over 16}\rangle\,.
\fe
In the Ising model ($r=1$), these are the free chiral fermions and the disorder operator $\mu(x)$ (See Section \ref{Sec:IsingExample}).  The lightest {\it non-degenerate Virasoro} primary in the defect Hilbert space is the disorder operator with  $\Delta={2r+1\over 8}$, if $r\le 5$, or the current algebra descendant $J_{-1}^a |0,\frac 12\rangle$ with $\Delta={3\over2}$ if $r\ge 6$.  Considered together with the $\bZ_2$ odd sector ${\cal H}^-$, we conclude that
\ie
(\A=+1)~~~~\Delta^{\rm ord/dis} _{\rm gap} = \begin{cases}
&{2r+1\over 8}\,,~~\text{if}~~1\le r\le5\,,\\
&\frac32 \,,~~~~\text{if}~~r\ge 6\,.
\end{cases}
\fe
On the other hand, $\Delta^+_{\rm gap}=1$.

\subsection{$C$ Series}

The $C_{r}=\widehat{\mathfrak{sp}(2r)}$ WZW model at level $k$ has central charge $c={ k (2r^2+r)\over  k+r+1}$.  The $\bZ_2$ center is anomalous  if and only if both $k$ and $r$ are odd.  The lightest even/odd current algebra primaries are
\ie
(\A=e^{i \pi k r})~~~~
&\bZ_2~\text{even}:~~\lambda= (0,1,0,\cdots,0)\,,~~~\Delta^+_{\rm gap}= {2r\over k+r+1} \,,\\
&\bZ_2~\text{odd}:~~\lambda = (1,0,\cdots,0,0)\,,~~~\Delta^-_{\rm gap}= {2r+1 \over 2(k+r+1)}\,.
\fe

\subsection{$D$ Series}

For the $D_{2n}$ WZW models, there are three $\bZ_2$ subgroups of the center. 
We focus on the first one, denoted by $A$ in Table \ref{Table:WZWZ2}.  
This $\bZ_2$, which is always non-anomalous $\A=+1$, can be uniformly discussed in both the $D_{2n+1}$ and the $D_{2n}$ cases. 
We will return to the other two $\bZ_2$ subgroups in the $D_{2n}$ WZW models in the special case of $k=1$.

The $D_{r}= \widehat{\mathfrak{so}(2r)}$ WZW model at level $k$ has central charge $c= {k(2r^2-r)  \over k+2r-2}$.  The lightest even/odd current algebra primaries under the $\bZ_2$ specified above  are
\ie
(\A=+1)~~~~
&\bZ_2~\text{even}:~~\lambda= (1,0,\cdots,0)\,,~~~\Delta^+_{\rm gap}= {2r-1\over k+2r-2} \,,\\
&\bZ_2~\text{odd}:~~\lambda = (0,\cdots,0,0,1), (0,\cdots,0,1,0)\,,~~~\Delta^-_{\rm gap}= {r (2r-1) \over 4(k+2r-2)}\,.
\fe

\paragraph{The free fermions}
The $(D_{r})_1=\widehat{\mathfrak{so}(2r)}_1$ WZW model with $c=r$ can be described as $2r$  free Majorana fermions summed over spin structures. 
There are four modules in the WZW model, corresponding to the affine weights $\hat \omega_0$, $\hat\omega_1$, $\hat\omega_{r-1}$, and $\hat\omega_r$. Their conformal weights $h$ are $h_{\hat\omega_0} =0\,,~~~h_{\hat \omega_1} = \frac 12\,,~~~ h_{\hat\omega_{r-1}} = h _{\hat \omega_r} = {r\over 8}$.  
  Their characters are
\ie
&\chi_{\hat \omega_0}   = \frac 12 \left( {\theta_3^r + \theta_4^r\over\eta^r} \right)\,,~~\chi_{\hat \omega_1} =  \frac 12 \left( {\theta_3^r - \theta_4^r\over\eta^r} \right)\,,~~
\chi\equiv \chi_{\hat \omega_{r-1}} = \chi_{\hat \omega_r}  =  \frac 12  {\theta_2^r \over\eta^r} \,.
\fe

There is a $\bZ_2$ symmetry that commutes with the current algebra which is always non-anomalous. The modules $\hat\omega_0$ and $\hat\omega_1$ are even under this $\bZ_2$, while the modules $\hat\omega_{r-1}$ and $\hat\omega_r$ are odd. That is
\ie
{\cal H}^+:~ |0,0\rangle,~|\frac 12,\frac 12\rangle\,,~~~~~{\cal H}^-:~ 2|\frac r8 ,\frac r8\rangle\,.
\fe 
The torus partition function $Z^{\cal L}$ is
\ie
Z^{\cal L} (\tau) = |\chi_{\hat\omega_0}(\tau)|^2  + |\chi_{\hat\omega_1}(\tau)|^2 -2|\chi(\tau)|^2  \,.
\fe
Under $S$, we obtain the defect Hilbert space partition function
\ie
Z_{\cal L}(\tau,\bar \tau) =  \chi_{\hat\omega_0} (\tau)\chi_{\hat \omega_1}(\bar\tau)+ \chi_{\hat\omega_1} (\tau)\chi_{\hat \omega_0}(\bar\tau)
+2  \left| 
\chi(\tau)
\right|^2\,.
\fe
The defect Hilbert space spectrum is
\ie
{\cal H}_{\cal L}:~|0,\frac 12\rangle, ~|\frac 12,0\rangle,~ 2 |\frac r 8, \frac r8\rangle\,.
\fe

The lightest {\it non-degenerate} {\it Virasoro} primary in the defect Hilbert space ${\cal H}_{\cal L}$ is either the current algebra descendant $J_{-1}^a |0,\frac 12\rangle$, or the current algebra primary $|\frac r8,\frac r8\rangle$, {\it i.e.} $(\Delta_{\rm gap})_{\cal L}= {r\over4}$ if $r\le 6$ and  $(\Delta_{\rm gap})_{\cal L}= {3\over2}$ if $r\ge 6$.  The gap in the odd sector ${\cal H}^-$, on the other hand, is always $r \over 4$. Hence the gap for this non-anomalous $\bZ_2$ int he $c=r$ $\widehat{\mathfrak{so}(2r)_1}$ WZW model is
\ie
(\A=+1)~~~~
\Delta_{\rm gap}^{\rm ord/dis} = \begin{cases}
\frac r4\,,~~~\text{if}~~1\le r\le 6\,,\\
\frac 32 \,,~~~\text{if}~~ 6 \le r\,,
\end{cases}
\fe
and $\Delta^+_{\rm gap} = 1$.   This explains the kink in Figure \ref{Fig:NonanomalousGeneral}.

For $\widehat{\mathfrak{so}(2r)}_k$ with $r=2n$, there are two other $\bZ_2$ subgroups denoted by $\widetilde A$ and $A\widetilde A$ in Table \ref{Table:WZWZ2}.  Let us discuss the second one $\widetilde A$.  The anomalies and gaps are the same for the third one $A\widetilde A$.  We will denote the $\bZ_2$ associated to $\widetilde A$ by $\widetilde {\bZ_2}$, which is anomalous if and only if both $k$ and $n$ are odd, {\it i.e.} $\A = e^{i\pi k n}$. 

In $\widehat{\mathfrak{so}(4n)}_1$, the $\widetilde \bZ_2$ acts on the current algebra primaries as
\ie
\widetilde \bZ_2:~~&|\frac 12,\frac12 \rangle \to -|\frac 12,\frac 12\rangle\,,~~~\\
&|h_{\hat \omega_{2n-1}},h_{\hat \omega_{2n-1}}\rangle\to (-1)^{n-1}|h_{\hat \omega_{2n-1}},h_{\hat \omega_{2n-1}}\rangle\,,~~\\
&
|h_{\hat \omega_{2n}},h_{\hat \omega_{2n}}\rangle\to (-1)^{n}|h_{\hat \omega_{2n}},h_{\hat \omega_{2n}}\rangle  \,.
\fe
The torus partition function with a spatial $\widetilde \bZ_2$ line inserted is $Z^{\widetilde {\cal L}} = |\chi_{\hat\omega_0}|^2 - |\chi_{\hat\omega_1}|^2$. 
The defect Hilbert space partition function is
\ie
Z_{\widetilde {\cal L}} (\tau ,\bar \tau) = \chi_{\hat\omega_0}(\tau) \chi(\bar\tau) +\chi(\tau)\chi_{\hat\omega_0}(\bar\tau)
+\chi_{\hat\omega_1}(\tau) \chi(\bar\tau) +\chi(\tau)\chi_{\hat\omega_1}(\bar\tau)\,.
\fe
Hence, the defect Hilbert space spectrum is 
\ie
{\cal H}_{\widetilde {\cal L}}:~|0,\frac n4\rangle,~~|\frac n4,0\rangle,~~
|\frac 12, \frac n4\rangle,~~|\frac n4,\frac 12\rangle\,.
\fe
According to the spin selection rule \eqref{spinselection}, $\widetilde \bZ_2 $ is anomalous if $n$ is odd (here $k=1$).

The various gaps for non-degenerate Virasoro primaries  
of the $c=2n$ $\widehat{\mathfrak{so}(4n)}_1$ WZW model for $\widetilde \bZ_2$ are
\ie
(\A=e^{i\pi n })~~
\widetilde{\Delta}^+_{\rm gap} = \text{Min}(2, \frac n2)\,,~~
\widetilde{\Delta}^-_{\rm gap}= \text{Min}(  1 , \frac n2) \,,~~
\widetilde{\Delta}^{\rm ord/dis}_{\rm gap} =  \text{Min}(1, {n+2\over4}, \frac n2 )\,.
\fe

\subsection{$E_7$}
\label{App:E7}

Finally, the $\widehat{\mathfrak{e}_7}$ WZW model at level $k$ has $c={133k \over k+18}$.  The center $\bZ_2$ is anomalous if and only if $k$ is odd, {\it i.e.} $\A = e^{i\pi k}$. 

 At level 1, there are two current algebra modules, the vacuum $|0,0\rangle$ and $|\frac 34,\frac 34\rangle$ with weight $\lambda = (0,0,0,0,0,1,0)$.  The latter is odd under the anomalous center $\bZ_2$.  By the spin selection rule \eqref{spinselection}, there cannot be any spin 0 operator in ${\cal H}_{\cal L}$, hence the only states in the defect Hilbert space are $|0,\frac 34\rangle$ and $|\frac 34,0\rangle$, which are degenerate.   The gaps for non-degenerate Virasoro primaries are
 \ie
(\A=-1)~~~ k=1:~~\Delta^+_{\rm gap} =2\,,~~~\Delta^-_{\rm gap} = \frac 32\,,~~~\Delta_{\rm gap}^{\rm ord/dis} = \frac 32\,,
 \fe
where the operator corresponding to $\Delta^+_{\rm gap} = 2$ is $J_{-1}^a\bar J_{-1}^b|0,0\rangle$. 

At higher level $k>2$, the lightest even and odd current algebra primaries are
\ie
(\A=e^{i\pi k})~~~~
&\bZ_2~\text{even}:~~\lambda = (1000000)\,,~~~\Delta^+ _{\rm gap} ={36\over k+18}\,,\\
&\bZ_2~\text{odd}:~~\lambda= (0000010)\,,~~~
\Delta^-_{\rm gap}=  {57\over2(k+18)} \,.
\fe

\subsection{(Almost) Saturating Examples}\label{App:Saturate}

There are many WZW models that saturate or almost saturate the numerical bootstrap bounds we present in Section \ref{Sec:General}.

\paragraph{Non-Anomalous $\Delta^+_{\rm gap}$}

\begin{itemize}
\item $c=1$:~~Ising$^2$,~~$\Delta^+_{\rm gap} =1$.
\item $c=\frac 32$:~~$\widehat{\mathfrak{so}(3)}_1 = \widehat{\mathfrak{su}(2)}_2$,~~$\Delta^+_{\rm gap}=1$.
\item $c=2$:~~$\widehat{\mathfrak{so}(4)}_1$,~~$\Delta^+_{\rm gap} =1$.
\item $c=\frac 52$:~~$\widehat{\mathfrak{so}(5)}_1=(B_2)_1$,~~$\Delta^+_{\rm gap} = 1$.
\item $c=8$:~~$\widehat{\mathfrak{so}(16)}_1=(D_8)_1$,~~$\widetilde{\Delta}^+_{\rm gap}=2$.\footnote{The tilde sign over $\Delta$ means that it is the gap with respect to the $\widetilde \bZ_2$ symmetry in the $D_8$ WZW model. }
\end{itemize}

\paragraph{Non-Anomalous $\Delta_{\rm gap}^{\rm ord/dis}$}

\begin{itemize}
\item $c = \frac n2 ,~n\in \mathbb{N},~1\le n\le 12$:~~$\widehat{\mathfrak{so}(n)}_1$,~~$\Delta_{\rm gap}^{\rm ord/dis}=  \frac n8$.
\end{itemize}

\paragraph{Anomalous $\Delta^+_{\rm gap}$}
\begin{itemize}
\item $c=1$:~~$\widehat{\mathfrak{su}(2)}_1$,~~$\Delta^+_{\rm gap}=2$.
\item $c={21\over5}$:~~ $\widehat{\mathfrak{sp}(6)}_1$=$(C_3)_1$, ~~$\Delta^+_{\rm gap} =\frac 65 $.
\item $c=5$:~~$\widehat{\mathfrak{su}(6)}_1=(A_5)_1$,~~$\Delta^+_{\rm gap}=\frac 43$.
\item $c=7$:~ ~$(E_7)_1$,~~$\Delta^+_{\rm gap} = 2$.
\end{itemize}

\paragraph{Anomalous $\Delta^-_{\rm gap}$}

\begin{itemize}
\item 
$c=1$:~~ $\widehat{\mathfrak {su}(2)}_1$,~~$\Delta^-_{\rm gap}=\frac 12$.
\end{itemize}

\section{More Examples of $c>1$ CFTs with $\bZ_2$ Symmetry}

\subsection{Monster CFT}

The (holomorphic) $c_L=24$ and $c_R=0$ Monster CFT has two non-anomalous $\bZ_2$ symmetries, usually denoted by $\bZ_{2A}$ and $\bZ_{2B}$.  Their $Z^{\cal L}$ are
\ie\label{monster}
Z^{2A}  (q) &=  {\eta(\tau )^{24}\over\eta(2\tau)^{24} } +2^{12} {\eta(2\tau )^{24}\over\eta(\tau)^{24} }  +24
=  {1\over q}  +4372 q+ 96256 q^2 +1240002q^3+{\cal O}(q^4)\,,\\
Z^{2B} (q)& =  {\eta(\tau )^{24}\over\eta(2\tau)^{24} }   +24 
= \frac 1q +276 q -2048 q^2 + 11202q^3 +{\cal O}(q^4)\,.
\fe
In $\cal H$, the $\bZ_2$ even and odd gaps are both $h_{\rm gap}^\pm  =  2$, for either $\bZ_{2A}$ or $\bZ_{2B}$.  

The modular $S$ transforms of \eqref{monster} are the defect Hilbert space partition functions $Z_{\cal L}$:
\begin{align}
Z_{2A} (q) &= 
2 ^{12} {\eta(\tau)^{24} \over \eta(\tau/2) ^{24}  }  +{\eta(\tau/2) ^{24}\over \eta(\tau)^{24}} +24 = {1\over q^{1/2} }
+4372 q^{1/2}  + 96256  q  + 1240002 q^{3/2} +{\cal O}(q^2)\,,\notag\\
Z_{2B}(q) &=
2^{12} {\eta(\tau)^{24} \over \eta(\tau/2)^{24}} +24
\notag\\
&= 24 + 4096 q^{1/2} +98304 q +1228800 q^{3/2} +10747904 q^2 +{\cal O}(q^{5/2})\,.
\end{align}
The gaps in the defect Hilbert space are $(h_{\rm gap })_{\cal L}={1\over2}$ for $\bZ_{2A}$, and $(h_{\rm gap})_{\cal L} = 1$ for $\bZ_{2B}$. 

We can take the tensor product of a holomorphic and an anti-holomorphic Monster CFT, and compare its gaps with our bootstrap bounds.   
We find that they are well within our numerical bounds in Section \ref{Sec:General}.

\subsection{Tensor Product Theories} 
Consider the tensor product of two copies of the same CFT. There is a non-anomalous $\bZ_2$ exchange symmetry. 
The defect Hilbert space ground state has weight $(h, \bar h) = ( {c \over 32}, {c \over 32} )$, where $c$ is the central charge of the product theory. 
Hence the universal bound on $\Delta_{\rm gap}^{\rm ord/dis}$ for a non-anomalous $\bZ_2$ cannot be stronger than  $c \over 16$.  This is indeed consistent with Figure \ref{Fig:NonanomalousGeneral}.

\bibliographystyle{JHEP}
\bibliography{z2}

\end{document}